# Physical-Layer Network Coding: Tutorial, Survey, and Beyond

Soung Chang Liew[*]    Shengli Zhang[**]    Lu Lu[*]

[*] Department of Information Engineering, The Chinese University of Hong Kong

[**] Department of Communication Engineering, Shenzhen University, China

email: {soung, ll007}@ie.cuhk.edu.hk; zsl@szu.edu.cn

**Abstract**

The concept of physical-layer network coding (PNC) was proposed in 2006 for application in wireless networks. Since then it has developed into a subfield of network coding with wide followings. The basic idea of PNC is to exploit the network coding operation that occurs naturally when electromagnetic (EM) waves are superimposed on one another. This simple idea turns out to have profound and fundamental ramifications. Subsequent works by various researchers have led to many new results in the domains of 1) wireless communication; 2) wireless information theory; and 3) wireless networking. The purpose of this paper is fourfold. First, we give a brief tutorial on the basic concept of PNC. Second, we survey and discuss recent key results in the three aforementioned areas. Third, we examine a critical issue in PNC: synchronization. It has been a common belief that PNC requires tight synchronization. Our recent results suggest, however, that PNC may actually benefit from asynchrony. Fourth, we propose that PNC is not just for wireless networks; it can also be useful in optical networks. We provide an example showing that the throughput of a passive optical network (PON) could potentially be raised by 100% with PNC.

## 1. Introduction

The concept of physical-layer network coding (PNC) was originally proposed in [1] in 2006 as a way to exploit the network coding operation [2] [3] that occurs naturally in superimposed electromagnetic (EM) waves. It is a simple fact in physics that when multiple EM waves come together within the same physical space, they add. This mixing of EM waves is a form of network coding, performed by nature.

In many wireless communication networks today, interference is treated as a destructive phenomenon. When multiple transmitters transmit radio waves to their respective receivers, a receiver receives signals from its transmitter as well as from other transmitters. The radio waves from the other transmitters are often treated as interferences that corrupt the intended signal. In Wi-Fi networks, for example, when multiple nodes transmit together, packet collisions occur and none of the packets can be received correctly.

As originally proposed in [1], PNC was an attempt to turn the situation around. By exploiting the network coding operation performed by nature, the "interference" could be put to good use. In a two-way relay channel (TWRC), for example, PNC can boost the system throughput by 100% [1].

The same idea as PNC for application in TWRC was also independently proposed in [4]. Ref. [1], however, went beyond TWRC to discuss the application of PNC in general network topologies. In addition, the implications of PNC for MAC (medium access control) protocols and network-layer designs were also discussed in [1]. The potential benefit of network coding taking into account the characteristics of the MAC (multiple-access channel) was investigated in [5] from the information-theoretic point of view. However, unidirectional multicast communication was the focus in [5], whereas bidirectional unicast communication was the focus in [1] and [4].

Since 2006, many researchers have made contributions that advance the understanding of PNC. The flavors of the research fall into three general categories: 1) communication-theoretic; 2) information-theoretic; and 3) networking-theoretic. The purpose of the present paper is fourfold. First, we give a brief tutorial on the basic concept of PNC. Second, we survey and discuss recent key results on the above three fronts. Third, we examine a critical issue in PNC: synchronization. It has been a common belief that PNC requires tight synchronization. We present some recent results suggesting that PNC may actually benefit from asynchrony. Fourth, we put forth the idea of "optical PNC". We provide an example showing that the throughput of a passive optical network (PON) could potentially be raised by 100% with optical PNC.

The remainder of this paper is organized as follows. Section 2 is a brief tutorial introducing the basic concept of PNC and the various relevant issues. Section 3 goes into the details of communication-theoretic studies of PNC. Results on asynchronous PNC and channel-coded PNC are discussed. Section 4 overviews some information-theoretic results of PNC and examines their implications. Section 5 considers MAC and network layer issues arising from PNC. In Section 6, we propose the idea of optical PNC. We conclude this paper in Section 7 by presenting our view on what future directions are worthwhile for PNC research.

## 2. A Brief Tutorial of PNC

The concept of PNC can be most easily illustrated with TWRC. TWRC is a three-node linear network in which two end nodes, nodes 1 and 2, want to communicate via a relay node $R$. There is no direct signal path between nodes 1 and 2. An example is a satellite network in which nodes 1 and 2 are the ground stations, and the relay $R$ is the satellite.

The half-duplex constraint is often imposed on wireless communication systems to ease engineering design. With the half-duplex constraint, a node cannot transmit and receive at the same time. With the half-duplex constraint, the relay in TWRC cannot receive from node 1 or node 2 and transmit to them at the same time. This means that each packet from node 1 to node 2 (and similarly, each packet from node 2 to node 1) must use up at least two time slots to reach its destination. Thus, the best possible packet exchange throughput is two packets for every two slots, one in each direction. That is, 1/2 packet per time slot per direction.

In the following, we examine the number of time slots needed for nodes 1 and 2 to exchange one packet with each other in various systems. In particular, we show that PNC can achieve the upper bound of 1/2 packet per time slot per direction.

**2.1. Non-network-coded Scheme (TS)**

Without the use of network coding, and with a design principle that tries to avoid interference, a total of four time slots are needed to exchange two packets, one in each direction. This is illustrated in Fig. 1. In this paper, we will simply refer to this non-network-coded scheme as the traditional scheme (TS). In time slot 1, node 1 transmits a packet $S_1$ to relay $R$; in time slot 2, relay $R$ forwards $S_1$ to node 2; in time slot 3, node 2 transmits a packet $S_2$ to relay $R$; and in time slot 4, relay $R$ forwards $S_2$ to node 1.

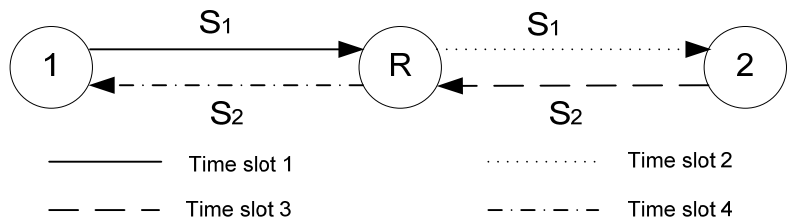

Fig. 1. Traditional non-network-coded scheme (TS).

**2.2. Non-physical-layer Network Coding Scheme (SNC)**

A straightforward way of applying network coding can reduce the number of time slots to three [6] [7]. We shall refer to this non-physical-layer network coding scheme simply as straightforward network coding (SNC)[1]. By reducing the number of time slots from four to three, SNC has a throughput improvement of 33% over TS.

Fig. 2 illustrates the idea of SNC. In time slot 1, node 1 transmits $S_1$ to relay $R$; and in time slot 2, node 2 transmits $S_2$ to relay $R$. After receiving $S_1$ and $S_2$, relay $R$ forms a network-coded packet $S_R$ as follows:

$$S_R = S_1 \oplus S_2 \qquad (1)$$

where $\oplus$ denotes the pairwise application of symbol-by-symbol XOR over $S_1$ and $S_2$. That is, if $S_1 = (a_1[1] + jb_1[1], \ldots, a_1[M] + jb_1[M])$, $S_2 = (a_2[1] + jb_2[1], \ldots, a_2[M] + jb_2[M])$, then $S_R = ((a_1[1] \oplus a_2[1]) + j(b_1[1] \oplus b_2[1]), \ldots, (a_1[M] \oplus a_2[M]) + j(b_1[M] \oplus b_2[M]))$, where $M$ is the number of symbols in a packet.

In time slot 3, relay $R$ broadcasts $S_R$ to both nodes 1 and 2. When node 1 receives $S_R$, it extracts $S_2$ from $S_R$ using the self information $S_1$ as follows:

$$S_1 \oplus S_R = S_1 \oplus (S_1 \oplus S_2) = S_2 \qquad (2)$$

Likewise, node 2 extracts $S_1$ from $S_2 \oplus S_R$.

Note that as with TS, SNC also tries to avoid simultaneous transmissions. That is, each node still transmits in a different time slot. Network coding is performed by the relay after receptions of the packets from nodes 1 and 2 in different time slots.

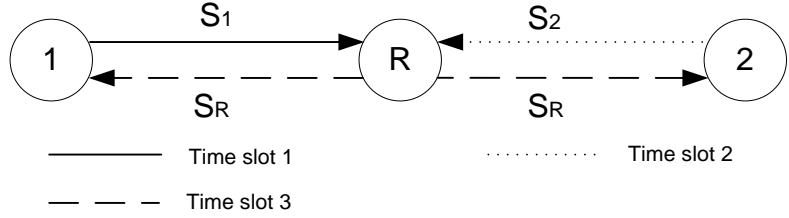

Fig. 2. Straightforward network coding scheme (SNC).

### 2.3. Physical-layer Network Coding Scheme (PNC)

PNC further reduces the number of time slots to two. It allows nodes 1 and 2 to transmit together and exploits the network coding operation performed by nature in the superimposed EM waves. By doing so, PNC can improve the performance of TS by 100%.

Fig. 3 illustrates the idea. In the first time slot, nodes 1 and 2 transmit $S_1$ and $S_2$ simultaneously to relay $R$. Based on the superimposed EM waves that carry $S_1$ and $S_2$, relay $R$ deduces $S_R = S_1 \oplus S_2$. Then, in the second time slot, relay $R$ broadcasts $S_R$ to nodes 1 and 2.

---

[1] This scheme is also called symbol-level network coding in some papers, although strictly speaking, many variants of PNC actually operate at the symbol levels. The main difference between PNC and SNC is whether the network coding operation occurs at the physical layer or at a higher layer.

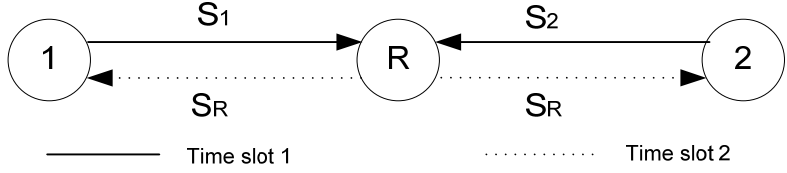

Fig. 3. Physical-layer network coding (PNC).

A key issue in PNC is how relay $R$ deduces $S_R = S_1 \oplus S_2$ from the superimposed EM waves. We refer to this process as "PNC mapping". More generally, PNC mapping refers to the process of mapping the received superimposed EM waves plus noise to some output packet for forwarding by the relay. PNC mapping could output a packet in a different form than $S_R = S_1 \oplus S_2$. Section 2.4 will discuss these other possibilities. All PNC mappings share the key requirement that nodes 1 and 2 must be able to deduce the information from the other node based on the output packet of relay $R$ and their self information.

For the discussion in this section, let us assume PNC mapping of $S_R = S_1 \oplus S_2$. In addition, all nodes use QPSK modulation for the transmitted signal. For simplicity, we further assume symbol-level and carrier-phase synchronization, and the use of power control, so that the packets from nodes 1 and 2 arrive at relay $R$ with the same phase and amplitude. We ignore noise in our simplified presentation for the time being.

In this paper, we use the uppercase letter to denote a packet and the corresponding lower-case letter to denote a symbol within the packet. For example, $S_1$ is a packet, and $s_1$ is a symbol within the packet.

Consider one particular symbol period. Suppose that nodes 1 and 2 modulate their symbols on RF frequency $\omega$, so that node $i$ send $\mathrm{Re}[(a_i + jb_i)e^{j\omega t}]$. The combined bandpass signal received by $R$ during one symbol period is

$$\begin{aligned}
y_R(t) &= s_1(t) + s_2(t) \\
&= [a_1 \cos(\omega t) - b_1 \sin(\omega t)] + [a_2 \cos(\omega t) - b_2 \sin(\omega t)] \\
&= (a_1 + a_2)\cos(\omega t) - (b_1 + b_2)\sin(\omega t)
\end{aligned} \qquad (3)$$

where $s_i(t)$, $i \in \{1,2\}$, is the bandpass signal transmitted by node $i$; and $a_i \in \{-1,1\}$ and $b_i \in \{-1,1\}$ are the corresponding QPSK modulated information bits. Note that for QPSK, $a_i = 1$ corresponds to bit 0, and $a_i = -1$ corresponds to bit 1, in the in-phase signal; likewise for $b_i$ in the quadrature-phase signal. With this definition, XOR becomes arithmetic multiplication: i.e., $a_1 \oplus a_2 \triangleq a_1 a_2$ and $b_1 \oplus b_2 \triangleq b_1 b_2$.

The baseband in-phase ($I$) and quadrature ($Q$) components corresponding to (3) are

$$\begin{aligned}
y_R^{(I)} &= a_1 + a_2 \\
y_R^{(Q)} &= b_1 + b_2
\end{aligned} \qquad (4)$$

Note that relay $R$ cannot extract the individual information symbols transmitted by nodes 1 and 2 from (4). This is because $y_R^{(I)}$ and $y_R^{(Q)}$ in (4) gives us two equations; however, there are four unknowns: $a_1$, $b_1$, $a_2$ and $b_2$.

In PNC, however, relay $R$ does not need the individual values of the four unknowns; it only needs to derive the two values, $a_1 \oplus a_2$ and $b_1 \oplus b_2$, to produce the PNC mapping $s_R \triangleq a_1 \oplus a_2 + j(b_1 \oplus b_2) \triangleq a_R + jb_R$. In particular, $a_1 \oplus a_2$ and $b_1 \oplus b_2$ can be derived from $y_R^{(I)}$ and $y_R^{(Q)}$. That is, we can find a PNC mapping function $f(\cdot,\cdot)$ such that $s_R = f(y_R^{(I)}, y_R^{(Q)})$.

Table 1 shows the PNC mapping for the in-phase component $a_R$; the mapping for the quadrature component $b_R$ is similar. For QPSK, $a_R = a_1 a_2$ should be set to $-1$ if $a_1 \neq a_2$, and to $1$ if $a_1 = a_2$. There are three possible values for $y_R^{(I)} = a_1 + a_2$: 0, 2, and $-2$. Since $y_R^{(I)} = 0$ when $a_1 \neq a_2$, and $y_R^{(I)} = -2$ or 2 when $a_1 = a_2$, the PNC mapping is as follows:

$$a_R = \begin{cases} -1 & \text{if } y_R^{(I)} = 0 \\ 1 & \text{if } y_R^{(I)} = -2 \text{ or } 2 \end{cases} \quad (5)$$

Table 1. PNC mapping of in-phase signal components.

| Symbol from node 1: $a_1$ | Symbol from node 2: $a_2$ | Composite symbol received at relay $R$: $y_R^{(I)} = a_1 + a_2$ | Mapping to symbol to be transmitted by relay $R$: $a_R$ |
|---|---|---|---|
| 1 | 1 | 2 | 1 |
| 1 | −1 | 0 | −1 |
| −1 | 1 | 0 | −1 |
| −1 | −1 | −2 | 1 |

After the PNC mapping, relay $R$ transmits the following signal to nodes 1 and 2 in time slot 2:

$$s_R(t) = a_R \cos(\omega t) + b_R \sin(\omega t) \quad (6)$$

The RF signal transmitted in time slot 2 in PNC is the same as the RF signal transmitted in time slot 3 in SNC. The key difference of the two systems lies in how they derive $(a_R, b_R)$. In PNC, $(a_R, b_R)$ is derived from $(a_1 + a_2, b_1 + b_2)$, which is the superimposed signal. In SNC, $(a_1, b_1)$ and $(a_2, b_2)$ are separately transmitted by nodes 1 and 2; and relay $R$ explicitly decodes $(a_1, b_1)$ and $(a_2, b_2)$ in order to form $(a_R, b_R)$.

We remark that the arithmetic sums in $(a_1 + a_2, b_1 + b_2)$ is also a form of network coding operation. In particular, it is the network coding performed by nature. In the above example, the relay transforms it to the XOR network coding operation $(a_1 \oplus a_2, b_1 \oplus b_2)$. In general, among many other possibilities, the relay could also retain $(a_1 + a_2, b_1 + b_2)$ as the PNC mapping to be used.

**2.4. Generalization of PNC**

The general concept of PNC is to make use of the mixing of superimposed EM waves that occurs in nature to realize a desired network coding operation. We refer to the mapping of the mixed signal to the desired network-coded signal as *PNC mapping*.

So far, we have assumed the desired network-coded signal is the XOR of the signals from nodes 1 and 2. As already mentioned, in general, PNC mapping is not limited to just the XOR mapping. For example, in the Analog Network Coding (ANC) in [8], the relay $R$ retains the additive mixing that occurs in nature and simply amplifies and forwards $(a_1 + a_2, b_1 + b_2)$ to the two end nodes. An idea similar to ANC was proposed in an earlier paper in the context of satellite communication [9]. The advantage of ANC is that it is simple to realize. The disadvantage is that the relay does not remove receiver noise at the relay, and noise is forwarded along with the signals $(a_1+a_2, b_1+b_2)$ to the end nodes. As a result, its fundamental performance is not as good as schemes in which the relay tries to clean up the noise.

Refs. [10] [11] showed that even when both nodes 1 and 2 use QPSK modulation, when the phases of the RF signals from the two end nodes are not exactly aligned, it is sometimes more desirable for the relay $R$ to use 5QAM (as opposed to QPSK) for the signal it transmits to nodes 1 and 2.

Ref. [12] classifies PNC mappings into two categories: finite-field PNC and infinite-field PNC. In finite-field PNC, the target signal is represented by a finite field. The XOR mapping and the QPSK-5QAM mapping are examples of finite-field PNC. In infinite-field PNC, the target signal is represented by an infinite field (e.g., a real number). Analog Network Coding (ANC) [8] is an example of infinite-field PNC. Regardless of whether finite- or infinite-field PNC is adopted, the key requirement is that nodes 1 and 2 must be able to extract the information from the other end node from the mapped signal transmitted by relay $R$. In Sections 3.1.1 and 3.1.2, we will further discuss various schemes of finite- and infinite-field PNC.

## 2.5. Important Issues in PNC

To ease exposition, we have ignored many important issues so far. This section gives an overview of these issues.

### 2.5.1. Consideration of Noise

Fundamental to all communication systems is the presence of noise. In our discussion thus far, we have ignored noise. Our throughput analysis has been based on time slot counting, assuming whatever is sent will be received correctly. If that were the case, we could have decreased the time-slot duration indefinitely while sending the same amount of information, giving us infinite throughput. A more in-depth analysis must include the consideration of noise. With noise, (4) becomes

$$y_R = a_1 + jb_1 + a_2 + jb_2 + w_R \qquad (7)$$

where $w_R$ is the noise typically modeled as a Gaussian random variable.

With noise, an issue in XOR PNC mapping, for example, is whether the bit error rate (BER) will increase relative to TS or SNC. It turns out that with QPSK modulation, the end-to-end BER between nodes 1 and 2 of PNC is comparable to that of TS, and is actually slightly better than that of SNC, as explained in the next three paragraphs.

Let $P_e$ be the BER of the classical QPSK point-to-point link. Thus, TS consists of four such one-hop links. In the TS, SNC, and PNC discussed below, for simplicity, let us assume equal transmit power for all nodes, equal channel gain in all directions, and equal receiver noise power at all nodes. Then, $P_e$ is the same for all the one-hop links in TS. The end-to-end BER of TS is therefore $2(1-P_e)P_e$. That is, a transmitted bit is in error under two situations: (i) there is a detection error in the first hop, but no detection error in the second hop; or (ii) there is no detection error in the first hop, but there is a detection error in the second hop.

The BER of the classical QPSK point-to-point link $P_e$ is plotted in Fig. 4. For comparison the BER of $a_R$ (or $b_R$) in the uplink of the XOR PNC system is also plotted in Fig. 4. We refer the interested reader to [1] for the derivation of the BER of $a_R$ and $b_R$. We emphasize that for PNC and SNC, the target signal at the relay is the XOR signal; thus, the BER is the BER of the decoded XOR, not the BER of the individual bits from the two ends. Therefore, using the same argument as in the previous paragraph, the end-to-end BER of PNC is approximately $2(1-P_e)P_e$, the same as in TS. Note, however, that PNC uses two time slots and TS uses four time slots. Thus, at equal BER, the throughput of PNC is twice that of TS.

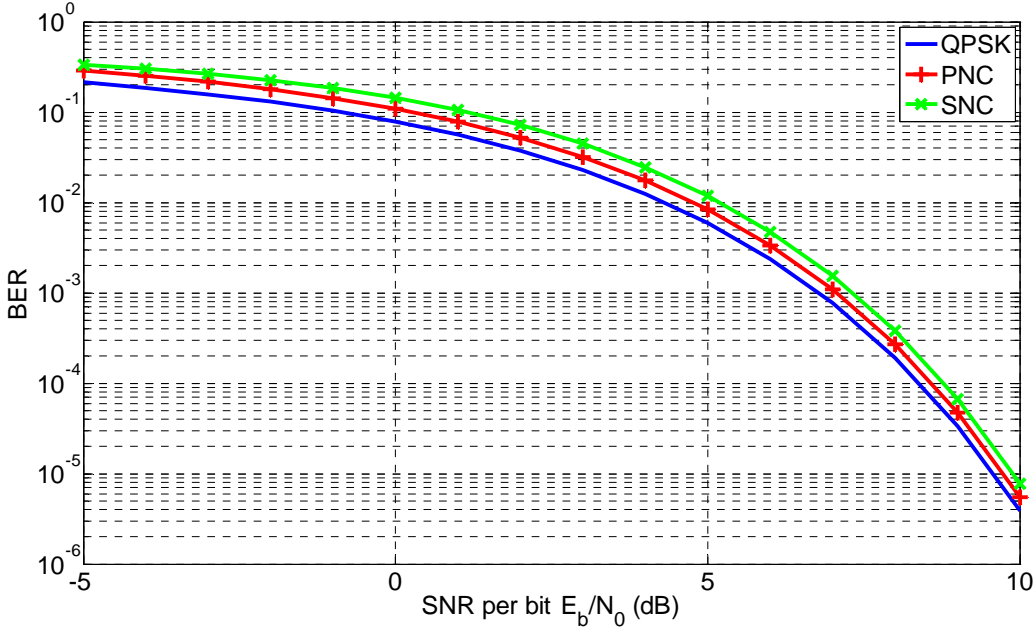

Fig. 4. BER of classical QPSK point-to-point link, uplink of SNC, and uplink of PNC.

For SNC, the BER at the relay is $2(1-P_e)P_e$. That is, there is an XOR decoding error if the bit from node 1 is decoded with error and the bit from node 2 is decoded correctly, and vice versa. That is, the uplink BER of SNR (also plotted in Fig. 4) is equal to the end-to-end BER of TS and PNC. The end-to-end BER of SNC is $2(1-P_e)P_e \cdot (1-P_e) + [1-2(1-P_e)P_e] \cdot P_e = 3P_e - 6P_e^2 + 4P_e^3 > 2(1-P_e)P_e$ for $P_e > 0$. We see that SNC not only have smaller throughput than PNC, it also has higher BER.

Our discussion above has focused on XOR PNC mapping. In general, investigations of other finite- and infinite-field PNC schemes should also take into the consideration of noise, especially when comparing their relative performance.

### 2.5.2. Channel Coding

In a communication system with noise, the use of channel coding is an important technique to ensure reliable transmission. Thus, nodes 1 and 2 in the PNC system may map their source packets $S_1$ and $S_2$ to channel-coded packets $C_1$ and $C_2$, respectively. An issue in PNC is how to integrate channel coding into the system. In general, channel coding can be applied on a link-by-link basis or an end-to-end basis. In the former, a relay performs channel decoding and re-encoding in addition to PNC mapping. In the latter, a relay only performs PNC mapping, and only the source performs channel coding and only the end receiver performs channel decoding.

In end-to-end channel-coded PNC, at the destination node, say node 1, after the channel-coded self information $C_1$ is removed from the received channel-coded signal, the remaining signal is just the channel-coded signal $C_2$ plus noise. The channel decoding operation at the end node is much like that in a traditional communication system. Because the relay does not clean up the relay noise by channel decoding and re-encoding, noise can accumulate from hop to hop. The noise accumulation can become severe in a more general setting in which there are many relays in between the two end nodes.

In link-by-link channel-coded PNC, channel decoding at the relay can be tricky (as compared to channel coding in ordinary point-to-point communication link). This is because the ultimate goal at the relay is recover $S_1 \oplus S_2$ (or more generally, other forms of PNC mapping on the source information), not the individual source information $S_1$ and $S_2$. After obtaining $S_1 \oplus S_2$, the relay can then channel-code $S_1 \oplus S_2$ before transmitting the channel-coded packet to the two end nodes. The two-step process is performed so as to clean up the noise before information forwarding. A subtlety of the first step $Y_R \to S_1 \oplus S_2$ is as follows. Note that $Y_R$ contains the arithmetic superposition of $C_1$ and $C_2$, not the arithmetic superposition of $S_1$ and $S_2$. In particular, the mapping $Y_R \to S_1 \oplus S_2$ involves both channel decoding and PNC mapping operations. Different ways of integrating such channel-decoding and PNC mapping at the relay may lead to different performance. Section 3.2 will delve into more elaborate discussions of the relevant issues.

### 2.5.3. Synchronization

The discussion of PNC thus far assumes perfect synchronization between nodes 1 and 2, so that their packets arrive at the relay with the packet boundary and symbol boundary aligned. In addition, the RF frequencies used by the two nodes are the same, and their relative phase offset is zero. An issue is whether there will be performance degradations (and if so, their extents) when PNC does not operate with such perfect synchronization.

Packet alignment is a MAC-layer scheduling issue. Compared with other kinds of synchronization, longer time scale is involved. MAC layer methods could be introduced to synchronize the transmissions of packets by nodes 1 and 2. However, even if nodes 1 and 2 transmit their packets simultaneously, it is possible for the packets to arrive at the relay with their symbol boundaries unaligned. Thus, a symbol from one node may overlap with two symbols from the other nodes. Symbol alignment is at a finer time scale than packet alignment and is therefore more challenging. There have been studies on how to align symbols of different transmitters at a common receiver [13]. This is a fundamental issue of relevance to many communication systems, not just PNC.

Even if symbols from the two nodes could be aligned, there would still be the issue of RF carrier frequency synchronization and relative phase offset. If the RF carrier frequencies at nodes 1 and 2 are derived from a common source, then there will be no frequency offset. If not, the frequency offset translates into a rotating phase offset: that is, the relative phase offset between the two nodes varies from symbol to symbol in a packet. In general, for a particular pair of symbols from the two end nodes, the baseband components in (7) become

$$y_R = a_1 + jb_1 + (a_2 + jb_2)e^{j\theta} + w_R \tag{8}$$

where $\theta$ is the relative phase offset between the pair of symbols. Phase offset $\theta$ is independent of time and is the same for all symbols in the packet if the RF frequencies of the two nodes are exactly the same; otherwise, $\theta$ will change incrementally in successive symbol pairs. The BER of PNC mapping will depend on $\theta$. For QPSK, $\theta = \pi/4$ has the worst BER while $\theta = 0$ has the best BER.

Section 3 will discuss the impact of the various kinds of asynchronies in more detail. In particular, we show some results indicating that asynchronies are not always bad. For example, in unchannel-coded PNC, phase asynchrony usually leads to a performance penalty. In channel-coded PNC, phase asynchrony may result in a performance reward rather than a performance penalty.

### 2.5.4. Non-symmetric Fading Channels and Channel Estimation

The discussion thus far assumes that the channels are symmetric. In general, nodes 1 and 2 can be at different distances from the relay. Also, the channels may undergo fading. Let $h_{iR}$ be the complex number denoting the channel gain for the uplink channel from node $i$ to relay $R$. For flat fading over the RF bandwidth of concern, the sampled symbol at the relay is

$$y_R = h_{1R}(a_1 + jb_1)j + h_{2R}(a_2 + jb_2) + w_R \tag{9}$$

The relay must estimate $h_{1R}$ and $h_{2R}$ in order to perform detection effectively. This estimation is typically done via known training symbols and/or pilots embedded in the packets. Channel estimation in PNC systems has also been an active area of research because unlike the point-to-point communication system in which only one channel gain needs to be estimated, two channel gains need to be estimated based on the simultaneously received signals.

If nodes 1 and 2 know $h_{1R}$ and $h_{1R}$ (say via feedback from the relay which estimates $h_{1R}$ and $h_{1R}$), they could multiply the symbol $(a_1 + jb_1)$ by $h_{1R}^*/|h_{1R}|^2$ and the symbol $(a_2 + jb_2)$ by $h_{2R}^*/|h_{2R}|^2$, respectively, before transmitting them. Doing so will yield the same equation as (7) for the received signal at the relay. In TWRC, for example, each time the relay broadcasts a packet to nodes 1 and 2, it could also embed $h_{1R}$ and $h_{2R}$ estimated by it within the broadcast packet. This way, nodes 1 and 2 can use this information to precode the symbols in the next packet transmitted to the relay. Generally, transmitter precoding by nodes 1 and 2 can lead to better performance.

There are two scenarios under which transmitter precoding is impractical. The first is the fast fading case in which the channel gains vary quickly so that between the transmission of one packet and the next, $h_{iR}$ has already changed substantially. The feedback $h_{iR}$ from relay $R$ to node $i$ is therefore not reflective of the actual $h_{iR}$ in the next time slot. The second is the bursty, sporadic traffic case in which the end nodes do not always have packets to transmit, and the relay is shared by many pairs of end nodes. In this case, a random access MAC protocol may be used to coordinate the packet exchange between the different node pairs. The intervening time between two successive packet exchange of a particular node pair may be long, and that $h_{iR}$ may have changed significantly since the last exchange.

In general, systems in which transmitter precoding is not used are simpler to implement and are applicable to a wider range of scenarios. For this reason, most research on PNC has assumed the end nodes do not pre-code. Note that precoding by the relay is a different story. If the channels are symmetric so that $h_{Ri} = h_{iR}$, the relay could use its knowledge on $h_{iR}$ (which has to be estimated anyway) to precode the signal it relays to the end nodes. Relaying occurs almost immediately after the transmission by nodes 1 and 2: therefore, if the channel gains do not change drastically from time slot to time slot, this strategy is still valid.

For non-flat fading over the transmission bandwidth, (9) is not valid and there will be inter-symbol interference. OFDM is a powerful technique for dealing with non-flat fading. The basic idea of OFDM is to carry the symbols on multiple sub-bands. If the sub-bands are narrow enough, the fading in each sub-band is flat. Thus, on each sub-band, (9) remains valid. In addition, OFDM provides a natural way to deal with the relative symbol offset between nodes 1 and 2 in PNC. In particular, any time-domain symbol offset will be translated to a phase term in the channel gain in the frequency domain so that (9) remains valid with $h_{iR}$ multiplied by a phase term. Because of its ability to deal with symbol offset and non-flat fading simultaneous, OFDM PNC is a popular system under investigation by many researchers. At the same time, there are also new challenges in the OFDM PNC system that do not exist in the traditional point-to-point OFDM system. An example is the estimation of the channel gains on the subcarriers within the OFDM PNC system, and the estimation of the RF carrier offsets of the two end nodes, based on the composite signal received from the two end nodes. The training symbols and pilots in the traditional point-to-point OFDM need to be redesigned for such estimations in the OFDM PNC system.

### 2.5.5. Information-Theoretic Capacity

At the most fundamental level, the performance of PNC should be analyzed from an information-theoretic perspective. This study falls into the domain of network information theory [14]. For TRWC, for example, of concern is the information exchange rates that can be achieved from node 1 to node 2, $R_{12}$, and from node 2 to node 1, $R_{21}$, subject to the noise.

It has been found that finite-field mapping schemes can achieve near information-capacity rates [15], whereas infinite-field mapping schemes such as ANC [8] cannot. In [16], it was shown that under Gaussian-noise channels, finite-field PNC with the use of lattice code can achieve rates within $1/2$ bit of the cut-set outer bound in TRWC.

In general, link-by-link channel-coded schemes have better exchange-rate performance than end-to-end channel-coded schemes. Within the category of link-by-link channel coded schemes, the relative performance of different schemes depend on the SNR regime of operation.

In the TWRC, the rates $R_{12}$ and $R_{21}$ could be uplink-limited (limited by the links from nodes 1 and 2 to relay $R$), downlink-limited (limited by the links from relay $R$ to nodes 1 and 2), or both uplink-and-downlink limited. Section 4 will present a detailed discussion of TWRC from an information-theoretic perspective.

### 2.5.6. General Network Topologies and Higher-layer Issues

The original proposal of PNC in [1] provided a brief discussion of its application in networks of general topologies, and the implications of PNC for higher-layer issues such as MAC scheduling and routing. The majority of subsequent PNC investigation, however, focused on TWRC.

It is straightforward to extend the TWRC scenario to a linear network scenario in which two end nodes exchange information via a chain of relays between them. With proper scheduling, as illustrated in Fig. 5 (see [1] for a more detailed description), the exchange throughput of $1/2$ packet per direction per time slot can be achieved (i.e., same as in TWRC).

For a general multihop network, there could be many end-to-end flows. Each flow is between two end nodes, and the intermediate nodes between the two nodes serve as the relays for the flow. If the flow is bidirectional and there are equal amounts of traffic in the two directions, then the bidirectional flow could make use of PNC. The two end nodes and the intermediate nodes traversed by the bidirectional flow look like the linear network in Fig. 5. A difference in a general network, however, is that the relay nodes may not be dedicated to that bidirectional flow alone. There could be many flows traversing a node, and therefore the transmission time of a node may needs to be divided among the multiple flows. Thus, in addition to the intra-flow scheduling such as that shown in Fig. 5, the inter-flow scheduling also needs to be considered.

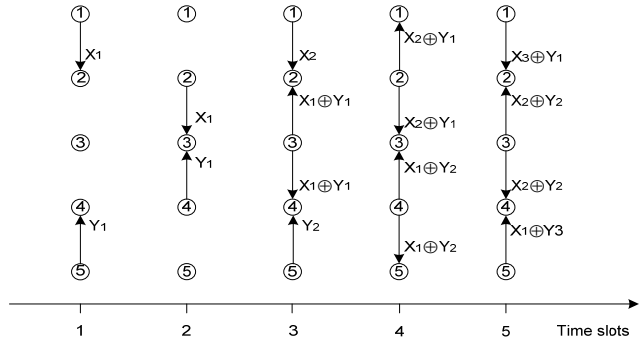

Fig. 5. PNC scheduling in linear chain.

PNC matches best with the equal-traffic bidirectional setting. In a network supporting general applications, we could also have unidirectional end-to-end flows. In addition, for some applications with bidirectional flows, the amounts of traffic in the two directions may not be equal. An example is a TCP file download session in which the TCP DATA mainly flows from the file server to the client, and there is relatively little traffic of TCP ACK in the other direction.

The concept of virtual path can be applied to the general setting[2]. We could establish many balanced-traffic bidirectional virtual paths to exploit PNC in the optimal way. Each balanced-traffic bidirectional virtual path is a linear PNC chain similar to that in Fig. 5. We could aggregate the traffic from multiple end-to-end flows onto each bidirectional virtual path. An example is shown in Fig. 6, in which we show the aggregation of two unidirectional flows. The overlapped portions of the two unidirectional flows could be carried on a bidirectional PNC virtual path. The non-overlapped portions could aggregate with other flows on other bidirectional PNC virtual paths.

---

[2] Chapter 7 of [17] contains a brief introduction of the concept of virtual paths and virtual circuits. Many other references on the topic of ATM networks also contain similar materials.

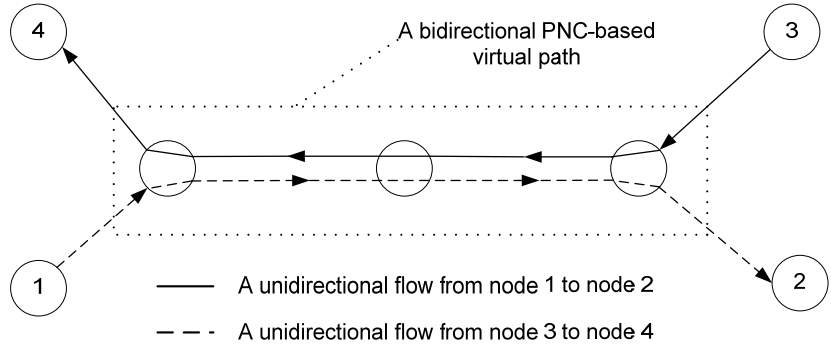

Fig. 6. PNC-based virtual paths.

Note that an end-to-end flow may traverse a sequence of bidirectional virtual paths, and each bidirectional virtual path may aggregate traffic from many flows. In particular, a bidirectional virtual path may aggregate traffic from different flows in a way that the aggregated traffic volumes in both directions of the virtual path are approximately equal, so that optimal use of PNC can be attained. Thus, the general principle is that for each unit of traffic in one direction, we try to find a unit of traffic (possibly from another flow) in the opposite direction for aggregation purposes.

Despite the best attempt, there could be still be some remainder traffic in one direction that can not be matched to traffic in the opposite direction. Thus, in addition to the bidirectional PNC virtual paths, the network could also form unidirectional virtual paths to carry such unmatched traffic. The unidirectional virtual paths could either use the traditional multi-hop method to schedule packet transmission along the path, or they could use unidirectional PNC scheduling. W refer the interested reader to [1] for details on unidirectional PNC, which in general could be more efficient than traditional unidirectional multi-hop scheduling.

In wired networks, the virtual paths that do not overlap do not mutually interfere. In wireless networks, non-overlapping virtual paths may still mutually interfere even if their nodes are within the proximity of each other. Thus, nearby virtual paths are not decoupled, and the scheduling of the transmission times of the nodes in a virtual paths must take into account the transmission times of the nodes in nearby virtual paths.

Besides scheduling, there is also the routing issue when deciding the route for an end-to-end flow. With the virtual path concept above, however, the problem becomes finding a sequence of virtual paths with enough unused capacity to carry the flow.

### 2.6. Concluding Remarks for Brief Tutorial

We have given a quick introduction to PNC and the issues involved. In the remaining sections, we will go into some of these issues at a deeper level. We conclude this brief tutorial by making some general observations about the status of PNC research to-date.

Among the three areas of PNC research (i.e., communication-theoretic, information-theoretic, and networking), networking issues have received the least attention today. Yet networking issue becomes important when we extend the application of PNC beyond TWRC. As the theoretical understanding of PNC in TWRC matures, we anticipate future research focus will move toward application of PNC in general topologies, with network and MAC-scheduling issues taking increasing important roles.

In addition, there have been little implementation and prototyping efforts for PNC. To our best knowledge, [8] remains the only work that attempted to prototype a PNC system. The simplest amplify-and-forward TWRC system was chosen for implementation in [8]. As the theoretical understanding of PNC matures, the implementation arena is likely to become a fertile ground for future research.

### 3. Communication-theoretic Studies

In this section, we delve into the communication-theoretic studies of PNC, focusing on TWRC. We start with unchannel-coded PNC in Section 3.1. Section 3.2 discusses channel-coded PNC. In particular, we discuss various ways in which channel coding can be integrated into the PNC framework. In both Sections 3.1 and 3.2, we will examine the synchronization issue.

### 3.1. Unchannel-coded PNC

In [1], the XOR PNC mapping was explored. Since then, many other schemes have been investigated. In general, these schemes can be divided to two categories according to the PNC mapping involved: PNC over finite set (PNCF) and PNC over infinite set (PNCI) [12][3].

Assuming symbol alignment for the time being, for a pair of symbols $(x_1, x_2)$ from nodes 1 and 2, the composite symbol received by node $R$ is

$$y_R = h_{1R}x_1 + h_{2R}x_2 + w_R \qquad (10)$$

where $h_{iR}$ is the channel gain from node $i$ to relay $R$, and $w_R$ is the Gaussian noise. We assume the transmit power have been factored into the $h_{iR}$. That is, $h_{iR} = \sqrt{P_i^{(t)}} g_{iR}$, where $P_i^{(t)}$ is the transmit power, and $g_{iR}$ is the actual gain of the channel.

PNC mapping attempts to map $y_R$ to a target symbol $z_R$ for broadcast back to nodes 1 and 2. In PNCF, $z_R$ is a symbol chosen from a finite set, and there are only a finite number of possibilities for $z_R$. For example, in XOR PNC mapping [1], the target symbol $z_R = z_{R,r} + jz_{R,i}$ where $z_{R,r}, z_{R,i} \in \{-1,1\}$. In PNCI, $z_R$ is a symbol chosen from an infinite set. For example, in amplify-and-forward PNC, i.e., ANC in [8], $z_R = h_{1R}x_1 + h_{2R}x_2$. Since $z_R$ is a real number, ANC belongs to PNCI.

Because of noise, the relay can only get an estimate for $z_R$, denoted by $\widehat{z}_R$. It is the estimate $\widehat{z}_R$ that is actually sent by the relay. That is, the relay sends $x_R = \widehat{z}_R$ to nodes 1 and 2.

#### 3.1.1. PNCF

In [1], the simplest PNCF scheme was considered. The constellations of $x_1$, $x_2$, and $z_R$ are all QPSK constellation. The QPSK-QPSK PNC mapping has good performance when $h_{1R}/h_{2R} = 1$, and this was the case assumed in our earlier presentation of PNC in Section 2.3.

When the relative phase offset of the signals from the two nodes is $\pi/4$, so that $h_{1R}/h_{2R} = e^{j\pi/4}$, there will be a significant performance penalty. Ref. [18] mentioned (without providing a proof) that the penalty could be as high as 6 dB. In actuality, the penalty could be even higher. This observation is quite alarming, and raises a question as to whether PNCF is viable in practice when the system is asynchronous. In Section 3.1.3, however, we will present results showing that this penalty can be significantly reduced when the symbols from the two end nodes arrive at the relay misaligned. Also, in Section 3.2.2, we will present results showing that the phase penalty in unchannel-coded PNC becomes a "phase reward" in channel-coded PNC. This leads us to believe that phase asynchrony is not a fundamental performance-limiting factor.

Ref. [10] showed that for QPSK $x_1$ and $x_2$, it is not always best to have QPSK $z_R$. The symbol-aligned case was studied. In particular, it was shown that when $h_{1R}/h_{2R} = \sqrt{2}e^{j\pi/4}$, a constellation map with at least five constellation points (e.g., 5QAM) for $z_R$ is needed in order that the end nodes can decode the symbol from the other node, even in the absence of noise. When the symbols are misaligned and/or when channel coding is incorporated into the PNC system, certain diversity effect will cause phase asynchrony to be a lot less detrimental (more on this in Sections 3.1.3 and 3.2.2) in QPSK-QPSK mapping. Thus, it is not clear that in a practical system, QPSK-5QAM mapping is necessary, especially in view of the fact that this will complicate the implementation of channel-coded PNC.

---

[3] In [12], PNCF and PNCI were defined as PNC over finite field and infinite field. More generally, the target PNC-mapped symbol $z_R$ at the relay need not be a field. Thus, in this paper we redefine PNCF and PNCI to be PNC over finite set and infinite set.

In our discussion of PNCF thus far, we treat $z_R$ as the target mapped symbol, and assume the decoding of the composite signal at $R$ is such that the decoded symbol is also drawn from the same set as $z_R$. That is both $z_R$ and $\hat{z}_R$ are elements belonging to the same finite set $F$. In estimation theory, even if the target $z_R \in F$, it is possible that $\hat{z}_R \notin F$ and that $\hat{z}_R$ is drawn from a infinite set. Refs. [12] and [19] discussed a number of such possibilities in detail. For example, in XOR PNC mapping, $z_R = x_1 \oplus x_2$, $x_1, x_2, z_R \in \{-1,1\}$, but the MMSE estimate of $z_R$ is a real number. Thus, when MMSE estimate is adopted, the relay actually sends out an analog rather than a discrete signal.

For MMSE, the relay transmits $x_R = \hat{z}_R = E[z_R | y_R] = P(z_R = 1 | y_R) - P(z_R = -1 | y_R)$, from which we can see that the relay actually forwards both the sign and the reliability of $x_R$; a large positive $x_R$, for example, means that the XOR value $x_1 \oplus x_2$ has a high probability of being 1. Similar to the water-filling algorithm, the symbol $x_R$ is transmitted with a high power when the reliability is high and with a low power when the reliability is low. From this view of point, MMSE estimate is a power allocation scheme for PNCF. The optimality of MMSE estimate for the one-way relay channel can be found in [20]. Due to this power allocation, MMSE estimate performs better than other schemes under a total power constraint.

Table 2. Examples of PNC mappings under each classification.

| Transmitted output, $x_R = \hat{z}_R$ / Target symbol, $z_R$ | Analog | Discrete |
|---|---|---|
| PNCF | Estimate (e.g., MMSE) [12] [19] | XOR [1] Denoising Map [10] |
| PNCI | ANC (i.e., linear MMSE) [8] Other estimates (e.g., MMSE) [12] | NA |

### 3.1.2. PNCI

In ANC, $z_R = h_{1R}x_1 + h_{2R}x_2$, what actually gets sent by the relay is $x_R = g \cdot y_R = g \cdot (h_{1R}x_1 + h_{2R}x_2 + w_R)$, where $g$ is the amplification applied by the relay before forwarding the signal. As explained in [12], ANC can be considered as using a linear MMSE estimator. In general, various different estimates in PNCI are also possible (e.g., general MMSE rather than linear MMSE [12]).

Ref. [12] showed that typically PNCF has better performance than PNCI when the uplink is good and the bottleneck is the downlink. Conversely, PNCI has better performance than PNCF when the downlink is good and the bottleneck is the uplink. This is not difficult to understand intuitively. When the uplink is good, uplink noise is relatively small compared with the uplink signals. For example, XOR mapping of $h_{1R}x_1 + h_{2R}x_2 + w_R$ to $x_1 \oplus x_2$ [1] is almost perfect. The small noise is removed by the hard decision of the mapping so that the relay does not expend transmit power on the uplink noise. The BER performance of the whole PNC set-up is reduced to the BER performance of a point-to-point communication system consisting of only the downlink. With PNCI, such as ANC [8], the noise at the relay will be amplified along with the signal before being sent by the relay to the end nodes. Some transmit power of the relay is used to carry the uplink noise. Thus, the BER performance will be worse than that of the BER performance of a point-to-point link consisting of the downlink alone. Ref. [12] presented results showing that when the uplink and downlink SNR are equal at 5dB, PNCF has slightly better BER performance than PNCI.

Conversely, when the uplink is bad, the XOR mapping of $h_{1R}x_1 + h_{2R}x_2 + w_R$ to $x_1 \oplus x_2$ is error-prone, while the downlink is not. Self information after the downlink transmission is not of much use when the XOR information transmitted by the relay is erroneous to start with. Meanwhile, the hard decision at the relay removes useful soft information that can be combined with the self information at the end nodes to improve final decoding. PNCI, such as ANC, passes along the soft information and defers the decision until after the transmission on the downlink is completed. Thus, PNCF tends to perform less well than PNCI when the downlink is good but the uplink is not.

The above discussion on PNCF versus PNCI has focused on unchannel-coded PNC. For channel-code PNC, an advantage of PNCF over PNCI is that PNCF is amenable to link-by-link channel coding using conventional channel codes. Application of channel coding to PNCI is not as straightforward.

### 3.1.3. Asynchrony Penalties

Let us now examine two types of asynchronies: symbol-level asynchrony and phase-level asynchrony. The situation is depicted in Fig. 7, where the baseband signals from nodes 1 and 2 are shown. The relative phase offset is embedded in the two complex numbers representing the channel gains, $h_{1R}$ and $h_{2R}$. Without loss of generality, we assume the signal of node 1 is ahead of the signal of node 2 by $\Delta$ symbol, $0 \leq \Delta < 1$. Note that if node 1 is ahead of node 2 by multiple symbols, we can define some null symbols at the head end of the packet of node 2 and at the tail end of the packet of node 1, essentially making the packets larger; our treatment below can be generalized to that situation with some minor modifications.

We will assume that relay $R$ can estimate $h_{1R}$ and $h_{2R}$ so that $h_{1R}$ and $h_{2R}$ are treated as known at the relay. However, we do not assume *a priori* knowledge of $h_{1R}$ and $h_{2R}$ at nodes 1 and 2. This means that they cannot perform precoding to remove the relative phase offset between $h_{1R}$ and $h_{2R}$.

Besides phase offset, there could be a frequency offset in the RF used by nodes 1 and 2. This frequency offset will translate to a rotating relative phase offset in $h_{1R}$ and $h_{2R}$ for successive symbols. If the frequency offset can be estimated, then the rotating relative phase offset can also be tracked. This basically means that different pairs of symbols from nodes 1 and 2 have different relative phase offsets, but these phase offsets can all be estimated and are therefore known. In the following discussion, for simplicity, we will assume a fixed relative phase offset throughout the whole packet.

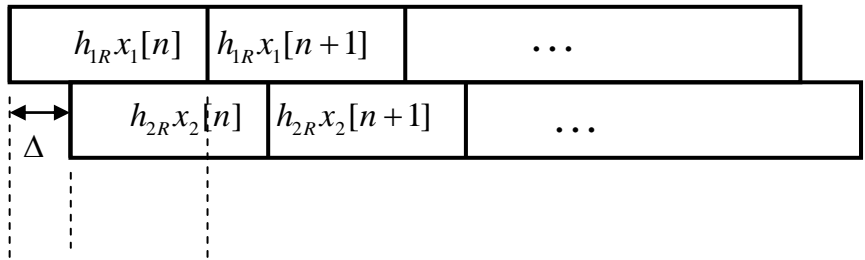

Fig. 7. Symbol offset between the signals from nodes 1 and 2 in TWRC.

In addition, for simplicity of exposition, we assume the use of the rectangular pulse for carrying the digital symbols in the analog domain. Let $X_i = (x_i[1], x_i[2], ..., x_i[N])$ be the packet from node $i$, $i = 1, 2$. At relay $R$, the received signal in continuous time in baseband is

$$y_R(t) = \sum_{n=1}^{N} \{h_{1R}x_1[n]p(t-n) + h_{2R}x_2[n]p(t-\Delta-n)\} + w_R(t), \tag{11}$$

where $h_{1R} = \sqrt{P_1}$ and $h_{2R} = \sqrt{P_2}e^{j\phi}$ ($\phi$ is the relative phase offset between nodes 1 and 2); $p(t)$ is the rectangular pulse ($p(t) = 1$ for $-1 \leq t < 0$ and $p(t) = 0$ otherwise); and $w_R(t)$ is the AWGN noise.

Note that we have assumed the normalization of the symbol duration so that it is one, and that the channels experience flat fading.[4] For simplicity, we further assume power control such that $P_1 = P_2 = P$. Eqn. (11) then becomes

$$y_R(t) = \sqrt{P} \sum_{n=1}^{N} \{x_1[n]p(t-n) + x_2[n]p(t-\Delta-n)e^{j\phi}\} + w_R(t), \quad (12)$$

In [21] and [22], suboptimal sampling were assumed: specifically, with respect to Fig. 7, only the overlapped part of $x_1[n]$ and $x_2[n]$ is sampled, and the useful signal in the overlap of $x_1[n+1]$ and $x_2[n]$ is not used. Furthermore, the joint effect of symbol and phase asynchronies were not considered. Here, we consider the use of an optimal maximum-likelihood decoding method based on the belief propagation (BP) algorithm. The method deals with symbol and phase asynchronies jointly.

We focus on the PNCF in which the goal of relay $R$ is to map $y_R(t)$ to $x_1[n] \oplus x_2[n]$, $n = 1,...,N$. Specifically, relay $R$ wants to minimize the error of the decoded $x_1[n] \oplus x_2[n]$ $\forall n = 1,...,N$, based on the observation of $y_R(t)$. We perform integration (matched filtering) on the overlapped symbols for a duration of $\Delta$ and a duration of $(1-\Delta)$ alternately, with a normalization factor $1/(\sqrt{P}\Delta)$ and $1/[\sqrt{P}(1-\Delta)]$, respectively. Samples $(2n-1)$ and $2n$ for $1 \le n \le N$, and sample $(2N+1)$ are then given by

$$\begin{aligned}
y_R[2n-1] &= \frac{1}{\Delta} \int_{(n-1)}^{(n-1)+\Delta} \left(x_1[n] + x_2[n-1]e^{j\phi} + w_R(t)/\sqrt{P}\right) dt \\
&= x_1[n] + x_2[n-1]e^{j\phi} + w_R[2n-1] \\
y_R[2n] &= \frac{1}{1-\Delta} \int_{(n-1)+\Delta}^{n} \left(x_A[n] + x_B[n]e^{j\phi} + w_R(t)/\sqrt{P}\right) dt \\
&= x_1[n] + x_2[n]e^{j\phi} + w_R[2n], \\
y_R[2N+1] &= \frac{1}{\Delta} \int_{N}^{N+\Delta} \left(x_B[N]e^{j\phi} + w_R(t)/\sqrt{P}\right) dt \\
&= x_2[N]e^{j\phi} + w_R[2N+1]
\end{aligned} \quad (13)$$

where $x_2[0] = 0$, and $w_R[2n-1]$ (also $w_R[2N+1]$) and $w_R[2n]$ are a zero-mean complex Gaussian noise with variance $N_0/(2P\Delta)$ and $N_0/[2P(1-\Delta)]$, respectively, for both the real and imaginary components. Here, $N_0/2$ is the double-sided power spectrum of the AWGN. For QPSK, $x_i[n] \in \{(1+j)/\sqrt{2}, (1-j)/\sqrt{2}, (-1+j)/\sqrt{2}, (-1-j)/\sqrt{2}\}$.

Let us write our sampled observations as $Y_R = (y_R[1],...,y_R[2N+1])$. We want to find the *a posteriori* probability $P(x_1[n] \oplus x_2[n] | Y_R)$, $n = 1,...,N$, so that the relay $R$ can compute $x_R[n] = \arg\max_{x_1[n] \oplus x_2[n]} P(x_1[n] \oplus x_2[n] | Y_R)$, $n = 1,...,N$ for the broadcast signal to nodes 1 and 2. Note that $P(x_1[n] \oplus x_2[n] | Y_R)$ can be found from the joint probability $P(x_1[n], x_2[n] | Y_R)$. Thus, we could focus our attention on finding $P(x_1[n], x_2[n] | Y_R)$. This is where the BP algorithm comes in handy.

Let us define a "joint symbol" as $(x_1[i], x_2[j])$, where $i = j$ or $(j+1)$. There are altogether $2N+1$ joint symbols and $2N+1$ sampled observations in $Y_R = (y_R[1],...,y_R[2N+1])$. The sample $y_R[i+j]$ gives us some information on the joint symbol $(x_1[i], x_2[j])$. Specifically, we have, for $1 \le n \le N$,

---

[4] As mentioned in Section 2.5.4, in frequency-selective channel, OFDM could be used so that essentially flat fading is experienced in each subcarrier. In this sense, the treatment here can be considered as that for a subcarrier within an OFDM PNC system.

$$P(x_1[n], x_2[n-1] | y_R[2n-1]) \propto \Delta \exp\left\{\frac{\left(y_R^{Re}[2n-1] - (x_1^{Re}[n] + e^{j\phi}x_2^{Re}[n-1])\right)^2}{N_0/(P\Delta)}\right\} \cdot$$

$$\exp\left\{\frac{\left(y_R^{Im}[2n-1] - (x_1^{Im}[n] + e^{j\phi}x_2^{Im}[n-1])\right)^2}{N_0/(P\Delta)}\right\},$$

$$P(x_1[n], x_2[n] | y_R[2n]) \propto (1-\Delta) \exp\left\{\frac{\left(y_R^{Re}[2n] - (x_1^{Re}[n] + e^{j\phi}x_2^{Re}[n])\right)^2}{N_0/[P(1-\Delta)]}\right\} \cdot$$

$$\exp\left\{\frac{\left(y_R^{Im}[2n] - (x_1^{Im}[n] + e^{j\phi}x_2^{Im}[n])\right)^2}{N_0/[P(1-\Delta)]}\right\},$$

and

$$P(x_2[N] | y_R[2N+1]) \propto \Delta \exp\left\{\frac{\left(y_R^{Re}[2N+1] - e^{j\phi}x_2^{Re}[N]\right)^2}{N_0/(P\Delta)}\right\} \cdot \exp\left\{\frac{\left(y_R^{Im}[2N+1] - e^{j\phi}x_2^{Im}[N]\right)^2}{N_0/(P\Delta)}\right\}. \quad (14)$$

Note that because of symbol misalignment, the successive sampled observations are correlated. For a given joint $(x_1[i], x_2[j])$, what we are interested in is not $P(x_1[i], x_2[j] | y_R[i+j])$ based on a local sample $y_R[i+j]$, but rather the *a posteriori* probability $P(x_1[i], x_2[j] | Y_R)$ based on the whole collection of samples $Y_R$. That is, samples other than $y_R(i+1)$ contains useful information on $(x_1[i], x_2[j])$. This is where the belief propagation (BP) algorithm enters the picture [23].

The idea of BP is that $P(x_1[i], x_2[j] | Y_R)$ for all $(i, j)$ pairs can be obtained from a sum-product algorithm [23] based on the individual observations $P(x_1[i], x_2[j] | y_R[i+j])$. BP is a general inference method for graph models. Interested readers are referred to [23] for a general tutorial on BP. The structure of the sum-product algorithm is given by the Tanner graph associated with the problem.

The Tanner graph for our problem is shown in Fig. 8. In Fig. 8, for brevity, we denote the joint symbol $(x_1[i], x_2[j])$ by $x^{i,j}$. The correlation between two adjacent joint symbols is modeled by the compatibility functions (i.e., check nodes) $\psi_o(x^{n,n-1}, x^{n,n})$ and $\psi_e(x^{n,n}, x^{n+1,n})$:

$$\psi_o(x^{n,n-1}, x^{n,n}) = \begin{cases} 1 & \text{if } x_1[n] \text{ in } x^{n,n-1} \text{ and } x^{n,n} \text{ are equal} \\ 0 & \text{otherwise} \end{cases}$$

$$\psi_e(x^{n,n}, x^{n+1,n}) = \begin{cases} 1 & \text{if } x_2[n] \text{ in } x^{n,n} \text{ and } x^{n+1,n} \text{ are equal} \\ 0 & \text{otherwise} \end{cases} \quad (15)$$

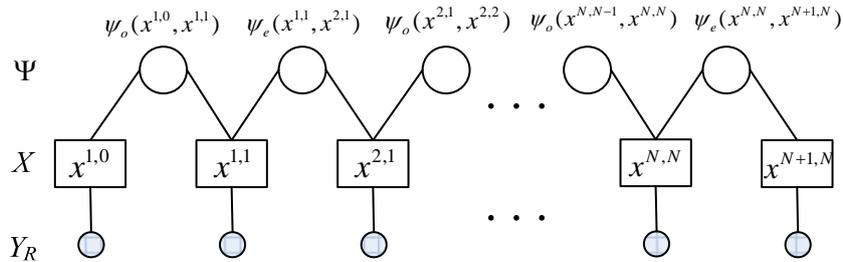

Fig. 8. Tanner graph for finding $P(x_1[i], x_2[j] | Y_R)$.

Once the Tanner graph is found, the standard BP procedure can be applied to generate the associated sum-product algorithm. In this paper, we will not go through this standard procedure. The reader is referred to [24] for further details.

We remark that BP is in general only an approximate algorithm. However, for Tanner graphs with a tree structure, BP yields the exact solution for the marginal probabilities, and it does so in one iteration of the sum-product algorithm [23]. Our Tanner graph in Fig. 8 has a tree structure. Therefore, BP is a very efficient algorithm for resolving symbol misalignment, and it yields the exact solution for $P(x_1[n], x_2[n] | Y_R)$. From $P(x_1[n], x_2[n] | Y_R)$, $x_R[n] = \arg\max_{x_1[n] \oplus x_2[n]} P(x_1[n] \oplus x_2[n] | Y_R)$, $n = 1,...,N$, can be readily obtained.

Let us look at some results for the QPSK case, in which $x_1[n]$, $x_2[n]$, and $x_R[n]$ are QPSK symbols. Fig. 9 plots the BER of $x_R[n]$ under various symbol and phase offsets. Note that for QPSK, each $x_R[n]$ contains two bits, an in-phase bit and a quadrature bit. The BER in Fig. 9 is the BER averaged over for 10,000 packets of 2,048 bits. In the figure, $E_b$ is the energy per bit in $x_i[n]$, $i = 1, 2$, which is equal to half the energy per symbol, $E_s/2$.

Fig. 9(a) and Fig. 9(b) plot the cases without and with symbol asynchrony, respectively. For both figures, the perfectly synchronized case of $\Delta = 0, \phi = 0$ is plotted for benchmarking purposes. As can be seen from Fig. 9(a), when symbols are aligned, the phase penalty can be as large as 6 to 7dB (when $\phi = \pi/4$). However, with symbol asynchrony, as can be seen from Fig. 9(b), the phase penalty reduces to within 1 dB. In other words, symbol asynchrony can ameliorate the phase-asynchrony penalty. This is attributed to certain "diversity" and "certainty propagation" effects, which we overview in the next three paragraphs. There reader is referred to [24] for elaboration.

If we examine the worst case in which $\Delta = 0$ and $\phi = \pi/4$, it turns out that certain combinations of the joint symbol $(x_1[n], x_2[n])$ can be decoded with confidence (e.g., $(x_1[n], x_2[n]) = ((1+j)/\sqrt{2}, (1+j)/\sqrt{2})$), while other combinations of $(x_1[n], x_2[n])$ are error prone (e.g., $(x_1[n], x_2[n]) = ((-1-j)/\sqrt{2}, (1+j)/\sqrt{2})$). This can be deduced from the constellation map [24] consisting of the 16 constellation points $x_1[n] + x_2[n]e^{j\phi}$ for the 16 combinations of $(x_1[n], x_2[n])$. Note that the observation $y_R[n]$ associated with $(x_1[n], x_2[n])$ is $x_1[n] + x_2[n]e^{j\phi}$ plus noise at the relay. Out of the 16 constellation points, eight are "good" constellation points with large Euclidean distances to adjacent constellation points, and eight are "bad" constellation points with small Euclidean distances at adjacent constellation points. The signal component in $y_R[n]$, $x_1[n] + x_2[n]e^{j\phi}$, is robust against noise for good constellation points, but vulnerable to noise for bad constellation points. The BER is dominated by the bad combinations when $\Delta = 0$ and $\phi = \pi/4$.

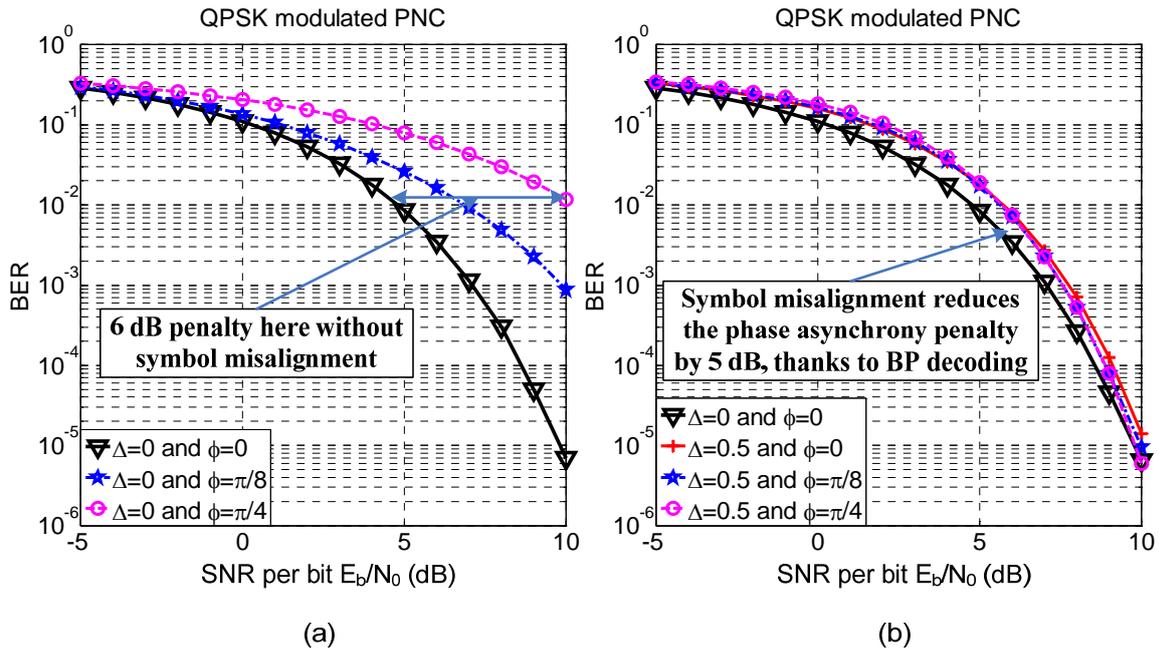

Fig. 9. BER of $x_R[n]$ for QPSK modulated PNC: (a) without symbol asynchrony ($\Delta = 0$); (b) with symbol asynchrony ($\Delta \neq 0$).

When $\Delta = 1/2$, even if $\phi = \pi/4$, the penalty is small, as shown in Fig. 9(b). This is due to two effects: diversity and certainty propagation. Even if the joint symbol $(x_1[n], x_2[n])$ corresponds to a bad constellation point, there is a chance that $(x_1[n+1], x_2[n])$ corresponds to a good constellation point. That is, thanks to symbol misalignment at the relay, each symbol from an end node is embedded in two joint symbols at the relay, and the system is more robust because of this diversity.

Now, symbol misalignment also produces a "certainty propagation" effect. To see this, consider the following. Even if $(x_1[n], x_2[n])$ and $(x_1[n+1], x_2[n])$ are bad constellation points, there is still a chance that $(x_1[n+1], x_2[n+1])$ is a good constellation point, and so on and forth. Once a good constellation point is encountered, then both the constituent symbols $x_1[\cdot]$ and $x_2[\cdot]$ in $(x_1[\cdot], x_2[\cdot])$ (not just the XOR of $x_1[\cdot]$ and $x_2[\cdot]$) can be decoded with high confidence. The certainties of $x_1[\cdot]$ and $x_2[\cdot]$ then propagate to the surrounding symbols via the BP algorithm [24], even if the joint symbols adjacent to $(x_1[\cdot], x_2[\cdot])$ are bad constellation points. This is because a bad constellation point is bad only when both the constituent symbols are unknown; once one of the constituent symbols is known with certainty, both the constituent symbols of the bad constellation point can be decoded with confidence [24]. In this way, the certainty can propagate a distance along the chain of joint symbols. We emphasize that certainty propagation depends on the maintenance of the joint probability $P(x_1[\cdot], x_2[\cdot] \mid y_R[\cdot])$ throughout the BP decoding process. If the joint probability is collapsed into the XOR probability $P(x_1[\cdot] \oplus x_2[\cdot] \mid y_R[\cdot])$ and a BP algorithm is run over this probability instead, the certainty propagation effect will vanish because of the lack of the knowledge of the constituent symbol.

We conclude our discussion of asynchrony in unchannel-coded PNC by remarking that although phase offset can be detrimental to PNC performance, symbol misalignment makes the system more robust against phase asynchrony. Generally, unless one deliberately attempts to align the symbol, most likely there will be some symbol misalignment in the system. Thus, phase offset may not be as bad as it seems. Also, if one has a mechanism to synchronize the symbol arrival times at the relay, it is actually a better strategy to intentionally desynchronize the timing so that $\Delta = 1/2$. The strategy of intentional symbol misalignment, however, requires further investigation when the signals are band-limited. For example, when pulse shaping is not rectangular, but say raised cosine, each sample $y_R[\cdot]$ may contain information from more than a pair of symbols even after matched filtering. The extent to which symbol offset is good in this case has not been carefully studied.

The use of channel coding will ameliorate phase asynchrony further. This is because with channel coding, the information on each source symbol is generally embedded in a number of channel-coded symbols through the channel coding process. So, if a particular channel-coded joint symbol at the relay has a bad constellation point, there is a chance that another channel-coded joint symbol that also contains information on a common source symbol is a good constellation point. That is, the diversity and certainty propagation effects also enter the picture in the channel-coded case. In Section 3.2.2, we will examine some results. We will see that channel coding can actually turn things around in the other direction so that phase penalty becomes a phase reward, even when the symbols are perfectly aligned at the relay.

### 3.2. Channel-coded PNC

For reliability, almost all communication systems make use of channel coding to protect the information being transmitted. An issue of interest is therefore how channel coding can be integrated into the PNC system. There are two possibilities: end-to-end channel coding and link-by-link channel coding. The discussion in this section focuses on TWRC.

For both end-to-end and link-by-link approaches, the two end nodes channel-code their packets before sending them out. Denote the source packets of nodes 1 and 2 by $S_1 = (s_1[n])_{n=1,\ldots,M}$ and $S_2 = (s_2[n])_{n=1,\ldots,M}$, respectively, After channel coding, they transmit $X_1 = C(S_1) = (x_1[n])_{n=1,\ldots,N}$, $X_2 = C(S_2) = (x_2[n])_{n=1,\ldots,N}$, respectively, to the relay. Here, we assume both end nodes use the same channel code $C(\cdot)$ with redundancy factor $N/M$.

End-to-end and link-by-link channel coding approaches differ in how the relay processes the received information. In the end-to-end approach, the relay does not try to perform channel decoding and re-encoding. For example, the relay may simply try to recover $x_1[n] \oplus x_2[n]$ in a symbol-by-symbol manner and pass the symbols along to the end nodes. In particular, the correlations among different symbols induced by channel coding are not exploited to improve the reliability of the detected $x_1[n] \oplus x_2[n]$. For the symbol-synchronous case, for example, the detection of $x_1[n] \oplus x_2[n]$ is based on solely on $y_R[n]$ and not on $y_R[m]$, $m \neq n$. At an end node, say node 1, the self information $(x_1[n])_{n=1,\ldots,N}$ is subtracted the received signal, leaving behind $(x_2[n])_{n=1,\ldots,N}$ plus noise; after that, channel decoding is applied to recover $(s_2[n])_{n=1,\ldots,M}$.

In the end-to-end approach, channel coding is transparent to the network-coding system. That is, channel coding can be considered as being applied at an upper layer above the network-coding system at only the end nodes. At the higher layer where channel coding and decoding are performed, the end nodes simply treat the PNC system as a bit pipe with certain bit error rate; thus, on an end-to-end basis, the system is similar to a traditional point-to-point channel.

Compared with the link-by-link approach, the end-to-end approach is simpler. However, it has two shortcomings: first, because the relay does not make use of the correlations among symbols, the detection of $x_1[n] \oplus x_2[n]$ is more error prone; second, in a system with multiple relays between nodes 1 and 2, the errors may accumulate because the relays do not clean up the errors.

In the link-by-link approach, the relay makes use of the correlations among successive symbols to recover the desired network-coded symbols with more accuracy. Potentially, channel coding and network coding functionalities can be integrated together for better performance. There are many subtleties and nuances on how this can be done. Our focus in this section is on the link-by-link approach. In particular, we discuss several options on how to integrate channel coding and network coding at the relay.

**System Model**

For a focus, we assume BPSK or QPSK symbols for $x_1[n]$ and $x_2[n]$. The system model as depicted in Fig. 10. Nodes 1 and 2 send $X_1 = C(S_1)$ and $X_2 = C(S_2)$, respectively. We assume $C(\cdot)$ is a linear code so that $X_1 \oplus X_2 = C(S_1 \oplus S_2) = C(S_1) \oplus C(S_2)$. The observed signal at the relay is $Y_R = (y_R[n])_{n=1,...,N}$, where $y_R[n] = h_{1R}x_1[n] + h_{2R}x_2[n] + w_R[n]$. For the time being, we assume the symbols are aligned. Section 3.2.2 will discuss symbol-asynchronous case.

At the relay, a Channel-decoding-Network-Coding operator $CNC(\cdot)$ [25] produces an estimate for $S_R = (s_R[n])_{n=1,...,N} \triangleq S_1 \oplus S_2 = (s_1[n] \oplus s_2[n])_{n=1,...,N}$ based on $Y_R = (y_R[n])_{n=1,...,N}$. That is, the estimate is $\widehat{S}_R = CNC(Y_R) = (\widehat{s}_R[n])_{n=1,...,N}$. We assume the relay uses the same channel code as the two end nodes. After obtaining $\widehat{S}_R$, the relay performs channel coding on $\widehat{S}_R$ to obtain $X_R = C(\widehat{S}_R)$ for broadcast to nodes 1 and 2.

In general, there are different designs for $CNC(\cdot)$ with different performances and implementation complexity. Much of our discussion on link-by-link channel coded PNC focus on $CNC(\cdot)$ because it is the critical component in the system responsible for noise cleaning and PNC mapping.

*PNC Mapping for the Channel-Coded PNC System*

Although the overall PNC mapping at the relay is $Y_R \to X_R$, conceptually, the process can be broken into two steps: (i) $Y_R \to \widehat{S}_R$ with $\widehat{S}_R = CNC(Y_R)$; and then (ii) $\widehat{S}_R \to X_R$ with $X_R = C(\widehat{S}_R)$. Step (ii) is similar to conventional channel coding. Step (i) is the new component introduced by PNC.

We also remark that for the downlink broadcast of $X_R$ to the end nodes, the situation at each end node is the same as that of point-to-point communication. Consider the reception at node 1. Suppose that there is no error in the uplink so that $\widehat{S}_R = S_R$. As long as node 1 decodes $S_R$ correctly, then after the subtraction of self information $S_1$ from $S_R$, $S_2$ is decoded correctly. Conversely, a symbol error in the decoded $S_R$ will also result in a corresponding symbol error in the decoded $S_2$. Thus, the main subtlety in the overall system lies in the operation of $CNC(\cdot)$ in the uplink communication. Therefore, our discussion in this section will focus on $CNC(\cdot)$ in the uplink.

In the next two subsections, we describe a number of different designs for $CNC(\cdot)$ assuming XOR mapping. As already mentioned, and worth emphasizing again, XOR mapping over BPSK or QPSK is only one form of PNCF. The target network coding in the $CNC(\cdot)$ designs discussed here can be replaced with another form of PNCF other than XOR. That said, the insights obtained from the comparison of different XOR designs also apply to non-XOR designs in general.

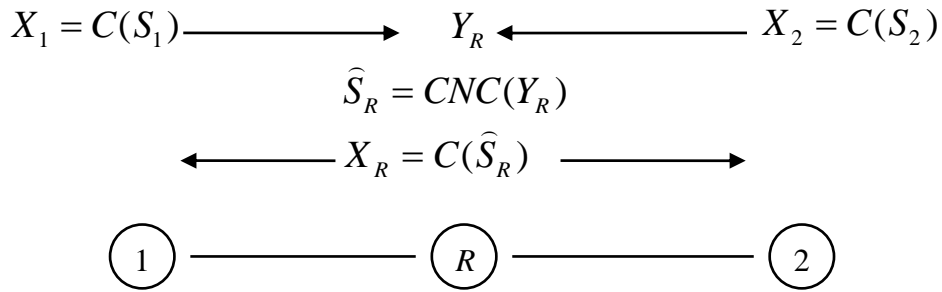

Fig. 10. System model of the relay of a link-by-link channel-coded PNC system.

### 3.2.1. Synchronous Channel-coded PNC

We begin with the simplest case in which perfect power control, precoding, and synchronization are applied so that $h_{1R} = h_{2R} = 1$; furthermore, the symbols of the two end nodes arrive at the relay perfectly aligned. In this case, $Y_R = (y_R[n])_{n=1,...,N}$, where $y_R[n] = x_1[n] + x_2[n] + w_R[n]$. There are various options to realize $CNC(Y_R)$. We will survey a few designs proposed in the literature, and then consider a new design. The discussion of the new design will lead to a general framework for $CNC(\cdot)$ in which the assumptions of perfect power control, precoding, and synchronization can be removed.

**MUD-XOR**

The first method for $CNC(\cdot)$ is depicted in Fig. 11, in which the operation $CNC(\cdot)$ is enclosed in the dashed. We refer to this scheme as $CNC_{MUD-XOR}(\cdot)$. The subscript *MUD-XOR* refers to the fact that we first use the multiuser detection technique (MUD) to channel-decode the individual source packets from the end nodes, $S_1$ and $S_2$ [26]; after that, we apply XOR network coding on the estimates $\hat{S}_1$ and $\hat{S}_2$ to obtain $\hat{S}_R = \hat{S}_1 \oplus \hat{S}_2$. For simpler reference, we will also refer to $CNC_{MUD-XOR}(\cdot)$ simply as *MUD-XOR*. There are different possibilities for the MUD component. One possibility, for example, is successive interference cancellation (SIC) [26] [27].

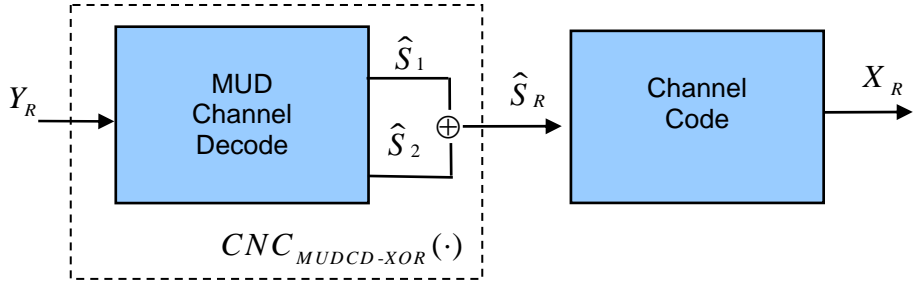

Fig. 11. $CNC_{MUD-XOR}(\cdot)$.

We remark that in *MUD-XOR*, the functionalities of channel decoding and network coding are disjoint operations. Specifically, we first channel decode the individual source information $S_1$ and $S_2$ before performing the XOR network coding operation on them.

In general, *MUD-XOR* is an overkill because for PNC, it is not necessary to decode $S_1$ and $S_2$ individually; only the XOR is needed. Ref. [28] showed that this is generally a suboptimal method. However, it can achieve the symmetric exchange capacity of TRWC at the low SNR region (also see Sections 4.1.1 and 4.1.2 of this paper) [28] [29]; the symmetric exchange capacity is defined as the capacity when nodes 1 and 2 want to transmit equal amount of information per unit time to each other. Generally, schemes like this work well in the low SNR regime, but not the high SNR regime.

**XOR-CD**

The second method for $CNC(\cdot)$ is depicted in Fig. 12. We refer to this scheme as $CNC_{XOR-CD}(\cdot)$, or simply as *XOR-CD*. The acronym *XOR-CD* refers to a two-step process, in which we first apply symbol-by-symbol PNC mapping on the channel-coded symbols to obtain information on the XOR: $x_1[n] \oplus x_2[n]$, $n = 1,...,N$; after that, we perform channel decoding on $X_1 \oplus X_2$ to obtain $S_R = S_1 \oplus S_2$. Note that in the first block in Fig. 12, we obtain the soft information in the form of the probability distributions of XOR of successive symbol pairs: $P(x_1[n] \oplus x_2[n] | y_R[n])$ for $n = 1,...,N$. This part is exactly the same as the PNC mapping in unchannel-coded PNC except that now we apply the mapping on the channel-coded symbols.

The first block in Fig. 12 is entirely a symbol-wise PNC operation that does not make use of the correlations among the successive symbols induced by channel coding. The second block in Fig. 12 is the channel-decode part. It exploits the correlations to obtain $\hat{S}_R = (\hat{s}_R[n])_{n=1,...,M}$. In both *MUD-XOR* and *XOR-CD*, the channel-decoding and network-coding operations are disjoint, albeit in different ways.

*XOR-CD* has a nice feature when a linear channel code is used. Specifically, if a linear channel code is used, the same channel-decode operation for point-to-point communication can be used here, as implied by the following results:

$$C(S_R) = C(S_1) \oplus C(S_2) = X_1 \oplus X_2 \Rightarrow S_R = C^{-1}(X_1 \oplus X_2) \tag{16}$$

The same channel decoder $C^{-1}(\cdot)$ can be used for input $P(x_1[n] \oplus x_2[n] | y_R[n])$ to obtain $\widehat{S}_R = (\widehat{s}_R[n])_{n=1,...,M}$.

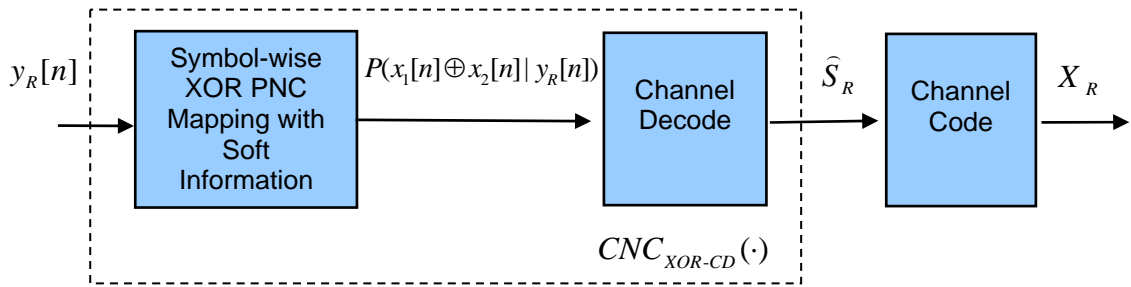

Fig. 12. $CNC_{XOR-CD}(\cdot)$.

This scheme was studied in [30]. It was also investigated in [31] in the context of an OFDM-PNC system. Ref. [25] pointed out that this scheme is suboptimal because the information contained in the observable $y_R[n]$ is embodied in $P(x_1[n] + x_2[n] | y_R[n])$, and some of the useful information for the operation $CNC(\cdot)$ is lost in the reduction of $P(x_1[n] + x_2[n] | y_R[n])$ to $P(x_1[n] \oplus x_2[n] | y_R[n])$ in the *XOR-CD* scheme.

Ref. [32] considered a scheme with a similar spirit as *XOR-CD*, using the nested lattice code instead of XOR. As in here, PNC Mapping is first applied on the channel-coded signal received before a separate channel-decoding operation. It is shown that the scheme can approach the exchange capacity of TWRC in the high SNR region. Further discussion of the results of the lattice-coded scheme [16] [33] can also be found in Sections 4.1.1 and 4.1.2 of this paper. Generally, schemes like this work well in the high SNR regime, but not in the low SNR regiem.

**AS-CNC**

The third method for $CNC(\cdot)$ is depicted in Fig. 13. This scheme was proposed in [25]. In this paper, we refer to this scheme as $CNC_{AS}(\cdot)$, or simply as $AS-CNC$ (in [25], this scheme is referred to as *ACNC*). The subscript *AS* reflects the fact that we first obtain the probability distributions of the *arithmetic sums* (AS) of the symbol pairs from nodes 1 and 2 in the first block; and then feed the probability distributions to the second block for joint network-coding channel-decoding operation. In particular, the channel-decoding and the XOR network-coding operations are integrated and not disjoint.

A key idea of $AS-CNC$ is as follows. A reason why *XOR-CD* does not work well in the low SNR region is that the XOR mapping in the first block in Fig. 12 loses useful information. $AS-CNC$, on the other hand, retains all useful information in the observation $Y_R$ in the first block. In the case where $y_R[n] = x_1[n] + x_2[n] + w_R[n]$, the network coding operation performed by nature is the arithmetic sum. Thus, if we adopt the arithmetic-sum (AS) network coding in the first block in Fig. 13, no information is lost. Mapping of the AS network-coded symbols to the desired XOR network-coded symbols is then done in conjunction with channel decoding in an integrated manner in the second block. In particular, the second block makes use of the soft information $P(x_1[n]+x_2[n] | y_R[n])$, $n=1,...,N$, to estimate $s_R[n] = s_1[n] \oplus s_1[n]$, $n=1,...,M$. A specially designed channel decoder has to be used for the second block in $CNC_{AS}(\cdot)$ (i.e., the original channel decoder $C^{-1}(\cdot)$ for point-to-point communication cannot be used even if $C(\cdot)$ is a linear channel code) [25].

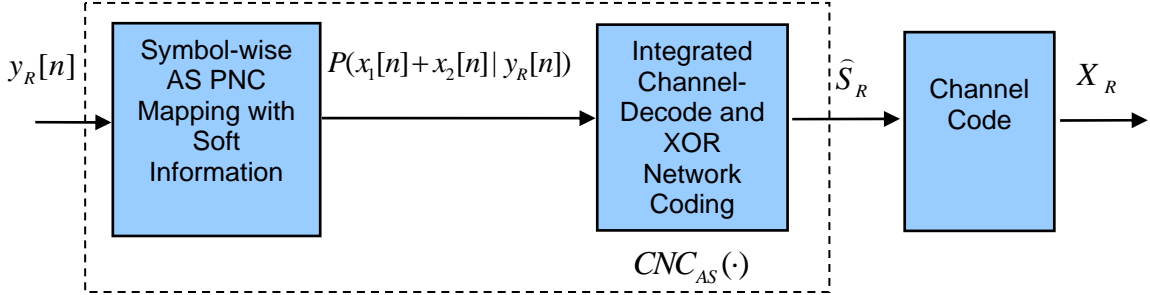

Fig. 13. $CNC_{AS}(\cdot)$

To summarize, in *MUD-XOR*, channel decoding to individual source information of the two end nodes is first applied, followed by XOR network coding of the source information. These two operations are disjoint. In *XOR-CD*, XOR network coding on the channel-coded information is first applied, followed by channel decoding of the XOR channel-coded information to XOR source information. These two operations are also disjoint. In $AS-CNC$, $XOR$ network coding and channel decoding are performed in an integrated manner. The two operations are not disjoint.

Fig. 14 shows some results adapted from [25] that compare the BER performance of the three schemes for various packet lengths. Here, the BER refers to the bit error rate of $s_R[n]$ at the output of the network-coded-channel-decoding operator, $CNC(\cdot)$. For $C(\cdot)$, the use of a repeat-accumulate (RA) code with a repeat factor of three is assumed. As can be seen, the relative performance of $AS-CNC$ is the best. The important insight here is not to lose information in $P(x_1[n]+x_2[n] | y_R[n])$ before the channel decoding process.

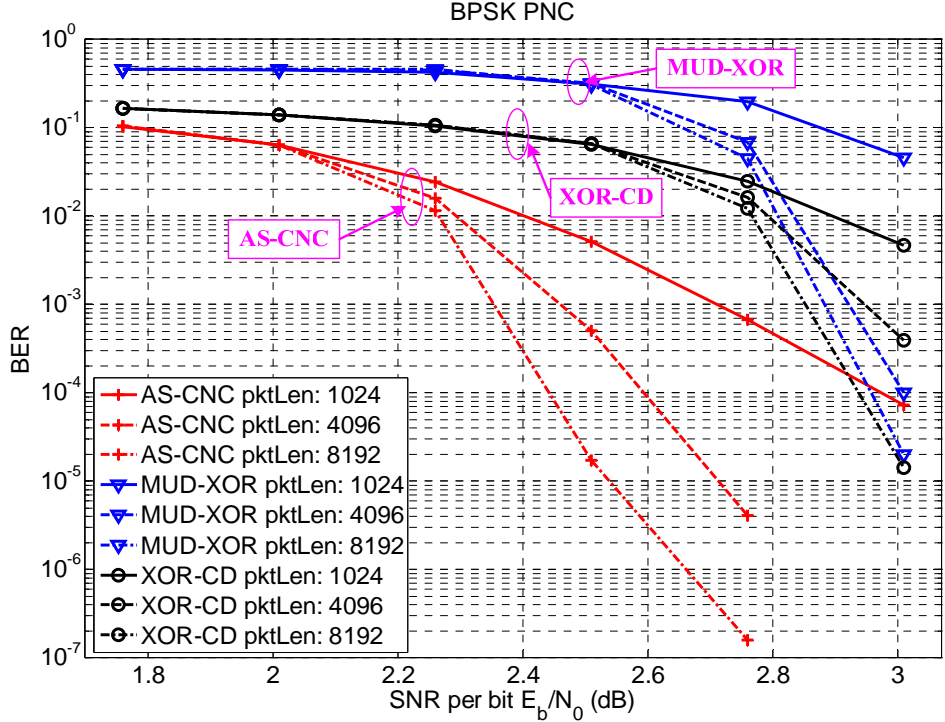

Fig. 14. Comparison of *MUD-XOR*, *XOR-CD*, and *AS−CNC*. Note: $E_b$ here is the energy per source bit; with channel-coding redundancy of three, energy per channel-coded bit is three times lower than $E_b$ here.

**Joint CNC**

Another possibility that does not lose information is to pass the joint probability distribution $P(x_1[n], x_2[n] | y_R[n])$ to the channel decoder. From $y_R[n] = x_1[n] + x_2[n] + w_R[n]$, $P(x_1[n], x_2[n] | y_R[n])$ can be computed. Note that $P(x_1[n], x_2[n] | y_R[n])$ does not lose information because $P(x_1[n] + x_2[n] | y_R[n])$ can be computed from $P(x_1[n], x_2[n] | y_R[n])$. On the other hand, $P(x_1[n] \oplus x_2[n] | y_R[n])$ in $CNC_{XOR-CD}(\cdot)$ loses information because $P(x_1[n] + x_2[n] | y_R[n])$ cannot be recovered from $P(x_1[n] \oplus x_2[n] | y_R[n])$.

In this paper, we refer to the CNC process that makes use of $P(x_1[n], x_2[n] | y_R[n])$, and that performs the CNC process as an integrated process rather than two disjoint channel decoding and network coding processes, as $CNC_{Jt}(\cdot)$, or simply as *Joint CNC*. It can be shown than *Joint CNC* and $AS-CNC$ have the same performance.

The reason for dealing with *Joint CNC* rather than $AS-CNC$ is not so much the former is more efficient than the latter. Rather, *Joint CNC* is amenable to a general framework in which the assumption of perfect power control, synchronization, and precoding can be removed. Section 3.2.2 will focus on this general framework of *Joint CNC*.

### 3.2.2. Asynchronous Channel-coded PNC

Let us now look at the case where there are symbol and phase asynchronies in the channel-coded PNC system.

**Asynchronous XOR-CD**

First of all, we note that the synchronous *XOR-CD* discussed in the preceding section can be extended for asynchronous operation as follows. We first use the asynchronous unchannel-coded PNC method discussed in Section 3.1.3 to obtain $P(x_1[n] \oplus x_1[n] | Y_R), n = 1,...,N$. This takes care of the asynchrony. Then, we feed $P(x_1[n] \oplus x_1[n] | Y_R), n = 1,...,N$, as the input to second block in *XOR-CD* (see Fig. 12). This approach is simple and will have good performance in the high SNR regime. However, it is not optimal in general. First, as in synchronous *XOR-CD*, the mapping in the first block loses information. Second, the correlations among successive symbols induced by symbol offset and the correlations among symbols induced by channel coding are exploited one after another rather than jointly. By this, we mean that the BP algorithm in Section 3.1.3 and the channel decoding are executed one after another. We will present some results of asynchronous *XOR-CD* later.

**Asynchronous Joint CNC**

In the following, we consider a framework for asynchronous *Joint CNC*. It turns out the framework can also be used to construct a design for asynchronous *MUD-XOR* as well. This will be discussed later.

Recall that the key idea in $AS-CNC$ is not to lose information in the CNC process. Instead of the conditional probability of arithmetic sum, we could also try to get $P(x_1[\cdot],x_2[\cdot]|y_R[\cdot])$ in the first block and use it as the input to the second block in Fig. 13. This approach is depicted in Fig. 15. Recall also that the asynchronous PNC decoding method in Section 3.1.3 (for unchannel-coded PNC) also take as input $P(x_1[\cdot],x_2[\cdot]|y_R[\cdot])$. This suggests the construction of a general Tanner graph for the second block in Fig. 15 that incorporates everything together.

The Tanner graph is shown in Fig. 16 and explained below. Our framework can be extended to the case in which power equalization is not performed, i.e., $|h_{1R}|\neq|h_{2R}|$. However, for simplicity, we assume $|h_{1R}|=|h_{2R}|=1$ so that we can focus on symbol and phase asynchronies only. We continue our discussion from the equations in (13).

From $y_R[\cdot]$ in (13), we can obtain the joint probability distributions $P(x_1[n],x_2[n-1]|y_R[2n-1])$, $P(x_1[n],x_2[n]|y_R[2n])$, $n=1,N$, and $P(x_2[N]|y_R[2N+1])$, assuming the phase offset $\phi$ can be estimated by the relay and is therefore known. The computation of these probabilities is done by the first block of the system diagram in Fig. 15. In the second block, we feed these joint probability distributions to a decoder that takes care of the handling of asynchrony, network coding, and channel decoding, in an integrated manner.

The design of the second block is embodied in the Tanner graph shown in Fig. 16 [34]. The inputs $P(x_1[n],x_2[n-1]|y_R[2n-1])$, $P(x_1[n],x_2[n]|y_R[2n])$, $n=1,N$, and $P(X_2[N]|y_R[2N+1])$ are fed to the bottom of the Tanner graph. That is, the $Y_R$ nodes of the Tanner graph feed these inputs as observations in the overall sum-product algorithm.

The bottom three rows of nodes in Fig. 16 are the same as Fig. 8, which was the design used to deal with symbol and phase asynchronies in unchannel-coded PNC. The part above the bottom three rows is related to the decoding of the source joint symbols. Fig. 15 assumes the use of the same RA code with repeat factor three at nodes 1 and 2. In the figure, the source joint symbols are $s^{n,n}=(s_1[n],s_2[n])$, $n=1,...,M$. The source joint symbols after 3-repeat and interleaving are $\tilde{s}^{n,n}=(\tilde{s}_1[n],\tilde{s}_2[n])$, $n=1,...,N$ where $N=3M$. The $\oplus$ in the figure are pairwise 2-tuple XOR: that is, $(a,b)\oplus(c,d)=(a\oplus b,c\oplus d)$. In other words, only the symbols of the same source mixed with each other, which is the case for channel coding at the end nodes 1 and 2.

A sum-product message passing algorithm can then be constructed based on the Tanner graph to estimate $P(s_1[n],s_2[n]|Y_R)$ [34]. We can then obtain

$$P(s_1[n]\oplus s_2[n]=a|Y_R)=\sum_{\substack{s_1[n],s_2[n]:\\s_1[n]\oplus s_2[n]=a}}P(s_1[n],s_2[n]|Y_R) \qquad (17)$$

We then set

$$\hat{s}_R[n]=\arg\max_a P(s_1[n]\oplus s_2[n]=a|Y_R) \qquad (18)$$

This corresponds to MAP (also ML) decoding of $s_1[n] \oplus s_2[n]$. The relay then performs channel coding on $\hat{S}_R = (\hat{s}_R[n])_{n=1,...,M}$ to obtain $X_R = (x_R[n])_{n=1,...,N}$ for broadcast to nodes 1 and 2.

We remark that strictly speaking, the computation of $P(s_1[n], s_2[n]|Y_R)$ using the sum-product algorithm is only approximate because the Tanner graph has loops. In general, the exact computation of $P(s_1[n], s_2[n]|Y_R)$ is a tough problem. Consequently, the $P(s_1[n] \oplus s_2[n]|Y_R)$ obtained is also approximate. Thus, this method is an approximate ML decoding of $s_1[n] \oplus s_2[n]$.

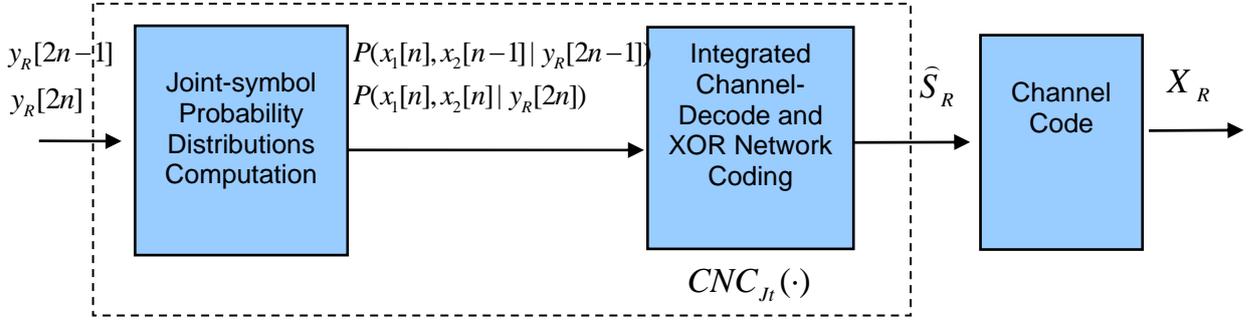

Fig. 15. $CNC_{Jt}(\cdot)$.

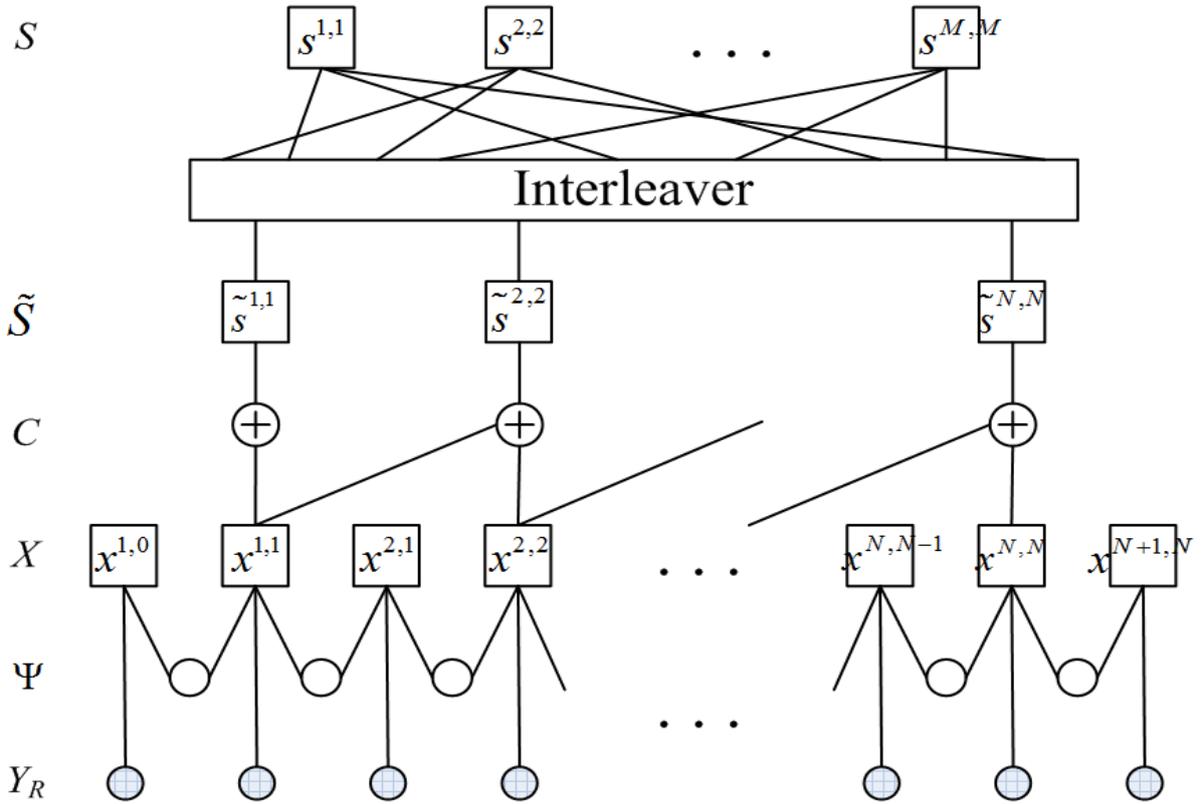

Fig. 16. Tanner graph for finding $P(s_1[n] \oplus s_2[n]|Y_R), n = 1,...,M$ based on $P(x_1[n], x_2[n-1]|y_R[2n-1])$, $P(x_1[n], x_2[n]|y_R[2n-1])$, $n = 1, N$, and $P(X_2[N]|y_R[2N+1])$, assuming the use of RA code with repeat factor three.

**Asynchronous MUD-XOR**

The BP framework for *Joint CNC* can also be used for a version of asynchronous *MUD-XOR*. To do so, we run the sum-product algorithm exactly as in *Joint CNC*. However, we compute $\hat{s}_R[n]$ differently. We first find the ML $(s_1[n], s_2[n])$ from

$$(\hat{s}_1[n], \hat{s}_2[n]) = \arg\max_{(c,d)} P(s_1[n] = c, s_2[n]) = d \mid Y_R) \qquad (19)$$

From (19), we compute

$$\hat{s}_R[n] = \hat{s}_1[n] \oplus \hat{s}_2[n] \qquad (20)$$

This $\hat{s}_R[n]$ is not the ML $s_1[n] \oplus s_2[n]$, but rather the $s_1[n] \oplus s_2[n]$ obtained from the ML $(s_1[n], s_2[n])$. Thus, in general, the symbol error rate will be higher.

We remark that this *MUD-XOR* is different from the *MUD-XOR* used to generate the results for the synchronous case in Fig. 14. In the *MUD-XOR* of Fig. 14, SIC was used. Here, we use the Tanner graph to decode $(s_1[n], s_2[n])$ jointly.

**Numerical Results**

We now look at some numerical results assuming QPSK modulation. We adopt the regular RA code with a coding rate of 1/3 in our simulations, as in as in [25]. In the graphs of BER versus SNR per bit to be presented shortly, for each data point, we simulated 10,000 packets of 4,096 bits. These 4,096 bits are divided into in-phase and quadrature parts, each having 2,048 bits.

**Asynchronous Joint CNC**

We will first examine the results of *Joint CNC*, our main interest. Thanks to its ML decoding of $s_1[n] \oplus s_2[n]$, *Joint CNC* has the best BER performance of $s_1[n] \oplus s_2[n]$. Therefore, it is most revealing as far as the *fundamental* effects of symbol and phase asynchronies are concerned.

In Fig. 17(a), we show the case in which there is no symbol offset. Recall that in unchannel-coded PNC, phase offset induces a penalty (see Fig. 9(a)). For the channel-coded case, instead of a phase penalty, there is a phase reward. With reference to Fig. 17(a), the BER of $\hat{s}_R[n]$ is actually smaller when $\phi \neq 0$. In Fig. 17(b), we show the case in which there is a symbol offset of $\Delta = 0.5$. The performance is better than when $\Delta = 0$. In addition, the phase reward is larger when $\Delta = 0.5$.

In general, for a given BER performance, the power spread due to different phase offsets is less than 1dB regardless of $\Delta$. Thus, we see that channel coding has the effect of desensitizing the system performance to the effect of phase offset significantly. That is, in addition to *phase reward*, there is also *phase robustness*.

The absence of phase penalty in channel-coded PNC can also be explained by the diversity and certainty propagation effects, as in the unchannel-coded PNC with symbol offset except that symbol offset is not required in channel-coded PNC to have these effects. With channel coding, the information on each source symbol is embedded in multiple channel-coded symbols. This is analogous to the situation in the symbol-misaligned unchannel-coded PNC where each source symbol pairs with two other symbols from the other end node.

A last observation is that, as expected, the BER performance with channel coding is much better than without channel coding. For fair comparison between unchannel-coded PNC and channel-coded PNC, in Fig. 17 (for the latter) we shift the curves by $10\log_{10} 3$ dB to the right to take into account that each bit is repeated 3 times in our RA channel coding; that is, the energy in Fig. 17 is energy per source bit, or the total energy of three channel-coded bits.

**Asynchronous XOR-CD**

For comparison, let us now look at the performance of asynchronous *XOR-CD*. Recall that for asynchronous *XOR-CD*, the PNC XOR processing is first performed on the channel-coded information (using the asynchronous unchannel-coded PNC method discussed in Section 3.1.3) to obtain $P(x_1[n] \oplus x_1[n] | Y_R)$, $n = 1,...,N$ ; after that, channel-decoding is performed on the soft information of $x_1[n] \oplus x_1[n]$ to obtain an estimate of $s_1[n] \oplus s_1[n]$ using the traditional channel decoder for point-to-point communication. The two processes are disjoint.

Fig. 18 shows the BER results of asynchronous *XOR-CD*. In general, we can see that this scheme, although less complex [5] than *Joint CNC*, has significantly worse performance. In addition, instead of phase reward, there is phase penalty. Its performance is far from what could be achieved fundamentally. The phase penalty is due to its suboptimality.

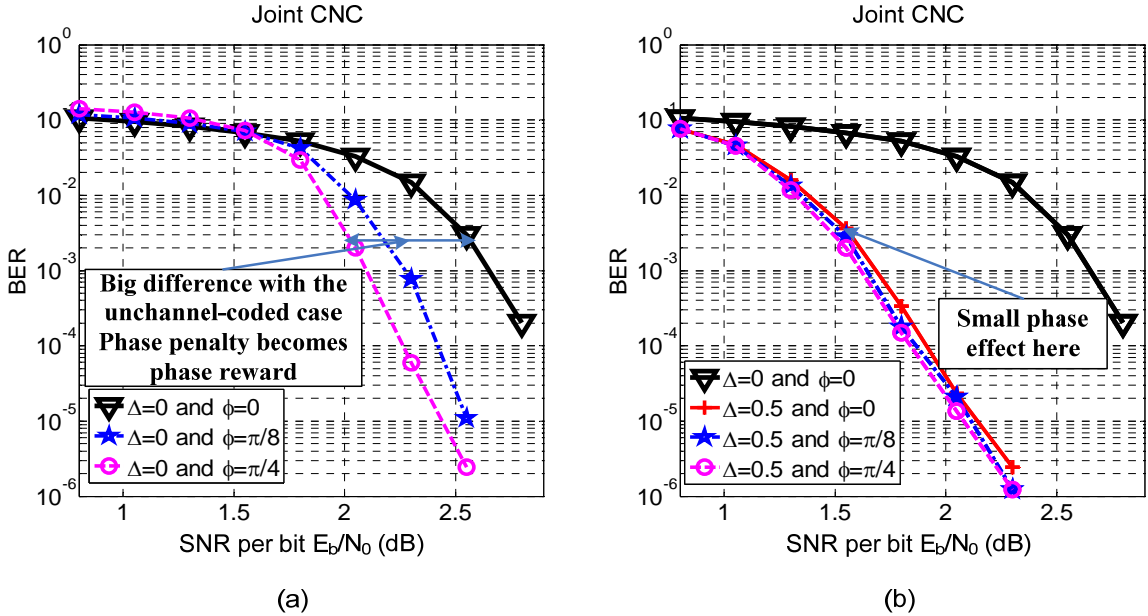

Fig.17. BER of $\hat{s}_R[n]$ in *Joint CNC* for QPSK modulated channel-coded PNC using RA code with repeat factor three : (a) without symbol asynchrony ($\Delta = 0$) ; (b) with symbol asynchrony ($\Delta \neq 0$). Note that $E_b$ is energy per source bit here.

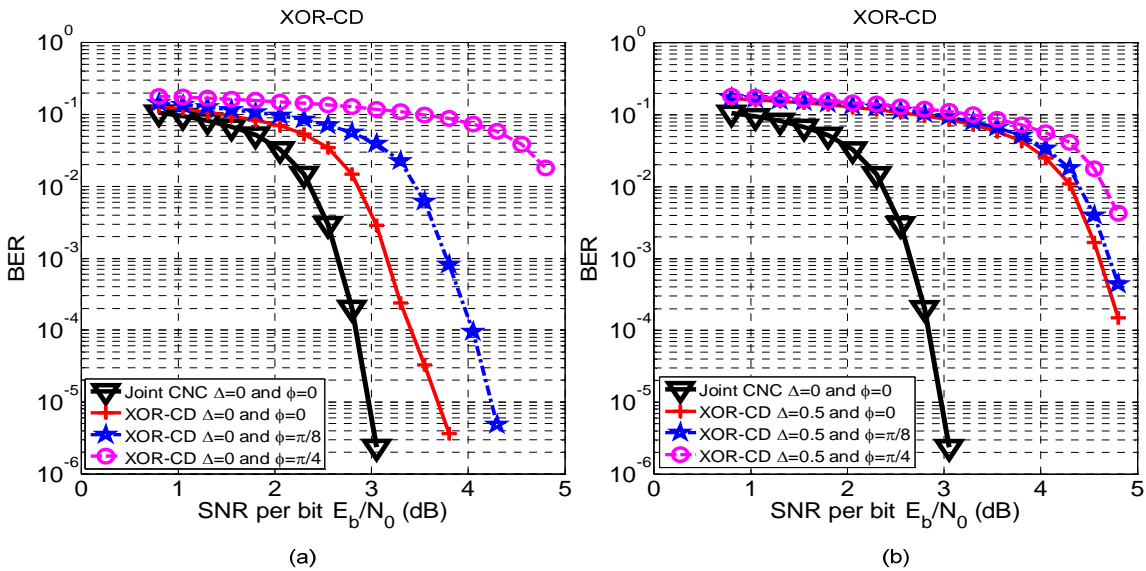

---

[5] The complexity of *Joint CNC* under QPSK modulation is due to the 16 combinations of $(x_1[n], x_2[n])$ and $(s_1[n], s_2[n])$ in the branches of the Tanner graph that the sum-product algorithm has to compute over. For *XOR-CD,* each branch has only 4 combinations in the channel decoding part, thanks to the XOR operation prior to channel decoding.

Fig. 18. BER of $\hat{s}_R[n]$ in *XOR-CD* for QPSK modulated channel-coded PNC using RA code with repeat factor three : (a) without symbol asynchrony ($\Delta = 0$) ; (b) with symbol asynchrony($\Delta \neq 0$). Note that $E_b$ is energy per source bit here.

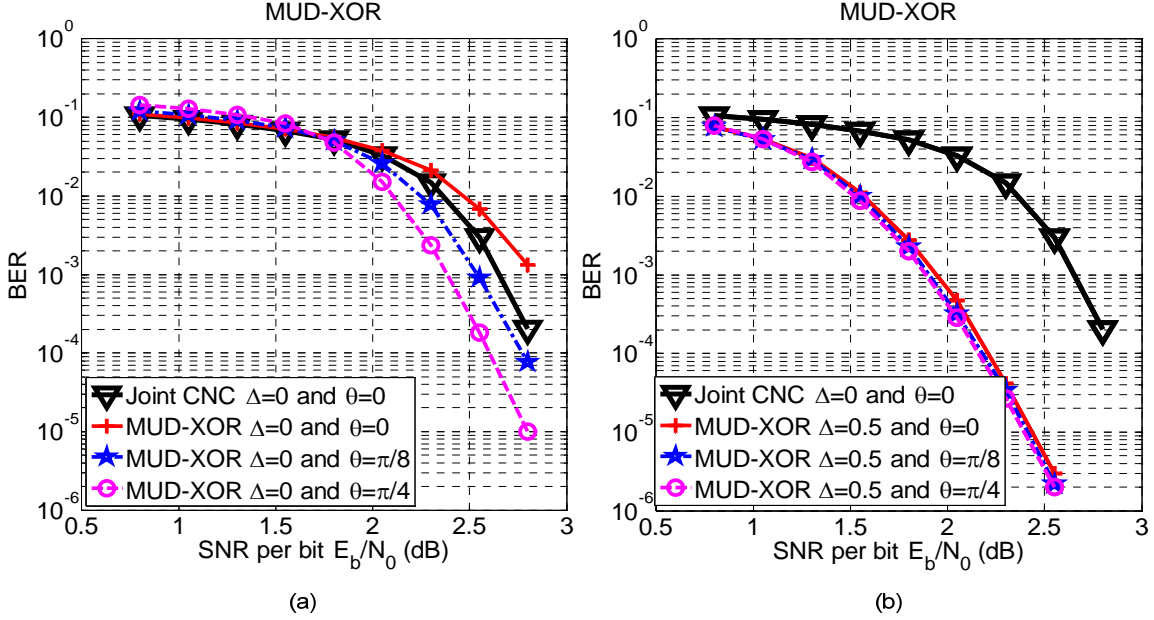

(a)  (b)

Fig 19. BER of $\hat{s}_R[n]$ in *MUD-XOR* for QPSK modulated channel-coded PNC using RA code with repeat factor three : (a) without symbol asynchrony ($\Delta = 0$) ; (b) with symbol asynchrony($\Delta \neq 0$). Note that $E_b$ is energy per source bit here.

**Asynchronous MUD-XOR**

We now look at the performance of asynchronous *MUD-XOR*. Recall that for our particular implementation of asynchronous *MUD-XOR*, the same decoding framework as asynchronous *Joint CNC* is used. Both compute $P(s_1[n] = c, s_2[n]) = d \mid Y_R)$ using the same Tanner graph; however, they use the computed $P(s_1[n] = c, s_2[n]) = d \mid Y_R)$ differently to obtain different estimates for $\hat{s}_R[n]$.

Fig. 19 shows the BER results of asynchronous *MUD-XOR*. Its performance is slightly worse than that of *Joint CNC*, but still much better than *XOR-CD*. In addition, similar phase reward and phase robustness as in *Joint CNC* are present in *MUD-XOR*.

### 3.3. To Probe Further

**Channel-coded PNC**

Refs. [35] and [36] considered a setup similar to that of Fig. 15, but with no symbol asynchrony. The issue being studied is whether the complexity of the ML decoding as embodied by the strategy in Fig. 15 can be reduced, with the assumption of the use of convolutional codes rather than RA codes, and BPSK modulation. It was shown that a reduced-state trellis can be constructed to reduce the complexity of the decoder with some performance penalty. We remark that because of the merging of multiple states into the XOR states during the decoding process of the reduced-state trellis, it is possible that the "certainty propagation effect" mentioned in our article here will vanish and that there will be a phase penalty associated with the reduced-state trellis approach. This conjecture remains to be further studied.

Ref. [37] also considered a similar setup as in Fig. 15, albeit with symbol synchronization. The use of Low-Density Parity-Check (LDPC) codes is assumed. As in our paper here, the basic idea is ML decoding. The qualitative results are similar to the results presented here.

In ANC, the relay aims to broadcast $h_{1R}x_1[n]+h_{2R}x_2[n]$, $n=1,...,M$, to the end nodes. In a simple end-to-end channel-coded ANC system, the relay estimates $h_{1R}x_1[n]+h_{2R}x_2[n]$ on a symbol-by-symbol basis. That is, symbols for different $n$ are estimated independently. This simple strategy does not exploit the correlations among different symbols induced by channel coding. How such correlations can be exploited to improve the estimate of $h_{1R}x_1[n]+h_{2R}x_2[n]$ was investigated in [38].

Ref. [39] investigated a system in which the relay explicitly decodes $S_1$ and $S_2$ in the uplink phase, but uses a combination of XOR and superposition coding for the downlink broadcast case. The focus is on how to deal with asymmetric channels to nodes and 2. Supposition coding has better performance than XOR in the broadcast phase when the downlinks are asymmetric.

**Synchronization Issues**

For unchannel-coded PNC, when the symbol offset is more than one symbol, say $L+\Delta$, the relay can do the following to form the network-coded packet. It can separately decode the head end and tail end of the non-overlapping parts, and then jointly decode the overlapping part using the scheme in Fig. 8 with an offset of $L$ symbols from the beginning of the packet from node 1. The network-coded packet of $N$ symbols can then be formed from the XOR of the head end and tail end, and the XOR of the overlapping part. That is, the final network-coded packet consists of two parts, as follows:

$$(x_1[1]\oplus x_2[N-L+1],..., x_1[L]\oplus x_2[N]) \quad (x_1[L+1]\oplus x_2[1],... , x_1[N]\oplus x_2[N-L])$$

Things are more complicated for channel-coded PNC when the symbol offset is $L+\Delta$. Ref. [40] considered the case when the convolutional channel-code is used, assuming $\Delta=0$. In practice, it may be unrealistic to assume that one can align the symbols so that $\Delta=0$ when one cannot even ensure the larger time-scale alignment of $L=0$. It will be interesting to study if the system in [40] can be modified incorporate nonzero $\Delta$. As for the RA code (or LDPC code) considered in our paper here, the Tanner graph will need to be modified. This will be an interesting subject for further study as well.

Ref. [31] is a first paper that explored OFDM PNC. Here, we explain the main essence and motivations. OFDM and PNC is an interesting combination in that symbol offset in the time domain is transformed to a phase term in each OFDM symbol of a subcarrier when the symbol offset is within the cyclic prefix (CP) of the OFDM system. Simply put, time-domain symbol offset is transformed to frequency-domain phase offset. If $\Delta$ is the time-domain symbol offset, the received signal (ignoring noise) on subcarrier $k$ can be shown to be

$$Y[k]=H_1[k]X_1[k]+e^{-j2\pi\frac{k}{N}\Delta}H_2[k]X_2[k], \quad k=0,..,N-1, \tag{21}$$

where $N$ is the number of subcarriers; $H_1[k]$ and $H_2[k]$ are the channel gains on subcarrier $k$ for the channels from node 1 and node 2 to the relay, respectively; and $X_1[k]$ and $X_2[k]$ are the signal transmitted by nodes 1 and 2 on subcarrier $k$. Thus, on each subcarrier, the symbols from the two end nodes are symbol-aligned with a random phase term. Note that the phase term $e^{-j2\pi\frac{k}{N}\Delta}$ is different for different subcarrier $k$; similarly, $H_i[k]$ could contain a random phase term that is different for different subcarrier $k$. The overall effect is such that there is a randomization of the phase offset across different subcarriers.

Recall that when symbols are aligned (in this case, the subcarrier symbols are aligned) and phase is not, different phase offset may lead to different performance (see Fig. 9 (a)). This means that some subcarriers will have good BER performance while others will have bad BER performance. The use of channel coding on top of the OFDM PNC system will be essential to average out the effect to ensure reliable communication. In summary, OFDM PNC is a natural design for an asynchronous PNC system that does not require deliberate tight-symbol level and phase synchronizations between nodes 1 and 2.

That said, RF carrier frequency offset is a potential issue. That is, if the RF carrier frequencies used by nodes 1 and 2 are not exactly the same, then there will be inter-subcarrier interference in the frequency domain. Specifically, in eqn. (21), we may not be able to ensure that the RHS is a function of subcarrier $k$ alone. The relay could perform signal processing such that the term due to node 1, $H_1[k]X_1[k]$, remains intact, but in that case the term due to node 2, $e^{-j2\pi\frac{k}{N}\Delta}H_2[k]X_2[k]$, will contain inter-subcarrier interferences from the signals on the other subcarriers of node 2. Similarly, if we eliminate inter-subcarrier interference for node 2, node 1 will have inter-subcarrier interference. This is a phenomenon that is absent in point-to-point link, in which if the RF offset between the transmitter and the receiver can be perfectly estimated, then inter-subcarrier can be eliminated altogether. For OFDM PNC, even if both the RF offsets of nodes 1 and 2 (with respect to the relay) can be estimated perfectly, only one of the inter-subcarrier interferences can be eliminated entirely. The strategy to deal with the phenomenon is an interesting topic for further research.

**MIMO PNC**

The use of MIMO in PNC systems can have two benefits: improved throughput performance and/or reduced processing complexity. Ref. [41] investigated MIMO PNC with linear detection to reduce processing complexity and to improve performance. The relay extracts the sum and difference of the two end packets, and then constructs the network-coded packet based on the sum and difference. Ref. [42] analyzed the use of optimal ML decoding at a relay with multiple antennas and concluded that the scheme is much better than the amplify-and-forward scheme.

Ref. [43] analyzed the symbol error rate of a system in which the two end nodes are equipped with two antennas and the relay has only one antenna, assuming the use of Alamouti space-time code. It shows that a diversity order of 2 can be achieved. Ref. [44] also studied a MIMO PNC system with space-time codes. The relay and the two end nodes are each equipped with two antennas.

Ref. [45] and Ref. [46] investigated MIMO ANC schemes in which the relay has multiple antennas and the end nodes have single antenna. Ref. [47] considered a cellular system in which a base station communicates with multiple users via a relay. The base station and the relay are equipped with multiple antennas while the end users are equipped with single antenna.

Refs. [48], [49], [50], [51], [52], [53], [54], and [55] studied systems in which there are multiple relays interconnecting the two end nodes. When there are multiple relays interconnecting the two end nodes, either a relay selection strategy could be adopted, or the multiple relays could work together as a distributed MIMO system.

**Channel Estimation**

Channel estimation and RF carrier frequency estimation are important topics in PNC and ANC systems because the detection at the relay and end nodes counts on the knowledge of the channels and RF carrier frequencies used by the end nodes. Note that the RF carrier frequencies of the end nodes may be offset by a small amount which may affect performance if ignored. If known by the relay, the RF carrier offset can be translated into known rotating phase offsets for successful received symbols in the time domain. For OFDM PNC, RF carrier offset is more problematic. Even if known, the carrier offset can cause inter-subcarrier interferences for at least one of the end node's signal (see discussion under "Synchronization Issues" above). In all cases, unless the RF carrier frequencies used by the transmission and reception at all nodes are synchronized or obtained from the same source, it is important for each receiver to estimate the RF carrier frequencies used by the transmitter.

Refs. [56] and [57] investigated several channel estimation methods for ANC, in which the relay just amplifies and forwards the composite signal to the end nodes, and channel estimation is done by the end nodes for the round-trip channel. Besides channel estimation, [57] also investigated the estimation of carrier frequency offset between the two end nodes.

Ref. [58] considered a two-phase channel estimation scheme for ANC in which the relay also participates in the channel estimation. The idea is to for the relay to reduce noise (denoise) the uplink channel estimation before forwarding the signal along, so that the channel estimation at the end nodes can be more accurate.

Refs. [59], [60], and [61] proposed several schemes for channel estimation when OFDM is used. The first two references considered both the estimation of the composite source-relay-source channels (i.e., end-to-end channels from node 1 to node 2 and vice versa) as well as individual channels between sources and relay; the third considered only the composite source-relay-source channels. Blind channel estimation of the composite channels in OFDM PNC was treated in [62].

Ref. [63] analyzed the performance of OFDM ANC systems in which the self information removal at the end nodes is imperfect. Imperfect self information removal could be due to imperfect channel estimation, imperfect frame synchronization, and carrier frequency offset.

**Multi-way PNC**

Beyond TWRC, it is also possible to apply PNC in systems in which more than two end nodes communicate via a relay. Ref. [64] investigated the use of a single relay to interconnect multiple pairs of nodes. To isolate the different pairs, each node pairs use a unique CDMA signature. Then, each node pair uses PNC for information exchange.

Instead of CDMA signatures, [65] made use of multiple antennas at the relay and end to provide the degrees of freedom needed for decoding. The setup studied in [65], referred to as the Y-channel, consists of three end nodes and a single relay. Instead of pairwise communication, each end node has independent information to be transmitted to the other two end nodes.

Refs. [66], [67], [68] and [69] also considered the setup with multiple end nodes and a single relay. Instead of pairwise communication or each node transmitting independent information to other nodes, a full-exchange broadcast setting is considered. In full-exchange broadcast, each node wants to broadcast the same information to all the other nodes.

**Other Topics**

We have not discussed channel-coded SNC thus far. A straightforward way to implement link-by-link channel-coded SNC is to treat the transmissions of the three time slots as being over three independent point-to-point links. In particular, channel decoding into $S_1$ and $S_2$ is performed independently in first two time slots at the relay. Then, $X_R = C(S_1 \oplus S_2)$ is sent in the third time slot. It is also possible to have end-to-end channel-coded SNC. Ref. [70] investigated simple symbol-by-symbol, memoryless processing without explicit channel decoding at the relay. The idea is for the relay to forward soft rather than hard network-coded symbols. Similar ideas for PNC systems are explored in [12], [19], and [71].

Ref. [72] considered the use of OFDM in ANC. It focuses on maximizing the sum capacity by power allocation and permutation of the OFDM tones. Ref. [73] studied the use of OFDM as well as the use of single carrier with frequency domain equalization (SC-FDE) in ANC when channel fading is frequency selective. A result is that SC-FDE has better BER performance than OFDM ANC, but a lower ergodic capacity.

Refs. [74], [75], and [76] investigated power allocation and consumption issues in PNC systems.

Refs. [77], [78], [79], [80] and [55] considered the use of non-coherent detection at the relay, in which the channel-state information is not available at the transmitter or receiver. To reduce the penalty due to noncoherent detection, Ref. [55] investigated a system in which multiple relays are installed between two end nodes, and a relay is selected from the multiple relay for amplify-and-forward relaying.

Refs. [81], [82], [83] explored the security implication of PNC.

## 4. Information-theoretic Studies

At the most fundamental level, the ultimate network capacity made possible by PNC has to be studied from an information-theoretic perspective. In this article, we restrict our information-theoretic discussions to TWRC. Furthermore, we assume the channels between the two end nodes and the relay are Gaussian channels. For simplicity, we assume the noise powers for all channels are normalized to be one.

### 4.1. Rate Region of Gaussian TWRC

### 4.1.1. Outer Bound for PNC Information Capacities

Let $C_{1R}$ and $C_{2R}$ be the information capacities of the uplink channels from node 1 and node 2 to relay $R$, respectively. In addition, let $C_{R1}$ and $C_{R2}$ be the information capacities of the downlink channels from relay $R$ to node 1 and node 2, respectively. In general, we have

$$C_{iR} = \frac{1}{2}\log_2(1+P_{iR})$$
$$C_{Ri} = \frac{1}{2}\log_2(1+P_{Ri})$$
(22)

where $P_{iR}$ is the power received by relay $R$ from node $i$, $i=1,2$; and $P_{Ri}$ is the power received by node $i$, $i=1,2$, from relay $R$.

Let $t_u \geq 0$ be the fraction of airtime dedicated to the uplink phase during which the relay receives from nodes 1 and 2, and $t_d = 1-t_u$ be the fraction of airtime dedicated to the downlink phase during which the relay transmits to nodes 1 and 2. Let $R_{12}(t_u)$ and $R_{21}(t_u)$ be the information rates from node 1 to node 2 and from node 2 to node 1, respectively, for a given $t_u$. Application of the cut-set bound yields

$$R_{12}(t_u) \leq \min[t_u C_{1R}, (1-t_u)C_{R2}] \triangleq U_{12}(t_u)$$
$$R_{21}(t_u) \leq \min[t_u C_{2R}, (1-t_u)C_{R1}] \triangleq U_{21}(t_u)$$
(23)

where $U_{12}(t_u)$ and $U_{21}(t_u)$ are the upper bounds for $R_{12}(t_u)$ and $R_{21}(t_u)$, respectively.

Fig. 20 shows the trace of a typical locus for the upper bounds $(U_{12}(t_u), U_{21}(t_u))$ as $t_u$ increases from 0 to 1 (the red arrowed lines). In the example, $C_{2R}/C_{R1} \geq C_{1R}/C_{R2}$. When $t_u$ is small, both $U_{12}(t_u)$ and $U_{21}(t_u)$ are limited by the uplink capacities, $t_u C_{1R}$ and $t_u C_{2R}$ respectively. As $t_u$ increases and reaches $t_u = C_{R1}/(C_{2R}+C_{R1})$, then $t_u C_{2R} = (1-t_u)C_{R1}$ but $t_u C_{1R} \leq (1-t_u)C_{R2}$. Beyond $t_u = C_{R1}/(C_{2R}+C_{R1})$, $U_{12}(t_u)$ is still uplink-limited, but $U_{21}(t_u)$ is downlink-limited. As $t_u$ increases further to $t_u = C_{R2}/(C_{1R}+C_{R2})$ and beyond, both $U_{12}(t_u)$ and $U_{21}(t_u)$ become downlink-limited. Thus, for $t_u < C_{R1}/(C_{2R}+C_{R1})$ and $t_u > C_{R2}/(C_{1R}+C_{R2})$, $U_{12}(t_u)$ and $U_{21}(t_u)$ increase and decrease together, and there is no trade-off between them.

Let $C_{12}$ and $C_{21}$ be the information capacities from node 1 to node 2, and from node 2 to node 1, respectively. The outer bound for $(C_{12}, C_{21})$ is therefore defined by $(U_{12}(t_u), U_{21}(t_u))$ in the interval $t_u \in [C_{R1}/(C_{2R}+C_{R1}), C_{R2}/(C_{1R}+C_{R2})]$. In Fig. 20, the outer bound is given by the three black lines. The shaded region is the region within which $(C_{12}, C_{21})$ must falls, defined by the following inequalities (besides $C_{12}, C_{21} \geq 0$):

$$\frac{1}{C_{21}} \geq \frac{1}{C_{2R}} + \frac{1}{C_{R1}}$$
$$\frac{1}{C_{12}} \geq \frac{1}{C_{1R}} + \frac{1}{C_{R2}}$$
$$\frac{C_{21}}{C_{R1}} + \frac{C_{12}}{C_{1R}} \leq 1$$
(24)

The interpretation of the first inequality in (24) is as follows. On the RHS, the term $1/C_{2R}$ is the minimum uplink time needed to transmit one bit from node 2 to relay $R$; and the term $1/C_{R1}$ is the minimum downlink time needed to transmit one bit from relay $R$ to node 1. The minimum time for the transport of one bit from node 2 to node 1 must be no less than the sum of these two terms; hence, the inequality. The second inequality in (24) is subject to similar interpretation.

The interpretation of the third inequality in (24) is as follows. This inequality is due to the black line with the negative slope in Fig. 20. On this line, $U_{21}$ is downlink-limited and $U_{12}$ is uplink-limited. Thus, we have $U_{21} = t_d C_{R1}$ and $U_{12} = t_u C_{1R}$. Since $t_u + t_d = 1$, we have $U_{21}/C_{R1} + U_{12}/C_{1R} = 1$. Noting that $C_{21} \leq U_{21}$ and $C_{12} \leq U_{12}$, we have the third inequality in (24).

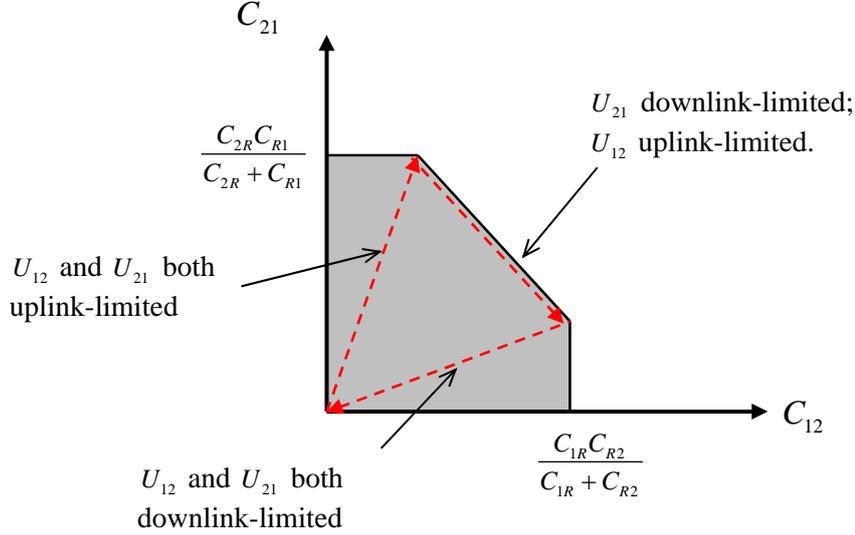

Fig. 20. Red arrowed lines: locus of $(U_{12}(t_u), U_{21}(t_u))$ as $t_u$ increases from 0 to 1. Black lines: outer bound for $(C_{12}, C_{21})$. This example assumes $C_{2R}/C_{R1} \geq C_{1R}/C_{R2}$.

In the case of $C_{2R}/C_{R1} < C_{1R}/C_{R2}$, the arrows reverse in direction as $t_u$ increases. The negative-sloped black line is one on which $U_{12}$ is downlink-limited and $U_{21}$ is uplink-limit. The first two inequalities in (24) remain the same, but the third inequality is replaced by

$$\frac{C_{21}}{C_{2R}} + \frac{C_{12}}{C_{R2}} \leq 1 \tag{25}$$

In the next few sections, we will discuss the extent to which the outer bound of TWRC can be approached by various PNC schemes.

### 4.1.2. Link-by-link Channel-coded PNC

We first consider link-by-link channel-coded PNC. We assume the use of nested lattice code. We adopt the results of the nested-lattice-coded scheme in [16] in this discussion. Conceptually, this scheme is similar to the scheme of $CNC_{XOR\text{-}CD}(\cdot)$ in Fig. 12 in that PNC mapping is first applied followed by channel decoding in a disjoint manner. For PNC mapping, in place of XOR over superimposed QPSK, we have MOD operation over nested lattice code in [16]. Rather than calling this scheme $CNC_{PNC_{LC}\text{-}CD}(\cdot)$, a mouthful, we simply refer to it as $PNC_{LC}$ in this section. We will make use of the results from [16] without getting into the details on how the results are obtained. Our goal is to compare the achievable rates with the outer bounds obtained in the preceding section.

With the use of self information, the achievable downlink rates from relay $R$ to nodes 1 and 2 are $R_{R1}^{PNC_{LC}} = C_{R1}$ and $R_{R2}^{PNC_{LC}} = C_{R2}$ respectively, where $C_{Ri}$, $i = 1, 2$, are the Shannon information capacities given in (22) [16]. Therefore, the achievable end-to-end information rates by $PNC_{LC}$ are

$$R_{12}^{PNC_{LC}}(t_u) = \min[t_u R_{1R}^{PNC_{LC}}, (1-t_u)C_{R2}]$$
$$R_{21}^{PNC_{LC}}(t_u) = \min[t_u R_{2R}^{PNC_{LC}}, (1-t_u)C_{R1}] \tag{26}$$

where $R_{1R}^{PNC_{LC}}$ and $R_{2R}^{PNC_{LC}}$ are the uplink capacities when $PNC_{LC}$ is used at the relay $R$.

Ref. [16] considered the uplink phase and showed that $(R_{1R}^{PNC_{LC}}, R_{2R}^{PNC_{LC}})$ can approach the cut-set bounds to within 1/2 bit as depicted in Fig. 21. Specifically, $R_{1R}^{PNC_{LC}} = \left[\frac{1}{2}\log_2\left(\frac{P_{1R}}{P_{1R}+P_{2R}}+P_{1R}\right)\right]^+$ and $R_{2R}^{PNC_{LC}} = \left[\frac{1}{2}\log_2\left(\frac{P_{2R}}{P_{1R}+P_{2R}}+P_{2R}\right)\right]^+$ are achievable; and it can be shown that $C_{1R} - R_{1R}^{PNC_{LC}} \leq 1/2$ and $C_{2R} - R_{2R}^{PNC_{LC}} \leq 1/2$ [16].

Fig. 21 also shows, in red dashed lines, the rates achievable by the MUD channel decoding system in which the relay $R$ explicitly decodes the individual information from nodes 1 and 2. This scheme is similar in spirit to $CNC_{MUD\text{-}XOR}(\cdot)$ in Fig. 11, except that we do not restrict ourselves to QPSK here. We will refer to this strategy as $PNC_{MUD}$ here.

In general, $PNC_{LC}$ has good performance in the high SNR region and $PNC_{MUD}$ has good performance in the low SNR region. With convex combination of $PNC_{LC}$ and $PNC_{MUD}$ (i.e., dividing the uplink time among the two strategies), the shaded area in Fig. 21 is the achievable rate region. In particular, the boundary of the shaded area is an inner bound for the achievable rates. For simplicity, for the rest of the discussion in this section, we will not consider such a combination.

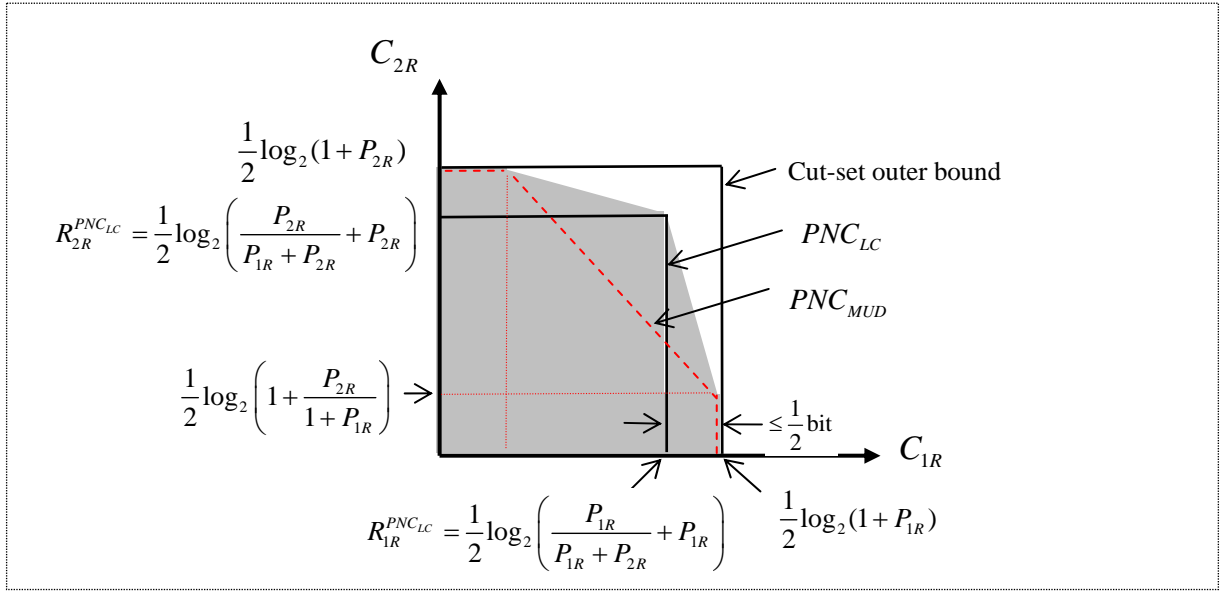

Fig. 21. Uplink capacities for channel-coded PNC.

Let us now consider the uplink and downlink phases together. Suppose that $t_u$ is such that both $U_{12}(t_u)$ and $U_{21}(t_u)$ in (23), as well as both $R_{12}^{PNC_{LC}}(t_u)$ and $R_{21}^{PNC_{LC}}(t_u)$ in (26), are downlink-limited. Then, $R_{12}^{PNC_{LC}}(t_u) = (1-t_u)R_{R2}^{PNC_{LC}} = (1-t_u)C_{R2}$; $R_{21}^{PNC_{LC}}(t_u) = (1-t_u)R_{R1}^{PNC_{LC}} = (1-t_u)C_{R1}$. The cut-set bound can be achieved exactly by PNC. Now, suppose that $t_u$ is such that both $U_{12}(t_u)$ and $U_{21}(t_u)$, as well as both $R_{12}^{PNC_{LC}}(t_u)$ and $R_{21}^{PNC_{LC}}(t_u)$ are uplink-limited. Then, $R_{12}^{PNC_{LC}}(t_u) \geq U_{12}(t_u) - t_u/2$; $R_{21}^{PNC_{LC}}(t_u) \geq U_{21}(t_u) - t_u/2$. In general, if $R_{ij}^{PNC_{LC}}(t_u)$ is downlink-limited, then the cut-set bound can be achieved; and if $R_{ij}^{PNC_{LC}}(t_u)$ is uplink-limited, then the cut-set bound can be achieved within $t_u/2$ bit.

Fig. 22 shows the locus of $(R_{12}^{PNC_{LC}}(t_u), R_{21}^{PNC_{LC}}(t_u))$ (blue dashed lines) versus the locus of $(U_{12}(t_u), U_{21}(t_u))$ (red dashed lines, reproduced from Fig. 20) as $t_u$ increases from 0 to 1. For all $t_u$, $U_{ij}(t_u) - R_{ij}^{PNC_{LC}}(t_u) \leq t_u/2$.

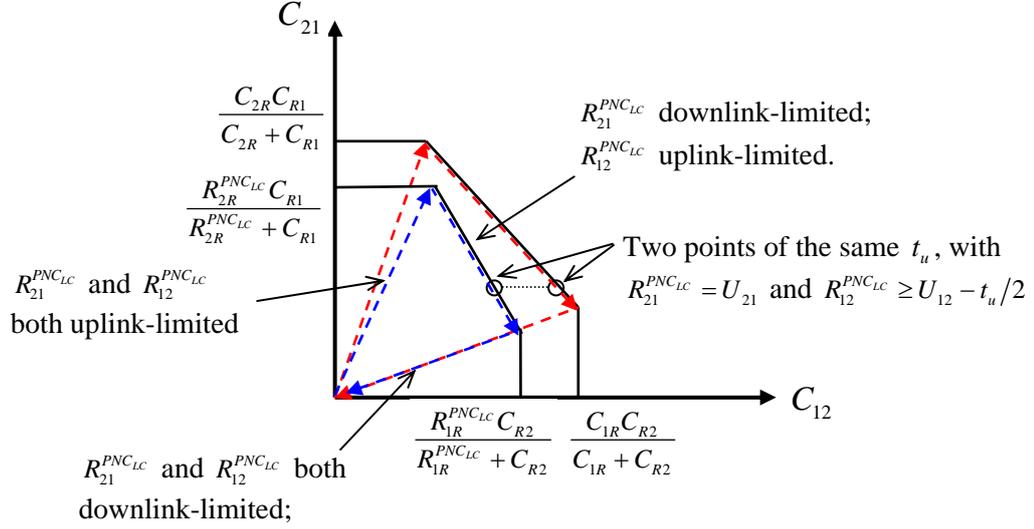

Fig. 22. Locus of $(R_{12}^{PNC_{LC}}(t_u), R_{21}^{PNC_{LC}}(t_u))$ (blue dashed lines) versus locus of $(U_{12}(t_u), U_{21}(t_u))$ (red dashed lines) as $t_u$ increases from 0 to 1. This example assumes $C_{2R}/C_{R1} \geq C_{1R}/C_{R2}$ and $R_{2R}^{PNC_{LC}}/C_{R1} \geq R_{1R}^{PNC_{LC}}/C_{R2}$.

*Achievable Symmetric Rates in Symmetric TWRC*

Let us now use the above results to analyze the case in which we target to achieve symmetric rates in both direction; i.e., $R_{12} = R_{21}$. We assume the simple homogeneous scenario in which all stations use the same transmit power and all channel gains are the same; i.e., $P_{1R} = P_{2R} = P_{R1} = P_{R2} = P$.

Let us consider $PNC_{LC}$ first. For the symmetric-rate case, we have $R_{12}^{PNC_{LC}}(t_u) = R_{21}^{PNC_{LC}}(t_u)$. Our goal is to find the $t_u$ that maximizes this transfer rate. To do so, we set the downlink rate to be the same as uplink rate: $(1-t_u^*)\log_2(1+P) = t_u^* \log_2(1/2+P)$, where $t_u^*$ is the optimal $t_u^*$. This gives

$$t_u^* = \frac{\log_2(1+P)}{\log_2(1+P) + \log_2(1/2+P)}$$

$$R_{12}^{PNC_{LC}}(t_u^*) = R_{21}^{PNC_{LC}}(t_u^*) = \frac{1}{2} \cdot \frac{\log_2(1+P) \cdot \log_2(1/2+P)}{\log_2(1+P) + \log_2(1/2+P)} \quad (27)$$

For this symmetric-rate case, the cut-set bound is

$$U_{12}(1/2) = U_{21}(1/2) = \frac{1}{4}\log_2(1+P) \quad (28)$$

Table 3 summarizes the results. Recall that noise power has been normalized to 1, so that $P$ is the SNR. As can be seen, as SNR increases, the gap between the achievable rate by $PNC_{LC}$ and the cut-set outer bound decreases quickly. At 10dB, the gap is less than 1%. In general, the larger the SNR, the smaller the gap.

For comparison, we also include the results for $PNC_{MUD}$. We assume that the downlink capacity $\left(\log_2(1+P)\right)/2$ can be achieved. For uplink, the rate $R_{1R}^{PNC_{MUD}} = R_{2R}^{PNC_{MUD}} = \left(\log_2(1+2P)\right)/4$.

Thus, the optimal $t_u$ and the achievable rate for $PNC_{MUD}$ are

$$t_u' = \frac{\log_2(1+P)}{\log_2(1+P)+0.5\log_2(1+2P)}$$
$$R_{12}^{PNC_{MUD}}(t_u') = R_{21}^{PNC_{MUD}}(t_u') = \frac{1}{4} \cdot \frac{\log_2(1+P) \cdot \log_2(1+2P)}{\log_2(1+P)+0.5\log_2(1+2P)} \quad (29)$$

From Table 3, we see $PNC_{MUD}$ may have better exchange rate than $PNC_{LC}$ at 0dB, the low SNR region. At higher SNR, $PNC_{LC}$ is better. At 0dB, the gap relative to the upper bound is 26% for $PNC_{LC}$, and 12% for $PNC_{MUD}$. By contrast, at 10dB, the gap is less than 1% for $PNC_{LC}$, and 22% for $PNC_{MUD}$. It can be easily shown that at extremely low SNR (as $P \to 0$), the performance gap of $PNC_{MUD}$ approaches 0 bits; and at extremely high SNR (as $P \to \infty$), the performance gap of $PNC_{LC}$ approaches 0 bits.

We could perform a similar analysis for *SNC* depicted in Fig. 2. Three time slots are need with each transmission taking the same amount of time. Each of the transmission can achieve the information capacity since each is a point-to-point link. Thus,

$$R_{12}^{SNC} = R_{21}^{SNC} = \frac{1}{6}\log_2(1+P) \quad (30)$$

As shown in Table 3, the performance of *SNC* is not as good as that of $PNC_{MUD}$ or $PNC_{LC}$ in general. The performance gap relative to the upper bound is 33% across all SNR.

Let us also take a look at the *TS* scheme in Fig. 1. Again, we assume link-by-link channel coding. By symmetry, each of the four transmissions uses a quarter of the airtime. The achievable rate is

$$R_{12}^{TS} = \frac{1}{8}\log_2(1+P) \quad (31)$$

The gap of *TS* compared with the upper bound is exactly 50%.

Let us now reflect on these information-theoretic results. In our earlier discussion in Section 2, based on slot counting, we mentioned that *PNC* can achieve 100% throughput improvement compared with *TS*. *SNC*, on the other hand, can achieve 33% throughput improvement. Strictly speaking, that is true only when the channels are highly reliable so that we do not have to worry about noise. That is, when the *SNR* is high. Our information-theoretic results here confirm that intuition. When *SNR* is high, $PNC_{LC}$ approaches the upper bound, which has twice the rate of *TS*. Similarly, when *SNR* is high, the information-theoretic rate of *SNC* is 33% that of *TS*. At the low *SNR* regime, $PNC_{MUD}$ can also approach the upper bound. Thus, at low SNR, the information rate of *PNC* is still twice that of *TS*.

As indicated in Table 3, there is still an appreciable gap for *PNC* in the mid SNR range. This appears to be a fertile ground for future research. Recall that $PNC_{MUD}$ and $PNC_{LC}$ are similar to *MUD-XOR* and *XOR-CD* in Fig. 11 and Fig. 12, respectively, in that the network coding and channel decoding operations are disjoint. They tend to work well only in the low and high SNR regimes, respectively. The designs in Fig. 13 and Fig. 15, on the other hand could work well under all SNR regimes; however, they are only limited to XOR over QPSK.

A gap that remains to be filled in the information-theoretic study is whether an upper-bound approaching scheme is available for all SNR regimes. For example, is there a lattice-code design similar to that of Fig. 15 that can work well under all SNR regimes? What is the ultimate fundamental capacity region for PNC-based TWRC?

Table 3: Achievable symmetric exchange rate for $PNC_{LC}$, $PNC_{MUD}$, $SNC$, $ANC$, and $TS$ in symmetric TWRC under equal power usage for all nodes.

| $P$ (dB) | 0 | 2 | 4 | 6 | 8 | 10 |
|---|---|---|---|---|---|---|
| $R_{12}^{PNC_{LC}}(t_u^*)$ | 0.185 | 0.299 | 0.424 | 0.559 | 0.704 | 0.856 |
| $R_{12}^{PNC_{MUD}}(t_u^{'})$ | 0.221 | 0.294 | 0.378 | 0.470 | 0.569 | 0.672 |
| $R_{12}^{SNC}$ | 0.167 | 0.228 | 0.302 | 0.386 | 0.478 | 0.577 |
| $R_{12}^{ANC}$ | 0.080 | 0.131 | 0.200 | 0.288 | 0.396 | 0.520 |
| $R_{12}^{TS}$ | 0.125 | 0.171 | 0.227 | 0.290 | 0.359 | 0.432 |
| $U_{12}(1/2)$ | 0.250 | 0.343 | 0.453 | 0.579 | 0.717 | 0.865 |
| $U_{12}(1/2) - R_{12}^{PNC_{LC}}(t_u^*)$ | 0.065 | 0.044 | 0.029 | 0.020 | 0.013 | 0.008 |
| $U_{12}(1/2) - R_{12}^{PNC_{MUD}}(t_u^{'})$ | 0.029 | 0.049 | 0.075 | 0.109 | 0.149 | 0.193 |
| $U_{12}(1/2) - R_{12}^{SNC}$ | 0.083 | 0.114 | 0.151 | 0.193 | 0.239 | 0.288 |
| $U_{12}(1/2) - R_{12}^{ANC}$ | 0.170 | 0.212 | 0.253 | 0.291 | 0.321 | 0.345 |
| $U_{12}(1/2) - R_{12}^{TS}$ | 0.125 | 0.171 | 0.227 | 0.290 | 0.359 | 0.432 |

### 4.1.3. End-to-end Channel-coded ANC

We now look the amplify-and-forward ANC scheme [8]. Recall that in this scheme the relay just amplifies and forwards the received signal. Thus, the symbol durations for the uplink and downlink phases are the same. This means $t_u = t_d = 1/2$. In addition, the relay is not involved in channel decoding and re-encoding. Channel coding can only be applied on an end-to-end basis.

Not being able to adjust the durations of $t_u$ and $t_d$ is one shortcoming of ANC because one loses a degree of freedom in the system design. A second shortcoming is that with only end-to-end channel coding, the relay cannot remove codeword errors incurred by the uplinks before forwarding the data.

Let us derive the information rate for the end-to-end channel-coded ANC. At the relay, the signal powers received from nodes 1 and 2 are $P_{1R}$ and $P_{2R}$, respectively. Recall that we normalize noise power to be one unit. Let $P_R^t$ be the transmission power of the relay. We have

$$P_R^t = \alpha \cdot (P_{1R} + P_{2R} + 1) \qquad (32)$$

for some amplification factor $\alpha$. At node 1, the received power is $P_{R1} = P_R^t |h_{R1}|^2$ where $h_{R1}$ is the channel gain for the channel from relay $R$ to node 1. Similarly, the received power at node 2 is $P_{R2} = P_R^t |h_{R2}|^2$. We assume the receiver noise powers at node 1 and node 2 are also one unit.

At node 1, after subtracting the self information, the signal power is $\alpha P_{2R} |h_{R1}|^2$. The noise power, including the accumulated at the relay and the noise at node 1, is $\alpha |h_{R1}|^2 + 1$. Hence, the SNR at node 1 (and similarly, SNR at node 2) is

$$SNR_1 = \frac{\alpha P_{2R} |h_{R1}|^2}{\alpha |h_{R1}|^2 + 1} = \frac{\alpha P_{2R} P_{R1}/P_R^t}{\alpha P_{R1}/P_R^t + 1} = \frac{P_{2R} P_{R1}}{P_{R1} + P_{1R} + P_{2R} + 1}$$
$$SNR_2 = \frac{P_{1R} P_{R2}}{P_{R2} + P_{2R} + P_{1R} + 1}$$
(33)

Since each of the uplink phase and downlink phase uses up half the airtime, we have

$$R_{21}^{ANC} = \frac{1}{4}\log_2\left(1 + \frac{P_{2R} P_{R1}}{P_{R1} + P_{1R} + P_{2R} + 1}\right)$$
$$R_{12}^{ANC} = \frac{1}{4}\log_2\left(1 + \frac{P_{1R} P_{R2}}{P_{R2} + P_{2R} + P_{1R} + 1}\right)$$
(34)

Let us now look at symmetric TWRC in which $P_{1R} = P_{2R} = P_{R1} = P_{R2} = P$. Then,

$$R_{12}^{ANC} = R_{21}^{ANC} = \frac{1}{4}\log_2\left(1 + \frac{P^2}{3P+1}\right)$$
(35)

Table 3 shows the numerical results for SNR ranges from 0dB to 10dB. We see that from an information-theoretic standpoint, end-to-end channel-coded *ANC* does not work as well as other link-by-link channel-coded PNC scheme. The performance gap relative to the upper bound ranges from 68% for 0dB to 40% for 10dB. It can be easily shown, however, that at extremely high SNR (as $P \to \infty$), the performance gap for ANC reaches a constant of 0.396 bits. Thus, percentage-wise, the performance gap goes to zero at extremely high SNR region.

### 4.2. Energy Implications for Gaussian TWRC

PNC was originally conceived as a capacity boosting device. In TWRC, SNC requires three transmissions for the exchange of a packet (one in each direction) between nodes 1 and 2. Traditional TS requires four transmissions. Thus, the energy saving for SNC is 25%. Although PNC saves an additional time slot, three transmissions are also needed. Thus, it may appear at first glance that there is no additional energy saving beyond that SNC. That turns out to be not the case.

In this section, we look at the issue more fundamentally from an information-theoretic perspective. The key idea is to convert capacity increase in PNC to energy saving. We argue that for the same exchange rate, PNC can achieve significant transmit energy saving in the high SNR regime. Our discussion here only considers the transmit energy. In a real system, there will be receiver energy and processing energy to consider as well. In that light, the discussion in this paper is only a preliminary foray into a new research direction.

Let us look at the symmetric TWRC again, in which the channel gains are all equal: $h_{1R} = h_{2R} = h_{R1} = h_{R2}$. However, we allow the transmit powers to be different for the uplink and downlink phases, but adjust the uplink time $t_u$ and downlink time $t_d$ such that node 1, node 2, and relay $R$, use the same energy. We target to have symmetric rates (i.e., equal rate in both directions).

Consider SNC. We assume that each of the three transmissions can achieve rate equal to Shannon capacity. Thus, $1/3$ of the airtime is dedicated to each phase. Thus, the energy expended by each node is $E^{SNC} = P^{SNC}/3$, where $P^{SNC}$ is the power of each node. For a given exchange rate target $R$, we have that

$$R = R_{12}^{SNC} = R_{21}^{SNC} = \frac{1}{6}\log_2\left(1 + P^{SNC}\right) = \frac{1}{6}\log_2\left(1 + 3E^{SNC}\right)$$
(36)

For PNC, we will consider both $PNC_{LC}$ and $PNC_{MUD}$ discussed in the preceding section. For $PNC_{LC}$ and $PNC_{MUD}$, in order that nodes 1 and 2 use the same energy as relay $R$, we allow $P_{1R}^{PNC_{LC}} = P_{2R}^{PNC_{LC}} \neq P_{R1}^{PNC_{LC}} = P_{R2}^{PNC_{LC}}$ and $P_{1R}^{PNC_{MUD}} = P_{2R}^{PNC_{MUD}} \neq P_{R1}^{PNC_{MUD}} = P_{R2}^{PNC_{MUD}}$.

We first explain the numerical method we use for $PNC_{LC}$. For a given rate requirement $R$, in order that the downlink rate and uplink rate are equal, we find $P_{1R}^{PNC_{LC}}$ and $P_{R1}^{PNC_{LC}}$ such that

$$R = t_u \cdot \frac{1}{2} \log_2 \left( \frac{1}{2} + P_{1R}^{PNC_{LC}} \right) = (1-t_u) \cdot \frac{1}{2} \log_2 \left( 1 + P_{R1}^{PNC_{LC}} \right) \tag{37}$$

Next, we compute the energies used by node 1 and the relay:

$$\begin{aligned} E_1^{PNC_{LC}} &= t_u P_{1R}^{PNC_{LC}} \\ E_R^{PNC_{LC}} &= (1-t_u) P_{1R1}^{PNC_{LC}} \end{aligned} \tag{38}$$

For numerical computation, we vary $t_u$ from 0 to 1 for a given $R$. For each $t_u$, we compute $P_{1R}^{PNC_{LC}}$ and $P_{R1}^{PNC_{LC}}$ according to (37). Then, we check if the two energies in (38) are equal. We identify the $t_u$ at which the equality $E_1^{PNC_{LC}} = E_R^{PNC_{LC}}$ is achieved, and this is the solution for a given $R$.

For $PNC_{MUD}$, corresponding to (37), we have $R = \frac{t_u}{4} \log_2 \left( 1 + 2P_{1R}^{PNC_{MUD}} \right) = \frac{(1-t_u)}{2} \log_2 \left( 1 + P_{R1}^{PNC_{MUD}} \right)$. Corresponding to (38), we have $E_1^{PNC_{MUD}} = t_u P_{1R}^{PNC_{MUD}} = (1-t_u) P_{R1}^{PNC_{MUD}} = E_R^{PNC_{MUD}} \triangleq E^{PNC_{MUD}}$. We can then easily get $t_u = 2/3$ and $2P_{1R}^{PNC_{MUD}} = P_{R1}^{PNC_{MUD}}$. This gives the following closed-form without the need for numerical computation:

$$R = \frac{1}{6} \log_2 \left( 1 + 3E^{PNC_{MUD}} \right) \tag{39}$$

Note that (39) is similar to (36). Thus, $PNC_{MUD}$ has the same energy performance as $SNC$.

For end-to-end channel-coded $ANC$, the power used by all three nodes should be equal in order that they use the same energy. For a target rate $R$, we have $R = \frac{1}{4} \log_2 \left( 1 + \frac{(P^{ANC})^2}{3P^{ANC} + 1} \right)$. In addition, $E^{ANC} = \frac{1}{2} P^{ANC}$. This gives

$$R = \frac{1}{4} \log_2 \left( 1 + \frac{(2E^{ANC})^2}{6E^{ANC} + 1} \right) \tag{40}$$

where $P^{ANC}$ and $E^{ANC}$ are the power and energy used by each node.

Finally, we consider $TS$ in Fig. 1. Half of the uplink time is given to the transmission of node 1, and half the downlink time is given to the transmission of relay $R$ to node 1. Thus, the rate is $R = \frac{t_u}{2} \cdot \frac{1}{2} \log_2(1 + P_{1R}^{TS}) = R = \frac{(1-t_u)}{2} \cdot \frac{1}{2} \log_2(1 + P_{R1}^{TS})$. Equating the energy usages of the nodes, we have $E_1^{TS} = \frac{t_u}{2} P_{1R}^{TS} = (1-t_u) P_{R1}^{TS} = E_{R1}^{TS} \triangleq E^{TS}$ (note that the relay transmits twice in the downlink phase, once to each end node). We use a numerical method similar to the one used for $PNC_{LC}$ above to find the solution.

In Table 4, we use *SNC* as the benchmark starting point. We vary $P^{SNC}$ from 0dB to 10dB and compute the resulting $R_{12}^{SNC}$ according to (36). We fix the target rate $R = R_{12}^{SNC}$, and then use the above methods to compute the energy requirements for $PNC_{LC}$, $PNC_{MUD}$, and *ANC*.

Table 4. Energies needed for $PNC_{LC}$, $PNC_{MUD}$, *SNC*, *ANC*, and *TS* in symmetric TWRC under equal energy usage for all nodes.

| $P^{SNC}$ (dB) | 0 | 2 | 4 | 6 | 8 | 10 |
|---|---|---|---|---|---|---|
| $R$ | 0.1667 | 0.2284 | 0.3020 | 0.3861 | 0.4783 | 0.5766 |
| $E^{SNC}, E^{PNC_{MUD}}$ (dB) | -4.771 | -2.771 | -0.771 | 1.229 | 3.229 | 5.229 |
| $E^{PNC_{LC}}$ (dB) | -1.522 | -1.024 | -0.342 | 0.648 | 1.900 | 3.264 |
| $E^{ANC}$ (dB) | 0.105 | 1.688 | 3.264 | 4.819 | 6.345 | 7.840 |
| $E^{TS}$ (dB) | -1.928 | 0.160 | 2.341 | 4.618 | 6.987 | 9.432 |

From Table 4, we see that in the high SNR regime (i.e., the high rate regime), $PNC_{LC}$ is most energy efficient. However, at the lower SNR (low rate) regime, $PNC_{MUD}$ and *SNC* have the most efficient energy usage. *ANC* is not energy efficient for the SNR range investigated here. In particular, it can be even less efficient than *TS* in the low SNR regime.

The discussion in this section is a preliminary exploration of the energy implication of PNC. In practical systems, it is not just transmissions that expend energy; information processing also uses up energy. A more careful study will be worthwhile for future research.

### 4.3. To Probe Further

The ANC rates in (35) are derived assuming the relay estimates $h_{1R}x_1[n] + h_{2R}x_2[n]$ for different $n$ independently. The rates for a non-memoryless ANC relay [38] would be better than the rates in (35). The exact information rates for non-memoryless ANC remain an open issue.

Ref. [45] investigated the capacity region of a MIMO ANC system in which the two end nodes have one antenna, and the relay has multiple antennas. Linear processing is assumed at the relay. The correlations among symbols within a packet due to channel coding are not exploited to boost performance.

There is a body of works related to a technique called compress-and-forward (CF) [84] [85] [86]. In CF, the relay first compresses the received packet $Y_R$ to $Y_R'$, and then encodes $Y_R'$ into a packet $X_R$ for broadcast to nodes 1 and 2. The end nodes first decode $Y_R'$ from the received signal, and then decode the packet from the other end based on $Y_R'$. The intuition of CF is that the relay tries to decrease (clean up) the uplink noise by some hard decision. The coding at the relay helps to achieve the downlink channel capacity (part of which is given to the remaining noise and part of which is given to the information). Since the hard decision can be adapted to the uplink noise, it may work well at different SNR for the uplink. However, it is suboptimal in the downlink. By a combination of amplified-and-forward, CF, and superposition coding, [85] showed that rates within $\log_2 3$ bit of the outer bound can be achieved.

To our best knowledge, the ultimate capacity region for PNC-based systems is still an open issue. The design in [16] is a design similar in spirit to the *XOR-CD* scheme in that the network coding operation and the channel decoding operation in the relay are disjoint. The network coding operation performed prior to channel decoding loses much useful information, and the scheme only approaches the outer bound in the high SNR regime.

Refs. [53], [87], [88], and [89] studied the case where the relay obtains explicit information on $S_1$ and $S_2$; and in the downlink broadcast phase, the relay needs to encode $(S_1, S_2)$ into $X_R$ so that nodes 1 and 2 can decode $S_2$ and $S_1$, respectively with side information. The explicit decoding of $S_1$ and $S_2$ at the relay in the uplink phase corresponds to the MUD scheme, which is embodied in the design of Fig. 11. This design is optimal in the low SNR regime of the uplink but not optimal in general. Designs similar in spirit to Fig. 15 and Fig. 16 do not aim to explicitly decode $S_1$ and $S_2$. Although the Tanner Graph in Fig. 16 is used to compute $P(s_1[n], s_2[n] | Y_R)$, which can be used to decode the ML $s_1[n]$ and $s_2[n]$ explicitly, the goal is not to do so. The goal is to get $P(s_1[n] \oplus s_2[n] | Y_R)$ (more generally a PNC mapping that is not necessarily XOR) from $P(s_1[n], s_2[n] | Y_R)$. From $P(s_1[n] \oplus s_2[n] | Y_R)$, the ML $s_1[n] \oplus s_2[n]$ can then be obtained. To our best knowledge, the ultimate capacity region of this kind of design is unknown.

Ref. [90] contains a rather approachable treatise on the use of nested lattice code for reliable PNC. In [91], an algebraic approach was taken and a class of PNC-compatible lattice partitions was found. The complexity of decoding lattice codes with a large alphabet cardinality is high; [92] investigated modified higher-order PAMs for binary-coded PNC.

We have not found any information-theoretic treatment that incorporates symbol asynchrony. This could be an interesting direction for future work.

## 5. Network-theoretic Studies

Thus far in this paper, our discussion of PNC has focused mostly on TWRC and the linear network, in which there are only two flows in opposing directions. In a general network, there could be multiple traffic flows. The application of PNC in a general network concerns issues not just at the physical layer, but also issues at the MAC layer and networking layer. When there are multiple flows, how to schedule the transmissions by different flows, how to route them, and what are the potential system throughput in a large network setting become an issue. In this section, we present some network-theoretic results of PNC. In Section 5.1, we demonstrate the advantage of PNC through asymptotic analysis of wireless networks with infinite flows [93] [94]. In Section 5.2, we present a distributed MAC protocol to allow PNC to be applied in a wireless network in a practical setting. Both sections are by no means comprehensive studies, and certainly other formulations and approaches are possible; our modest goal here is to convey two flavors of PNC research in the networking domain. Among the three tracks of PNC research – i.e., communications, information-theory, and networking – issues related to the networking track are perhaps the least understood thus far.

### 5.1. Asymptotic Performance Analysis:

In this section, we discuss the feasibility and throughput superiority of PNC in 1-D and 2-D wireless networks with $N$ nodes. In our analysis, half the nodes are randomly selected as the source nodes and the remaining other half are the sink nodes. We assume we want to deliver equal amounts of traffic for all the $N/2$ flows. We derive the asymptotic throughput per flow for large $N$.

We adopt the well accepted physical interference model in [95]. Specifically, we assume that the transmission of a link is successful if and only if the SIR is more than a given threshold, say 10dB. In addition, we impose the half-duplex constraint in our system model (i.e., a node cannot transmit and receive at the same time).

We first investigate the 1-D regular topology in which the nodes are regularly spaced. We show that the throughput of PNC can approach the theoretical upper bound asymptotically as $N$ goes to infinity. We then extend the analysis to the 1-D random topology in which nodes are not regularly spaced. We show that the asymptotic results remain valid for the 1-D random topology if a hierarchical routing scheme is adopted. We further extend our analysis to the 2-D regular and random topologies. Overall, our analysis indicates that the use of PNC can boost throughput by a fixed factor compared with the traditional store-and-forward scheme.

**Topology 1: Regular 1-D Network**

Consider a regular 1-D linear network topology in which the $N$ nodes are regularly spaced such that the distance between any two adjacent nodes is a constant, as shown in Fig. 23. There are altogether $N/2$ flows in the network with half the nodes randomly chosen as the sources and half the nodes randomly chosen as the destinations. Two issues are of interest: 1) how to apply PNC to the $N/2$ flows; and 2) what is the asymptotic throughput per flow for such a PNC system as $N \to \infty$.

To address the first issue, we apply a concept similar to the concept of virtual paths and virtual circuits in ATM broadband networks (as a reference, see Chapter 7 of [17], or other books on broadband networking). In an ATM broadband network, the traffic of end-to-end flows between end users is carried on virtual circuits. A virtual circuit may traverse a number of virtual paths from source to destination. Embedded in each virtual path could be a number of virtual circuits. Thus, a virtual path may aggregate the traffic from multiple virtual circuits. It turns out that this aggregation technique is what we need to apply PNC in a general multihop network.

For our $N/2$ flows in the 1-D network, the basic idea is as follows. First, we note that some flows are in one direction and some flows are in the opposite direction. If we observe $\alpha$ units of traffic (possibly from different flows) along a sequence of links, say links $i, i+1, ..., j-1, j$; and also $\alpha$ units of traffic along the sequence of links, $j, j-1, ..., i+1, i$ in the opposite direction, then we could aggregate the traffic in the two opposing directions for transport over a bidirectional PNC virtual path. We can then apply PNC scheduling as in Fig. 5 to this PNC bidirectional virtual path to reduce the number of time slots needed. An issue, of course, is the optimal way to form PNC virtual paths out of the $N/2$ flows. In what follows, instead of seeking an optimal algorithm, we simply use a heuristic algorithm for our asymptotic study. And this heuristic can achieve the optimal throughput per flow asymptotically.

In our heuristic, instead of trying to derive the PNC virtual paths of many flows in one shot, we incrementally aggregate traffic of only two flows in opposing directions. We refer to such a "mini bidirectional virtual path" consisting of the traffic of only two opposing flows as a "PNC unit"; an overall "PNC bidirectional virtual path" consists of a bundle of mini bidirectional virtual paths on the same sequence of nodes. An example of a PNC unit could simply be a three-node TWRC we have discussed in the previous section. More generally, a PNC unit is a linear chain of multiple nodes, for which the PNC mechanism can be used.

To form PNC units, we introduce the concept of "packings". We refer to Fig. 23 for illustration. A *right packing* consists a sequence of non-overlapping right-bound flows from left to right in the 1-D network. Similarly, a *left packing* consists a sequence of non-overlapping left-bound flows from right to left. In Fig. 23, Flow 1 belongs to a right packing, and Flows 2 and 3 belong to a left packint. In general, we can form many "tight" right packings in a greedy way. We start from the left of the 1-D network and look to the right. Upon encountering the source node of the first right-bound flow, we include it into a right packing. Then, we continue to look to the right beyond the destination of the first right-bound flow to find the next right-bound flow. We repeat this until we reach the right end of the 1-D network. We then have a right packing consisting of a sequence of non-overlapping right-bound flows. To form the next right packing, we repeat the above procedure for the remaining right-bound flows. We do this until we have a set of right packings. Left packings are constructed similarly by progressing from right to left.

A *dual packing* is formed by matching a right packing and a left packing. Suppose that we have $R$ right packings and $L$ left packings. Then the number of dual packing is $M = \min(R, L)$. The numbers of unmatched right and left packings are respectively $R - M$ and $L - M$, one of which must be zero; equivalently, the number of unmatched unidirectional flows is $R + L - 2M$

Fig. 23 shows an example of a dual packing. Flows 2 and 3 are part of a right packing in the 1-D network, and Flow 1 is part of a left packing. Note that some of the nodes are traversed by both a right-bound flow and a left-bound flow. We call these nodes the *common nodes*, and the other nodes the *non-common nodes*. A sequence of adjacent common nodes in between two non-common nodes (an ellipse in Fig. 23) forms a *PNC unit*. A sequence of adjacent non-common nodes in between two common nodes (a rectangle in Fig. 23 contains two common nodes at the boundary and a sequence of non-common nodes in between) may or may not have traffic flowing over them. When there is traffic, the traffic is in one direction only, and the traditional multi-hop store-and-forward scheme can be used to carry the unidirectional traffic.

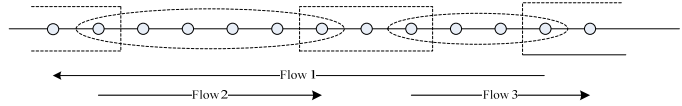

Fig. 23. A linear network containing an example of a dual packing formed by a right packing and a left packing. An ellipse corresponds to a PNC unit. The nodes between two adjacent ellipses (including the terminal nodes of the ellipses) are grouped together by a rectangle.

The set of dual packings yield a set of mini virtual bidirectional flows (each corresponding to a PNC unit) and some residual unidirectional virtual flows. Thus, our construction results in two types of entities: 1) PNC units that can exploit the bidirectional PNC mechanism; 2) unidirectional flows consisting of the above residual unidirectional virtual flows and the aforementioned $R + L - 2M$ unmatched unidirectional flows. Our scheduling strategy is described below.

Recall that we want to deliver equal amount of traffic for all the $N/2$ flows. Our scheduling is frame-based. Each frame consists of $F$ time slots. Within the $F$ time slots, each of the $N/2$ flows will deliver one packet from its source to its destination. The throughput per flow is therefore $1/F$.

A frame is divided into two intervals as follows:
1) The first interval is dedicated to the PNC units. Note that with $M$ dual packings, $2M$ time slots are needed in the worst case. To see this, we note that in the worst case, the PNC units in different dual packings use different time slots to transmit. The PNC units in the same dual packing can be scheduled to use the same two time slots because they are non-overlapping in space; for each PNC unit, by pipelining (as in Fig. 5), within two time slots a packet will reach each end node [6].
2) The second interval is dedicated to the aforementioned unidirectional flows. They will make use of the traditional multihop scheme for data transport.

As argued in [93], the frame length is dominated by interval 1 asymptotically as $N$ goes to infinity.

*Theorem 1:* With PNC, we can approach the upper bound of the per-flow throughput of the 1-D regular network. Specifically, the throughput per flow is the following with high probability.

$$\lambda_p = 4/N - \varepsilon = 4/N - O(1/\log(N))/N \tag{41}$$

That $4/N$ is an upper bound in the half-duplex 1-D network can be seen from the fact it corresponds to a situation in which the bottleneck link is busy all the time[7]. For a detailed proof on how PNC can approach this upper bound, see [93].

A corollary of *Theorem 1* is that PNC can improve the asymptotic throughput of the 1-D network by a factor of 2 and 1.5 relative to the traditional transmission scheme and the SNC scheme, respectively [93]. Note that this is the same improvements as observed in TWRC with only two flows.

---

[6] Two caveats are in order. The first is that according to our construction, there could be "trivial" PNC units with two nodes only. In this case, the PNC relay mechanism is not needed, and each node gets to transmit directly to the other node. Regardless of whether the PNC unit is trivial or not, two time slots are needed for the bidirectional flows. The second caveat is that there could be two PNC units in the same dual packing next to each other. For example, suppose nodes 1, 2, and 3 form a PNC unit, and nodes 4, 5, 6 forms another. To avoid conflict, the scheduling of the transmissions on these two PNC units should be such that nodes 1, 3, 4 and 6 transmit in one time slot while nodes 2 and 5 transmits in another time slot. Again, two time slots are needed. The SIR in linear network is much larger than the threshold and the PNC schedule is feasible under the 10dB SIR threshold [93].

[7] To see this intuitively, consider that there $N/2$ flows. If we examine the "bottleneck" link at the middle of the 1-D network, asymptotically there are $N/8$ flows having traffic crossing from left to right of this bottleneck link, and $N/8$ flows having traffic crossing from right to left; the other $N/4$ flows do not have traffic crossing the bottleneck links. Thus, by the half duplex constraint, at least $N/4$ time slots are needed for the traffic that crosses the bottleneck link. Therefore, there must be at least $N/4$ time slots within each frame.

For simplicity, we have assumed that the source and destination nodes of the $N/2$ flows are distinct. *Theorem 1* can be extended to a case in which the source and destination of each flows are randomly selected among the $N$ nodes with equal probability. In other words, a node may be the source or destination of multiple flows, or it may be not an end node at all. Theorem 1' below is the corresponding modified version of Theorem 1.

*Theorem 1'*: With PNC, we can also approach the upper bound of the per-flow throughput of the 1-D regular network when the source and destination nodes are randomly selected from the $N$ nodes with equal probability. Specifically, the throughput per flow is the following with high probability.

$$\lambda_{PNC} = 4/N - \varepsilon = 4/N - O(1/\log(N))/N \tag{42}$$

The proof of this theorem is similar to that of *Theorem 1* and it is omitted here.

**Topology 2: Random 1-D Network**

We now extend the schemes in the regular 1-D network to the random 1-D network where the $N$ nodes are randomly distributed over the line. Specifically, each node is randomly placed on the 1-D line with uniform distribution, and the placements of the nodes are independent.

We propose a transmission scheme inspired by [96] and [97]. In this scheme, we form a hierarchical network in which some nodes are selected to be routing nodes. In selecting the routing nodes, we ensure that they are almost evenly located to form a regular 1-D network structure. This requirement is imposed by the requirement of PNC, and it does not exist in the setting in [97], which considers SNC only.

Define the length of the linear network as one unit. We divide the line evenly into $N/\log(N)$ bins so that the length of each bin is $\log(N)/N$, as in [97]. As an extension, we further divide each bin into $\log(N)/\log\log(N)$ subbins, and the length of each subbin is $\log\log(N)/N$. With an approach similar to that in [98], we can prove the following lemma.

*Lemma 1:* With high probability, there is at least one node in each subbin as $N$ goes to infinity.

We select one node in the middle subbin of each bin as the routing node. We can prove the following lemma.

*Lemma 2*: According to our subbin construction scheme, the distance between any two adjacent routing nodes is lower- and upper-bounded by $\frac{\log N}{N}\left[1 - \frac{\log\log N}{\log N}, 1 + \frac{\log\log N}{\log N}\right]$. As $N$ goes to infinity, the upper bound and the lower bound converge. In other words, the routing nodes form a regular linear network for large $N$.

The transmission schedule is divided into two phases as in [97]. In the first phase, the nodes in each bin transmit their own traffic to the routing node and the routing node broadcasts the received information to the nodes in the same bin with one-hop transmissions. In the second phase, the transmission scheme for the regular 1-D network is adopted by all the routing nodes for the transport of inter-bin traffic.

We first argue that the time used for the first phase is of order $\log(N)$ with high probability. Using the Chernoff Bound, the probability that the number of nodes in each bin is more than $2\log(N)$ can be shown to be less than $1/N$. Therefore, the probability that the bin with maximum nodes has more than $2\log(N)$ nodes is less than $\frac{N}{\log(N)} \cdot \frac{1}{N} = 1/\log(N)$, which goes to zero as $N$ goes to infinity. By noting that the time of the first phase is negligible compared with the time used in the second phase, we can prove the following theorem by focusing on the time used for the second phase.

*Theorem 2:* With PNC, we can approach the upper bound of the per-flow throughput of the 1-D random network with high probability:

$$\lambda_{PNC}(N) = \left(\frac{4}{N/\lceil \log(N) \rceil} - \frac{O(1/\log(N/\lceil \log(N) \rceil))}{N/\lceil \log(N) \rceil}\right) \Big/ \lceil \log(N) \rceil = \frac{4}{N} - \frac{O(1/[(\log(N) - \log \lceil \log(N) \rceil])}{N} \quad (43)$$

We will not go into the details of the proof of Theorem 2. The proof approach is as follows. We divide the $N/2$ flows randomly into $\lceil \log(N) \rceil$ groups, with each group having $K = N/(2\lceil \log(N) \rceil) \leq N/(2\log(N))$ flows. We then apply Theorem 1' by setting $N$ in Theorem 1' to $2K$. Application of Theorem 1' on all the $\lceil \log(N) \rceil$ groups yields the above result.

**Topology 3: Regular 2-D Network**

We now extend the regular 1-D results to the regular 2-D case as shown in Fig. 24, where $N$ nodes are uniformly located at the cross points of the grid network. To ensure that the SIR is above a target threshold, the transmissions in the horizontal lines and vertical lines can be performed in orthogonal time slots. Consider the horizontal lines (similar schedule applies for the vertical lines). The first two time slots are dedicated to transmissions on lines $1, J+1, 2J+1, \ldots$; the next two time slots are dedicated to transmissions on lines $2, J+2, 2J+2, \ldots$; and so on. The separation $J$ must be large enough to meet the target SIR requirement. As shown in [93], for a typical value of $\alpha = 4$, the SIR is about 13.5dB, 12.3dB, and 10.0dB for $J$ = 5, 4, and 3 respectively. With an assumed 10dB target, $J = 3$ is enough to guarantee successful transmissions.

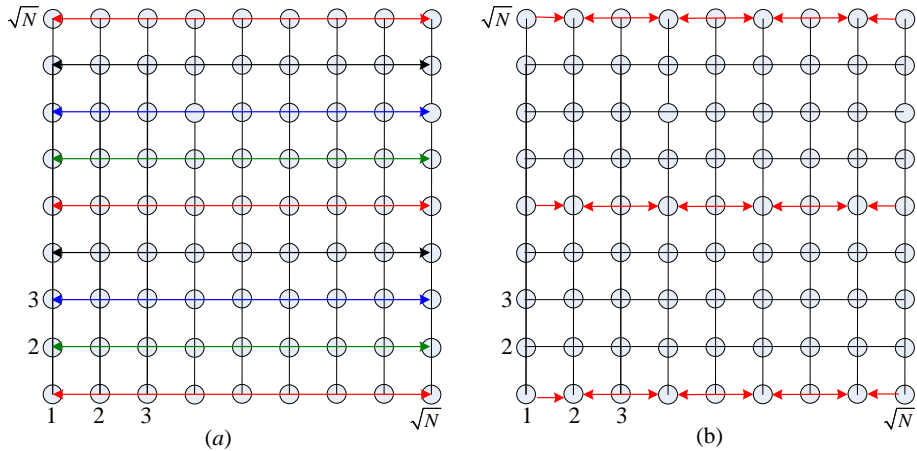

Fig. 24. (a) A 2-D grid network with one bidirectional flow in each row. The rows transmitting together have the same colors. Two rows of the same color are separated by $J - 1 = 3$ rows. (b) Scheduling for one group of active rows in a specific time slot (red rows).

Let us now investigate the application of PNC in the 2-D grid network with a more general traffic pattern. Here we apply a simple routing scheme, as in [97]. For a source-destination pair $(x_s, y_s)$-$(x_d, y_d)$, the data will be first forwarded vertically to the node at $(x_s, y_d)$ before being forwarded horizontally to the destination $(x_d, y_d)$. The horizontal and vertical transmissions are separated into two different time intervals. For horizontal (or vertical) transmissions, the scheduling within each row (column) is the same as that in 1-D topology and the scheduling among different rows (columns) is the same as above.

Consider the horizontal transmission. We argue that the number of flows in each row is almost bounded to the average value $\sqrt{N}/2$ with high probability as $N$ goes to infinity. Using the Chernoff Bound and based on the assumption of random source/destination selection procedure, we can prove that the number of sources is almost equal to the number of destinations in each row (or column), both of which are bounded as $\sqrt{N}\left[0.5-\sqrt{\frac{\log(N)}{\sqrt{N}}}, 0.5+\sqrt{\frac{\log(N)}{\sqrt{N}}}\right]$ with probability more than $1-1/N$. Then, it can be proved that the maximum number of source/destination nodes in any of the $\sqrt{N}$ lines is less than $\sqrt{N}(0.5+\sqrt{\frac{\log(N)}{\sqrt{N}}})$ with probability $1-1/\sqrt{N}$, which goes to 1 as $N$ goes to infinity. According to the result in *Theorem 1'*, the per-flow PNC throughput is four times the reciprocal of the number of nodes within the row (or column), $4/\sqrt{N}$, with high probability. Since the horizontal transmissions and vertical transmissions are scheduled in different time intervals and in each interval every $J$ lines (columns) transmit simultaneously, we have

*Theorem 3*: With PNC, and with the source and destination nodes randomly selected from the $N$ nodes with equal probability, the per-flow throughput of the *2-D grid network* can approach the following for large $N$ with high probability:

$$\lambda_{PNC}(N) = \frac{4}{\sqrt{N}} \cdot \frac{1}{J} \cdot \frac{1}{2} = \frac{2}{J\sqrt{N}} \tag{44}$$

where $J$ is determined by the SIR threshold. A typical value of $J$ is 3 (under an SIR threshold of 10dB and path-loss exponent of 4).

For comparison purposes, let us look at the per-flow throughput under the traditional multihoping scheme, TS, and under the upper-layer network coding scheme, SNC. With the routing/scheduling strategy and the corresponding throughput analysis in [97], it was shown that TS and SNC can achieve the following throughputs, respectively:

$$\lambda_{TS}(N) = \frac{4}{(1+\Delta)\sqrt{N}} \cdot \frac{1}{3} \cdot \frac{1}{2} = \frac{2}{9\sqrt{N}}$$
$$\lambda_{SNC}(N) = \frac{4}{(1+\Delta/2)\sqrt{N}} \cdot \frac{1}{3} \cdot \frac{1}{2} = \frac{1}{3\sqrt{N}} \tag{45}$$

where $\Delta+1$ is the distance between the receiver and the nearest interfering node. Note that the throughput in (45) is obtained under the protocol interference model [95]; under the physical interference model (as in our analysis for PNC above), the throughputs of TS and SNC could be worst than in (45). Thus, here we are giving TS and SNC an advantage when comparing them to PNC; yet they compare unfavorably with PNC, as detailed in the next paragraph.

Due to the different scheduling scheme in PNC, there is a factor $J$ in Theorem 3 that is different from $\Delta$. But $J$ and $\Delta$ do play a similar role in that they impose a separation requirement between nodes that transmit simultaneously in order to meet an SIR target. For the 10dB SIR threshold, we need to set $\Delta = 2$ in (45) and $J = 3$ in (44). We can then conclude that PNC can achieve a throughput improvement factor of 3 and 2 relative to the traditional transmission scheme and the SNC scheme, respectively. We remark that the improvement factors under the 2-D network are larger than those under the 1-D network, which are 2 and 1.5, respectively.

**Topology 4: Random 2-D Network**

The idea behind the analysis of the random 2-D network is similar to that of the random 1-D network. We first divide the region into small grids of size $\frac{\log\sqrt{N}}{\sqrt{N}} \times \frac{\log\sqrt{N}}{\sqrt{N}}$ and then divide each small grid into subgrids of size $\frac{\log\log\sqrt{N}}{\sqrt{N}} \times \frac{\log\log\sqrt{N}}{\sqrt{N}}$. We then select one node in the middle subgrid as a routing node and we can prove that all the routing nodes form a regular 2-D network asymptotically.

The transmission is divided into two phases. The first phase is dedicated to the local transmission within each small grid, where one-hop transmissions are adopted. The second phase is dedicated to the transmission among routing nodes, which uses the strategy as discussed in the 2-D regular network. As with the random 1-D network, it can be proved that the time used for the first phase is negligible compared with that used in the second phase. The set-up in the second phase is such that it is equivalent to a regular 2-D network with $\frac{\sqrt{N}}{\log\sqrt{N}} \times \frac{\sqrt{N}}{\log\sqrt{N}}$ nodes and $N/2$ flows in total. Similar to the argument in *Theorem 1'*, we randomly divide all the flows into $\lceil \log^2\sqrt{N} \rceil$ groups and the traffic of the flows are transmitted group by group. Then we have the following theorem:

*Theorem 4*: With PNC, and with the source and destination nodes randomly selected from the $N$ nodes with equal probability, the per-flow throughput of the 2-D *random network* can approach the following for large $N$ with high probability:

$$\lambda_{PNC}(N) = \frac{4}{\sqrt{N}} \cdot \frac{1}{J} \cdot \frac{1}{2} = \frac{2}{J\sqrt{N}} \tag{46}$$

where $J$ is determined by the SIR threshold. A typical value of $J$ is 3 (under an SIR threshold of 10dB and path-loss exponent of 4).

We remark that the result that $\lambda_{PNC}(N) = \Theta(1/\sqrt{N})$ in Theorem 4 is consistent with the prior results on the non-PNC system making use of traditional multihopping technique. The seminal paper [95] established that the throughput per flow in the traditional wireless network subject to SIR constraint is $O(1/\sqrt{N})$. The paper, however, only demonstrated that throughput of $\Omega(1/\sqrt{N\log N})$ is achievable when the sources and destinations are randomly placed. The gap between the upper bound and lower bound was closed by [96], which provides a hierarchical routing scheme (the hierarchical scheme we assume here is similar in spirit to that scheme although not exactly the same) that can achieve $\Omega(1/\sqrt{N})$ throughput. Thus, the throughput per flow is also $\Theta(1/\sqrt{N})$ in the traditional wireless network. The advantage of PNC lies in the smaller constant factor to $1/\sqrt{N}$ rather than order improvement.

## 5.2. Practical Protocol Design

The transmission strategy in Section 5.1 provides much insight on the application of PNC in wireless networks. For example, selecting the routing nodes (relays) for forwarding is analogous to the hierarchical routing strategy in mesh networking in which cluster heads are responsible for forwarding traffic between different clusters of nodes.

The centralized TDMA MAC protocol in Section 5.l, however, may not be practical in two settings. First, when the number of nodes $N$ is large, the complexity of centralized routing and scheduling may become unmanageable. Second, in many practical scenarios, the traffic from the flows is not constant and may be bursty in nature.

In this section, we consider a distributed MAC similar to that in IEEE 802.11 to coordinate transmissions in a distributed manner. We borrow the protocol ideas in [99] and the synchronization ideas in [13] to present an opportunistic protocol based on 802.11 for applying PNC in WiFi access network.

Consider the simple two-hop relay network in WLAN as shown in Fig. 25. A cluster of clients are connected to the AP via a wireless relay. We assume that traffic may be generated by the AP or the clients in an unpredictable manner. Thus, the initialization of a transmission may come from the AP or a client. To make use of PNC, whenever possible we would like to merge two transmissions in opposing directions together into an overlapped transmission. Thus, before a node (the AP or a client) begins transmitting a full-length DATA packet, it will send a probe in the form of an RTS (request-to-send) to the relay to look for an opportunity to merge the transmission of the DATA packet with a DATA packet in the opposite direction.

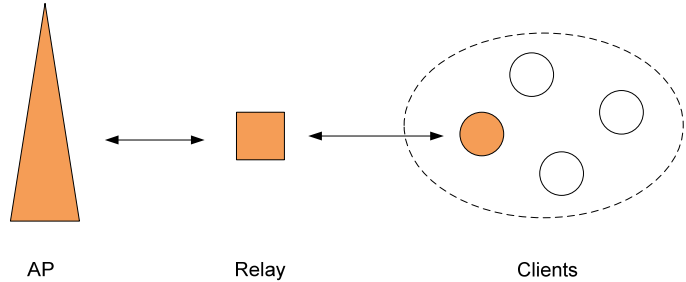

Figure 25. A relay network where a cluster of clients are connect to the AP via a relay.

With reference to Fig. 25, suppose that the AP and the colored (filled) client have a packet to transmit to each other. A sequence of events in our opportunistic protocol is depicted in Fig. 26 and described as follows. With a MAC protocol akin to IEEE 802.11, either the AP or the client may initiate the RTS first. In Fig. 26, we assume it is the AP that initiates the RTS. The ID of the final destination (in this case, the address of the colored client) is embedded in the RTS. Upon receiving the RTS, the relay broadcasts a CTS (clear to send) packet as in the protocol used in IEEE 802.11. Within the CTS, in addition to the destination ID of the client, a time stamp is also inserted by the relay. The time stamp contains the time instant at which the CTS is transmitted.

As soon as the AP receives the CTS from the relay, it first calculates the propagation time (from the relay to the AP) by subtracting the time stamp in the CTS from the receiving time. Here, we assume a global time reference among all the nodes, which can be realized with the help of GPS. We denote this propagation time by $d_1$. Then the AP sends its DATA packet to the relay with a delay of $T - d_1$ after the CTS is received, where $T$ is a value larger than any possible propagation delay. For example, $T$ could be the SIFS used in 802.11 [100] plus some constant value. At the same time, when the target client receives the CTS from the relay, it can calculate the propagation time from the relay to itself, $d_2$. Since the client also has a packet for the AP in our example, it sends its DATA packet to the relay with a delay of $T - d_2$. As a result, the two packets should both arrive at the relay together. In general, if OFDM PNC is used [31], very tight synchronization is not needed and difference between the delay of the AP and the delay of the client needs to be within the cyclic prefix (CP) of the OFDM system only.

When the data transmission is finished, the relay can network-code the two packets to obtain the new network -coded packet. If the network-coded packet is correctly obtained, the relay broadcasts it to the AP and the client after a SIFS delay.

After correctly receiving the network-coded packet, both the AP and the client send an ACK packet to the relay with a delay of $T - d_1$ and $T - d_2$. Since the two ACK packets will arrive in a synchronized way, the relay can check whether it receives the supposition of the two ACK packets. If so, the exchange of one packet is finished. Note that the overlapped ACKs could be detected using the MUD technique or by detecting/correlating some unique signatures pre-allocated to the AP and the client that are embedded into the preambles of the ACKs.

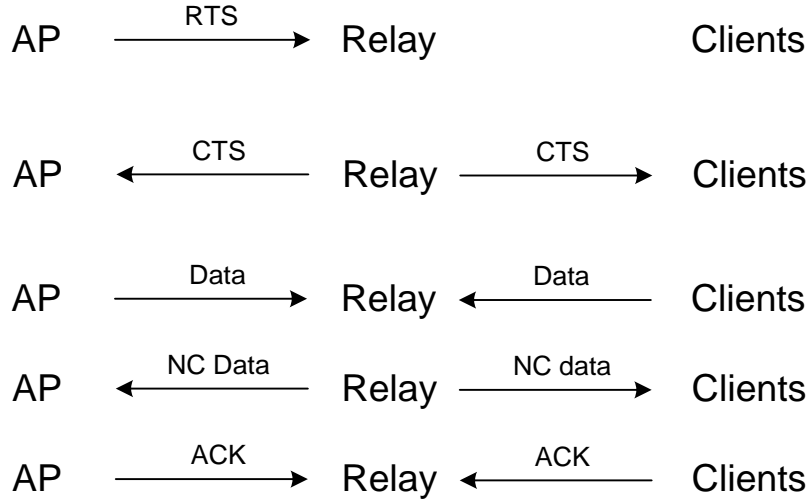

Figure 26. An opportunistic transmission protocol for PNC.

The above protocol is an opportunistic protocol in that a bidirectional PNC transmission will be used only when there is an opportunity to do so. If only the AP has a packet for the target client, ordinary one-way transmission will be adopted.

The protocol could be further improved by allowing a different client to transmit to the AP while the AP transmits to its target client. That is, if the AP's target client does not have a packet for the AP, but a different client has a packet for the AP, then this other client can also try to combine its transmission with the transmission of the AP to the relay. This feature requires the clients within the region to be able to overhear each other so that they can learn about each other's self information. There are also details to work out, including the possibility of two clients having packets to the AP. The coordination is an interesting topic for future research.

### 5.3. To Probe Further

The asymptotic performance of PNC in large 2-D wireless networks under random unicast traffic was given in [94]. Unlike our treatment here which assumes the more realistic physical interference model for the PNC analysis, the pairwise protocol interference model was assumed in [94].

In [101], the physical interference model was adopted with a highway system as in [96]. The highway system we consider in this paper is slightly different in that the routing nodes (nodes in the highway) form a 2-D regular network while that in [96] is not regular. The asymptotic throughput per flow obtained in [101] is somewhat inferior to what we have here, although they are of the same order.

In [102], the broadcast setting is analyzed for 2-D regular networks. Specifically, the times needed to broadcast $b$ blocks of data from one source node to all other nodes in the 2-D grid and hexagonal networks were analyzed, and broadcast throughput with PNC is 2.5 times that the traditional multihop scheme in the grid network, and 2 times in the hexagonal network. The result also implies that PNC can approach the broadcast throughput upper bound under the half-duplex constraint in which a node can be transmitting at most half the time when relaying the broadcast information.

Ref. [103] showed that PNC can not improve the complexity of scheduling wireless links under the physical interference model, i.e., the complexity is still NP complete as in the scheduling problem in traditional multihop network.

Ref. [104] formulated an optimization problem to solve the scheduling problem in PNC relay networks, assuming unicast traffic. Formulation in a multichannel network and multiradio wireless nodes was considered in [105].

### 6. Optical PNC

PNC was originally proposed for application in wireless networks. It is based on the observation that network coding operation is implicit in many natural phenomena. Whenever two quantities in nature, $x$ and $y$, meet to produce a third quantity, $z = f(x, y)$, a form of network coding operation occurs. For example, $z$ could be the amplitude of an EM wave when $x$ and $y$ are amplitudes of coherent EM waves; or $z$ could be sum power of two EM waves when $x$ and $y$ are the powers of noncoherent EM waves. In general, $x$ and $y$ could be other physical quantities, including acoustic waves.

We end this paper by proposing to apply PNC in the optical domain. Since light is also a form of EM wave, PNC for lightwave communication is just a small step from PNC for wireless communication. For fiber-optic communication, the channel gain is more stable and the issue of fading is not a main concern. Also, full duplexity can be more readily implemented by isolating the transmitted signal from the received signal through two different optical fibers, or on the same fiber through a directional coupler at the transceiver. Conceivably, it could even be easier to realize PNC in fiber-optic communication than in wireless communication, especially for noncoherent optical systems. We provide an example of optical PNC in this section.

The passive optical network (PON) [106] is a network architecture of much interest in the optical communication community. PONs have also been commercially deployed in the field. A PON consists of an optical line terminal (OLT) at the central office, a passive optical splitter-combiner, and a number of optical network units (ONUs) at or near the end users' premises. Downstream signals are broadcasted by the OLT to the ONU, and each ONU filters out all the signals except its own signal; encryption can be used at the upper layer to prevent eavedropping. Upstream signals from the ONUs to the OLT make use of a multiple access protocol for access of the shared medium. The time-division-multiple-access (TDMA) protocol is popular protocol being used.

For generality, we consider here the star topology as shown in Fig. 27. Here, node 1 could be the OLT, and nodes 2 to N could be the ONUs. More generally, for our purpose here, we do not assign specific roles to the nodes. We simply have a system in which there are $N$ nodes wanting to exchange information with each other. Each node is connected to the splitter-combiner through an output fiber and an input fiber. In principle, the input and output fibers could also be the same physical fiber with signals travelling in opposing direction; a directional coupler us inserted at a node to isolate the transmitted and received signals [8].

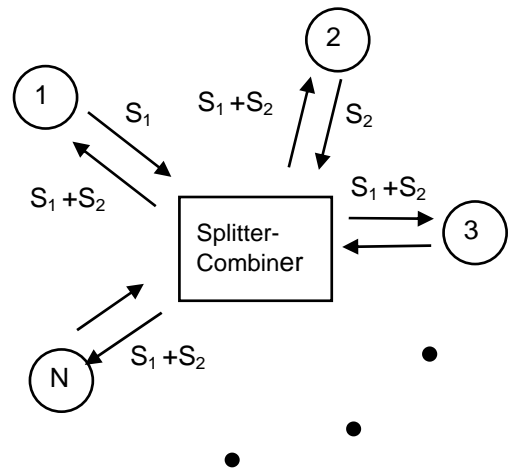

Fig. 27. Optical PNC for application in passive optical network.

---

[8] In the single-fiber system, each node is connected only to an input of the splitter-combiner (star coupler). A loop-back path is provided by interconnecting two outputs of the splitter-combiner, say outputs 1 and 2. The signal on one output is looped back from the other output for broadcast to all the inputs of the splitter-combiner. In this way, the transmitted and received signals of a node travel on the same fiber in opposite directions.

At the splitter-combiner, the signals from its inputs are combined and broadcasted to all the nodes. For example, if node 1 sends $S_1$ and node 2 sends $S_2$, all the nodes receive $S_1 + S_2$. We see immediately that PNC has a role to play here. In particular, it allows full-duplex communication between two nodes in the network at any given time. Suppose that $S_1$ is targeted for node 2, and $S_2$ is targeted for node 1. With the pure TDMA system, nodes 1 and 2 will send $S_1$ and $S_2$ in two different time slots. With PNC, nodes 1 and 2 will send $S_1$ and $S_2$ together, and they use self information to extract $S_2$ and $S_1$, respectively, from $S_1 + S_2$. The potential throughput increase, as in wireless, is 100%.

Compared with wireless communication, a subtlety is as follows. In wireless communication, half-duplex constraint is often imposed. This is because the signal being transmitted is of much higher power than the signal being received if simultaneous transmission and reception are allowed, and it is not easy to extract the received signal in practice. For the optical star, the transmitted signal at a node is isolated from the received signal through two different optical fibers (or a directional coupler on a single fiber). For node 1, in the received composite signal $S_1 + S_2$, the power of $S_1$ may not be much larger than that of $S_2$. This makes it easier to implement full-duplexity in the optical star network.

When the traffic is bursty and unpredictable, the use of TDMA is not efficient. A random multiple-access protocol (e.g., carrier-sense-multi-access) can be used to coordinate the access of the shared medium among the nodes. A MAC protocol for bidirectional communication that employs PNC can be as follows. Suppose that node 1 acquires the channel and sends a burst of data to node 2, and this burst triggers a burst of data from node 2 to node 1 in the reverse direction (many real-world applications are bidirectional in that when there is data in one direction, there is also data in the reverse direction at the same time). Without PNC, the two bursts must acquire the channel separately using the random multiple-access protocol. With PNC, once node 1 acquires the shared medium and sends a data burst to node 2, node 2 will detect that it is the recipient of the data; and if node 2 also has data for node 1, it can simultaneously send the data burst to node 1. Thus, channel acquisition is bidirectional.

The observant reader may notice wavelength-division-multiple-access (WDMA) could achieve the same effect. Specifically, the signal from node 1 to node 2 and the signal from node 2 to node 1 can be carried on two different wavelengths. Thus, simultaneous bidirectional communication is also possible with WDMA. However, if multiple wavelengths were available, PNC could still achieve better throughput. Instead of using the two wavelengths for one bidirectional flow; the two wavelengths could be used for two bidirectional flows instead, hence doubling the throughput.

While there have been many works on wireless PNC since [1] [1], to our knowledge, there has not been any work on optical PNC. This is a first proposal suggesting that PNC can be extended to the optical domain. We have considered a rather simple optical network. There could be other optical scenarios in which PNC is useful. The optical domain appears to be a futile ground for future PNC research.

## 7. Conclusions

In this paper, we have introduced PNC and overviewed some recent research results. The recent works are categorized into three domains: 1) wireless communication, 2) wireless information theory, and 3) wireless networking. Within the three domains, we have further grouped works into various sub-domains. We believe researchers in the field will find our survey and categorization useful as a reference in their future investigations. We have attempted to list and categorize the major works in the area. Despite that, it is likely that we have also missed some important works.

To date, most works on PNC have focused on two-way relay channel (TWRC). The theoretical understanding of TWRC is maturing. An unresolved issue is the information-capacity region of TWRC. The use of multiuser detection (MUD) technique could approach the capacity at the low SNR regime, and the use of nested lattice code and the corresponding detection technique could approach the capacity at the high SNR regime[9]. Neither technique, however, gives the ultimate information capacity of PNC TWRC for all SNR. As outstanding challenge is the information capacity of TWRC and an efficient implementation to achieve it.

Beyond TWRC, there have also been many investigations on its extension to the multi-way relay channel (MWRC) in which a relay (or a system of relays) interconnect more than two end nodes. We believe the theoretical understanding of MWRC will also be maturing shortly.

In both TWRC and MWRC, there are at most two hops between two communicating end nodes. In a general multi-hop network, MAC-layer and network-layer issues will take on an increasingly important role, particularly with regard to complexity management when there are many simultaneous flows in the network. We have outlined the potential use of the concept of virtual paths in PNC to manage complexity at the network layer. We have also briefly discussed MAC-level issues by means of an MAC scheduling protocol. By and large, these have been high-level discussions and many details remain to be ironed out. Compared with the large volume of communication-theoretic and information-theoretic PNC investigations, there have been far fewer works with a networking flavor. A reason could be that the problem formulations at the MAC and network layer are not as clear cut as those at the lower layers. There are many outstanding challenges at the higher layers.

Another futile ground for future PNC research is implementation and prototyping. Besides [8], we are not aware of successful implementations of PNC. There is a gap between theory and implementation at the moment. We believe that beyond the simple amplify-and-forward ANC scheme in [8], PNC systems with better performance can also be demonstrated, and worthwhile issues for further investigations can be identified through the prototyping efforts.

Finally, although PNC was originally conceived for application in wireless networks, network coding operations abound in nature. In fact, any physical phenomenon in which an effect (output) is the outcome (function) of a number of causes (inputs) can be exploited in the network coding construct. In this paper, we have attempted to extend the application of PNC to optical networks. The application of PNC could potentially be extended to many other domains.

## Acknowledgements

This work was partially supported by grants from the University Grants Committee of Hong Kong Special Administrative Region, China (Project No. AoE/E-02/08 and Project No. 414507). This work was also partially supported by NSF of China (Project No. 60902016) and NSF of Guangdong (Project No. 10151806001000003). We thank Raymond Yeung and Silas Fong for their valuable suggestions and comments that remove some errors and improve the presentation of the results in this paper.

## References

[1] S. Zhang, S. C. Liew, P. P. Lam, "Hot Topic: Physical-layer Network Coding," *ACM MobiCom '06,* pp. 358-365, Sept. 2006.
[2] R. Ahlswede, N. Cai, S.-Y. R. Li, R. W. Yeung, "Network Information Flow," *IEEE Trans. Inform. Theory*, vol. 46, no. 4, pp. 1204-1216, July 2000.

---

[9] Although [16] showed that the use of nested lattice code could approach the cut-set bound in Gaussian channel to within 1/2 bit per channel use, the 1/2 bit as a fraction of the information capacity is still quite significant in the low SNR regime.

As mentioned in the main body of this paper, the detection technique at the relay in [16] (and, as far as we know, all other papers that study lattice-code PNC) consists of two parts. It first maps the superimposed codewords from the two end nodes into a codeword of lower cardinality compared with the superimposed codewords. It then channel-decodes the latter using a standard lattice decoder. The mapping in the first part is done before channel decoding and it loses useful information. As a result, the scheme is suboptimal in the low SNR regime. This loss of useful information could be illustrated and understood intuitively by examining the QPSK example in Section 3.2. In Fig. 12, the mapping of the superimposed channel-coded symbols into XOR channel-coded symbols before channel decoding causes suboptimal performance of *XOR-CD* in Fig. 12. By not separating the channel decoding from PNC mapping, *Joint CNC* in Fig. 13 could achieve better performance. However, the complexity of the design in *Joint CNC* is high when higher-order constellations than QPSK are used. An outstanding issue is whether there are efficient implementations that achieve similar performance as *Joint CNC*.

[3] S.-Y. R. Li, R.W. Yeung, N. Cai, "Linear Network Coding," *IEEE Trans. Inform. Theory*, vol. 49, no.2, pp. 1204-1216, Feb. 2003.

[4] P. Popovski, H. Yomo, "The Anti-packets Can Increase the Achievable Throughput of a Wireless Multi-hop Network," *IEEE ICC '06*, pp. 3885-3890, June 2006.

[5] B. Nazer, M. Gastpar, "Computing over Multi-access Channels with Connections to Wireless Network Coding," *IEEE ISIT '06*, pp. 1354-1358, July 2006.

[6] Y. Wu, P. A. Chou, and S. Y. Kung, "Information Exchange in Wireless Networks with Network Coding and Physical Layer Broadcast," *Proc. 39th Annual Conf. Inform. Sci. and Systems (CISS)*, 2005.

[7] C. Fragouli, J. Y. Boudec, J. Widmer, "Network Coding: An Instant Primer," *ACM SIGCOMM Computer Communication Review*, pp. 63-68, vol. 36, no. 1, Jan. 2006.

[8] S. Katti, S. S. Gollakota, D. Katabi, "Embracing Wireless Interference: Analog network Coding," *ACM SIGCOMM '07*, pp. 397-408, Aug. 2007.

[9] M. Dankberg, "Paired Carrier Multiple Access (PCMA) for Satellite Communications" *Pacific Telecommunications Conference*, Jan. 1998.

[10] T. Koike-Akino, P. Popovski, V. Tarokh, "Denoising Maps and Constellations for Wireless Network Coding in Two–way Relaying Systems," *IEEE Globecom '08*, Nov. 2008.

[11] T. Koike-Akino, P. Popovski, V. Tarokh, "Optimized Constellations for Two-way Wireless Relaying with Physical Network Coding," *IEEE J. Select. Areas Commun.*, vol. 27, no. 5, pp. 773-787, June 2009.

[12] S. Zhang, S. C. Liew, Lu Lu, "Physical Layer Network Coding Schemes over Finite and Infinite Fields," *IEEE Globecom '08*, Nov. 2008.

[13] H. Rahul, H. Hassanieh, D. Katabi "SourceSync: A Cooperative Wireless Architecture for Exploiting Sender Diversity," *ACM SIGCOMM '10*, pp. 171-182, Aug. 2010.

[14] T. M. Cover, J. A. Thomas, *Elements of Information Theory*, $2^{nd}$ *Edition*, Wiley 2006.

[15] R. Knopp, "Two-Way Wireless Communication via Relay Station," *GDR-ISIS Meeting*, ENST, Paris, Mar. 2007.

[16] W. Nam, S.Y. Chung, Y.H. Lee "Capacity of the Gaussian Two-way Relay Channel to within 1/2 bit," *IEEE Trans. Info. Theory*, vol. 56, no. 11, pp. 5488-5495, Nov. 2010.

[17] T. T. Lee, S. C. Liew, *Principles of Broadband Switching and Networking*, Wiley 2010.

[18] Y. Hao, D. Goeckel, Z. Ding, D. Towsley, K. K. Leung, "Achievable Rates for Network Coding on the Exchange Channel," *IEEE MILCOM '07*, Oct. 2007.

[19] T. Cui, T. Ho, and J. Kliewer, "Relay Strategies for Memoryless Two-way Relay Channels: Performance Analysis and Optimization", *IEEE ICC'08*, pp. 1139-1143, May 2008.

[20] K. S. Gomadam, S. A. Jafar, "Optimal Relay Functionality for SNR Maximization in Memoryless Relay Networks", *IEEE J. of Select. Areas Commun.,* vol. 27, no. 2*,* pp. 390-401, Feb. 2007.

[21] S. Zhang, S. C. Liew, P. P. Lam, "Physical-layer Network Coding," http://arxiv.org/abs/0704.2475, Apr. 2007.

[22] S. Zhang, S. C. Liew, P. P. Lam, "On the Synchronization of Physical-layer Network Coding," *IEEE ITW '06*, pp. 404-408, Oct. 2006.

[23] J. S. Yedidia, W.T. Freeman, Y. Weiss, "Understanding Belief Propagation and Its Generalizations," *Technical Report TR2001-22*, MERL, 2001.

[24] L. Lu, S. C. Liew, S. Zhang, "Optimal Decoding Algorithm for Asynchronous Physical-Layer Network Coding," *IEEE ICC '11*, June 2011.

[25] S. Zhang, S. C. Liew, "Channel Coding and Decoding in a Relay System Operated with Physical-Layer Network Coding," *IEEE J. Select. Areas Commun.* vol. 27, no. 5, pp. 788-796, June 2009.

[26] M. C. Reed, C. B. Schlegel, P. D. Alexander, J. A. Asenstorfer, "Iterative Multiuser Detection for CDMA with FEC: Near-Single-User Performance," *IEEE Trans. Commun.*, vol. 46, no. 12, pp. 1693-1699, Dec. 1998.

[27] X. Wang, H. V. Poor, "Iterative (Turbo) Soft Interference Cancellation and Decoding for Coded CDMA*", IEEE Trans. Commun.*, vol. 47, no. 7, pp. 1046-1061, July 1999.

[28] S. Zhang, S. C. Liew, H. Wang, X. Lin, "Capacity of Two-Way Relay Channel," *Access Network,* Springer*,* vol. 37, Nov. 2009.

[29] B. Nazer, and M. Gastpar, "Computation over Multiple-access Channels," *IEEE Trans. Inform. Theory*, vol. 53, no. 10, pp. 3498-3516, Oct. 2007.
[30] P. Popovski, H. Yomo, "Physical Network Coding in Two-way Wireless Relay Network," *IEEE ICC '07*, pp. 707-712, June 2007.
[31] F. Rossetto, M. Zorzi "On the Design of Practical Asynchronous Physical Layer network coding", *IEEE SPAWC '09*, pp. 469-473, June 2009.
[32] K. Narayanan, M. P. Wilson, A. Sprintson "Joint Physical Layer Coding and Network Coding for Bi-directional Relaying," *Annual Allerton Conf. '07*, pp. 254-259 Sept. 2007.
[33] B. Nazer, M. Gastpar, "Lattice Coding Increases Multicast Rates for Gaussian Multiple-access Networks," *Annual Allerton Conf. '07*, pp. 1089-1096, Sept. 2007.
[34] L. Lu, S. C. Liew, "Asynchronous Physical-layer Network Coding," manuscript under preparation.
[35] D. To, J. Choi, "Convolutional Codes in Two-way Relay Networks with Physical-layer Network Coding*,*" *IEEE Trans. Wireless Commun.*, vol. 9, no. 9, pp. 2724-2729, Sept. 2010.
[36] D. To, J. Choi, "Reduced-State Decoding with Two-way Relay Networks with Physical-layer Network Coding," *IEEE ITW '10*, Aug. 2010.
[37] D. Wubben, Y. Lang, "Generalized Sum-product Algorithm for joint Channel Decoding and Physical-Layer Network Coding in Two-way Relay System," *IEEE Globecom '10*, Dec. 2010.
[38] S. Zhang, S. C. Liew, Q. Zhou, L. Lu, W. Wang, "Non-memoryless Analog Network Coding in Two-Way Relay Channel, " *IEEE ICC '11*, June 2011.
[39] J. Liu, M. Tao, Y, Xu, X. Wang, "Superimposed XOR: A New Physical Layer Network Coding Scheme for Two-way Relay Channels," *IEEE Globecom '09*, Nov. 2009.
[40] D. Wang, S. Fu, K. Lu, "Channel Coding Design to Support Asynchronous Physical Layer Network Coding," *IEEE Globecom '09*, Nov. 2009.
[41] S. Zhang, S. C. Liew, "Physical Layer Network Coding with Multiple Antennas," *IEEE WCNC '10*, Apr. 2010.
[42] Z. Zhou, B. Vucetic, "An Optimized Network Coding Scheme in Two-way Relay Channels with Multiple Relays," *IEEE PIMRC '09*, pp. 1717-1721, Sept. 2009.
[43] D. To, J. Choi, I. M. Lim, "Error Probability Analysis of Bidirectional Relay Systems using Alamouti Scheme," *IEEE Comm. Letters*, vol. 14, no. 8, pp. 758-760, Aug. 2010.
[44] N. Xu, S. Fu, "On the Performance of Two-way Relay Channels using Space–time Codes," *Int. J. Commun. Syst.*, Jan. 2011.
[45] R. Zhang, Y. Liang, C. C. Chai, "Optimal Beamforming for Two-Way *Multi-Antenna Relay* Channel with Analogue Network Coding," *IEEE J. Select. Areas Commun.*, vol. 27, no. 5, pp. 699-712, June 2009.
[46] R. Annavajjala, A. Maaref, J. Zhang, "Multiantenna Analog Network Coding for Multihop Wireless Networks," *Int. J. Digital Multimedia Broadcasting*, vol. 2010, Article ID 368562, 10 pages, 2010.
[47] Z. Ding, I. Krikidis, J. Thompson, K. K. Leung, "Physical-layer Network Coding and Precoding for the Two-way Relay Channel," *IEEE Trans. Signal Processing*, vol. 59, no. 2, pp. 696-712, Feb. 2011.
[48] N. Ning, C. Ling, K. Leunng, "Active Physical-layer Network Coding for Cooperative Two-way Relay Channels," *IEEE SECON Workshop '09*, June 2009.
[49] S. A. K. Tanoli, I. Khan, N. Rajatheva, F. Atachi, "Advances in Relay Networks: Performance and Capacity Analysis of Space-time Analog Network Coding," *Eurasip J. Wirel. Commun. Netw.*, vol. 2010, Article ID 232754, 10 pages, 2010
[50] Q. F. Zhou, Y. Li, F. C. M. Lau, B. Vucetic, "Decode-and-Forward Two-way Relaying with Network Coding and Opportunistic Relay Selection, " *IEEE Trans. Commun.,* vol. 58, no. 11, pp. 3070-3076, Nov. 2010.
[51] Z. Yi, I. M. Kim, "Optimum Beamforming in the Broadcasting Phase of Bidirectional Cooperative Communication with Multiple Decode-and-forward Relays," *IEEE Trans. Wireless Commun.*, vol. 8, no. 10, pp. 5806, 5812, Dec. 2009.
[52] P. Hu, C. W. Sung, K. W. Shum, "Joint Channel-network Coding for the Gaussian Two-way Two-relay Network," *Eurasip J. Wirel. Commun. Netw.*, vol. 2010, Article ID 708416, 13 pages, 2010.

[53] B. Rankov, A. Wittneben, "Spectral Efficient Protocols for Half-duplex Fading Relay Channels", *IEEE J. Select. Areas Commun.*, vol. 25, no. 2, , pp. 379-389, Feb. 2007.
[54] S. Kim, J. Chun, "Network Coding with MIMO Pre-equalizer using Modulo in Two-way Channel," *IEEE WCNC '08*, Mar. 2008.
[55] L. Song, G. Hong, B. Jiao, M. Debbah, "Joint Relay Selection and Analog Network Coding using Differential Modulation in Two-way Relay Channels," *IEEE Trans Vehicular Tech.,*, vol. 59, no. 6, pp. 3070-3076, July 2010.
[56] F. Gao, R. Zhang, Y. C. Liang, "Optimal Channel Estimation and Training Design for Two-way Relay Networks," *IEEE Trans. Comun.*, vol. 57, no. 10, Oct. 2009.
[57] G. Wang, F. Gao, C. Tellambura, "Joint Frequency Offset and Channel Estimation Methods for Two-way Relay Networks," *IEEE Globecom '09*, Nov. 2009.
[58] B. Jiang, G. Gao, X. Gao, "Channel Estimation and Training Design for Two-way Relay Networks with Power Allocation," *IEEE Trans. Wireless Commun.*, vol. 57, no. 10, pp. 2022-2032, June 2010.
[59] F. Gao, R. Zhang, Y. C. Liang, "Channel Estimation for OFDM Modulated Two-way Relay Networks," *IEEE Trans. Signal Processing*, vol. 57, no. 11, pp. 4443-4455, No. 2009.
[60] T. Sjodin, G. Gacacin, F. Adachi, "Two-slot Channel Estimation for Analog Network Coding based on OFDM in a Frequency-selective Fading Channel," *IEEE VTC Spring 2010*, May 2010.
[61] W. Yang, Y. Cai, J. Hu, W. Yang, "Channel Estimation for Two-way Relay OFDM Networks," *Eurasip J. Wirel. Commun. Netw.,* vol. 2010, Article ID 186182 , 6 pages, 2010.
[62] X. Liao, L. Fan, F. Gao, "Blind Channel Estimation for OFDM Modulated Two-way Relay Network," *IEEE WCNC '10*, Apr. 2010.
[63] H. Gacanin, F. Adachi, "The Performance of Network Coding at the Physical Layer with Imperfect Self Information Removal," *Eurasip J. Wirel. Commun. Netw*, Article ID 659291 , 8 pages, 2010.
[64] M. Chen, A. Yener, "Multiuser Two-way Relaying: Detection and Interference Management Strategies," *IEEE Trans. Wireless Commun.*, vol. 9, no. 8, pp. 4296-4305, Aug. 2009.
[65] N. Lee, J. B. Lim, J. Chun, "Degrees of Freedom of MIMO Y Channel: Signal Space Alignment for Network Coding," *IEEE Trans. Info. Theory*, vol. 56, no. 7, pp. 3332-3342, July 2010.
[66] Y. E. Sagduyu, D. Guo, R. Berry, "Throughput Optimal Control for Relay-assisted Wireless Broadcast with Network Coding," *IEEE SECON Workshop '08*, June 2008.
[67] Y. E. Sagduyu, D. Guo, R. Berry, "On the Delay and Throughput of Digital and Analog Network Coding for Wireless Broadcast," *CISS '08*, pp. 534-539, Mar. 2008.
[68] F. Gao, T. Cui, B, Jiang, X, Gao, "On Communication Protocol and Beamforming Design for Amplify-and-Forward N-way Relay Networks," *IEEE Int.Workshop on Computational Advances in Multi-sensor Adaptive Processing*, pp. 109-112, Dec. 2009.
[69] A. U. T. Amah, A. Klein, "Beamforming-based Physical-layer Network Coding for Non-regenerative Multi-way Relaying," *Eurasip J. Wirel. Commun. Netw.*, vol. 2010, 12 pages, Jan. 2010.
[70] S. Zhang, Y. Zhu, S. C. Liew, "Soft Network Coding in Wireless Two-Way Relay Channels," *J. Communication and Networks*, *Special Issues on Network Coding*, vol. 10, no. 4, Dec. 2008.
[71] T. Cui, T. Ho, and J. Kliewer, "Memoryless Relay Strategies for Two-way Relay Channels," *IEEE Trans. Commun.*, vol. 57, no. 10, pp. 3132-3143, Oct. 2009.
[72] C. H. Ho, R. Zhang, Y. C. Liang, "Two-way Relaying over OFDM: Optimized Tone Permutation and Power Allocation," *IEEE ICC '08*, pp. 3908-3912, May 2008.
[73] H. Gacanin, F. Adachin "Broadband Analog Network Coding," *IEEE Tran. Wireless Commun.*, vol. 9, no. 5, pp. 1577-1583, May 2010.
[74] W. Shin, N. Lee, J. B. Lim, C. Shin, "An Optical Transmit Power Allocation for Two-way Relay Channel using Physical-layer Network Coding," *IEEE ICC '09,* June 2009.
[75] C. Li, S. He, L. Yang, W. P. Zhu, "Joint Power Allocation for Multicast Systems with Physical-layer Network Coding," *Eurasip J. Wirel. Commun. Netw.,* vol. 2010, Article ID 423234, 9 pages, 2010.
[76] M. Zhao, Y. Zhou, D. Ren, Yixian Yang, "A Minimum Power Consumption Scheme for Two-way Relay with Physical-layer Network Coding,"$2^{nd}$ *IEEE Int. Conf. Network Infrastructure and Digital Content*, pp. 704-708, Sept. 2010.

[77] T. Cui, F. Gao, C. Tellambura, "Differential Modulation for Two-way Wireless Communications: A Perspective of Differential Network Coding at the Physical Layer," *IEEE Trans. Commun.*, vol. 57, no. 10, pp. 2977- 2987, Oct. 2009.

[78] M. C. Valenti, D. Torrieri, T. Ferrett, "Noncoherent Physical-layer Network Coding using Binary CPFSK Modulation," *IEEE MILCOM '09*, Oct. 2009.

[79] L. Song, Y. Li, A. Huang, B. Jiao, A. Vasilakos, "Differential Modulation for Bidirectional Relaying with Analog Network Coding," *IEEE Trans. Signal Processing*, vol. 58, no. 7, pp. 3933-3938, July 2010.

[80] J. H. Sorenson, R. Krigslund, P. Popovski, T. K. Akino, T. Larsen, "Physical Layer Network Coding for FSK Systems," *IEEE Commun Lett.*, vol 13, no. 8, pp. 597-599, Aug. 2009.

[81] K. Lu, S. Fu, Y. Qian, T. Zhang, "On the Security Performance of Physical-layer Network Coding," *IEEE ICC '09*, June 2009.

[82] M. Hay, B. Saeed, C. H. Lung, A. Sribuvasab, "Co-located Physical-layer Network Coding to Mitigate Passive Eavedropping," *8$^{th}$ Ann. Int. Conf. Privacy, Security, and Trust*, Aug. 2010.

[83] W. Wang, D. Pu, A. Wyglinski, "Detecting Sybil Nodes in Wireless Networks with Physical Layer Network Coding," *IEEE/IFIP Int. Conf. Dependable Systems and Networks (DSN)*, pp. 21-30, June 2010.

[84] C. Schnurr, T. J. Oechtering, S. Stanczak, "Achievable Rates for the Restricted Half-duplex Two-way Relay Channel," *41st Asilomar Conf. on Signals, Systems and Computers*, pp. 1468-1472, Nov. 2007.

[85] S. Avestimehr, A. Sezgin, and D. Tse, "Approximate Capacity of the Two-way Relay Channel: A Deterministic Approach," *Annual Allerton Conf.*, pp. 1582-1589, Sept. 2008.

[86] S. J. Kim, N. Devroye, P. Mitran, and V. Tarokh, "Achievable Rate Regions for Bi-directional Relaying," http://arxiv.org/abs/0808.0954, Aug. 2008.

[87] T. J. Oechtering, C. Schnurr, S. Stanczak, "Broadcast Capacity Region for Two-phase Bidirectional Relaying," *IEEE Trans. Inform. Theory*, vol. 54, no. 1, pp. 454-458, Jan. 2008.

[88] L. L. Xie, "Network Coding and Random Binning for Multi-user Channels," *IEEE 10$^{th}$ Canadian Workshop on Inform. Theory*, pp. 85-88, June 2007.

[89] S. J. Kim, P. Mitran, V. Tarokh, "Performance Bounds for Bidirectional Coded Cooperation Protocols," *IEEE Trans. Inform. Theory*, vol. 54, no. 11, pp. 5235-5241, Nov. 2008.

[90] B. Nazer, M. Gastpar, "Reliable Physical Layer Network Coding," *Proceedings of IEEE*, vol. 99, no. 3, pp. 438-460, Mar. 2011.

[91] C. Feng, D. Silva, F. R. Kschischang, "An Algebraic Approach to Physical-layer Network Coding," *IEEE ISIT '10*, pp. 1017-1021, June 2010.

[92] H. J. Yang, Y. Choi, J. Chun, "Modified Higher-order PAMS for Binary-coded Physical-layer Network Coding," *IEEE Commun.. Letters*, vol. 14, no. 8, pp. 689-691, Aug 2010.

[93] S. Zhang, S. Liew, "Applying Physical Layer Network Coding in Wireless Networks," *Eurasip J. Wirel. Commun. Netw*, vol. 2010, Article ID 870268, 12 pages, 2010.

[94] K. Lu, S. Fu, Y. Qian, H.-H. Chen, "On Capacity of Random Wireless Networks with Physical-layer Network Coding," *IEEE J. Select. Areas Commun.*, vol. 27, no. 5, pp. 763–772, 2009.

[95] P. Gupta, P. R. Kumar, "The Capacity of Wireless Networks," *IEEE Trans. Inform. Theory*, vol. 46, no. 2, pp. 388–404, Mar. 2000.

[96] M. Franceschetti, O. Dousse, D.N.C. Tse, P. Thiran,"Closing the Gap in the Capacity of Wireless Networks via Percolation Theory," *IEEE Trans. Inform. Theory*, vol. 53, no. 3, pp. 1009-1018, Mar. 2007.

[97] J. Liu, D. Goeckel, D. Towsley, "Bounds of the Gain of Network Coding and Broadcasting in Wireless Networks," *IEEE Infocom '07,* pp. 724-732, May 2007.

[98] P. Kumar, "A Correction to the Proof of a Lemma in the Capacity of Wireless Networks," *IEEE Trans. Inform. Theory*, vol. 49, p. 3117, Nov. 2003.

[99] S. Boppana, J. Shea, "Impact of Overlapped Transmission on the Performance of TCP in Multihop Ad Hoc Networks," *IEEE MILCOM '08*, Nov. 2008.

[100] IEEE Std 802.11-1997, *IEEE 802.11 Wireless LAN Medium Access Control (MAC) and Physical Layer (PHY) Specifications*.

[101] T. Zhang, K. Lu, A. Jafari, S. Fu, "On the Capacity Bounds of Large-Scale Wireless Network with Physical-Layer Network Coding under the Generalized Physical Model," *IEEE ICC '10*, May 2010.
[102] C. Chen, K. Cai, H. Xiang, "Scalable Ad hoc Networks for Arbitrary-cast: Practical Broadcast-relay Transmission Strategy Leveraging Physical-layer Network Coding," *Eurasip J. Wirel. Commun. Netw*, vol. 2010, Article ID 621703, 15 pages, 2008.
[103] O. Goussevskaia, R. Wattenhofer, "Complexity of Scheduling with Analog Network Coding," *ACM FOWANC '08*, pp. 77-84, 2008.
[104] H. M. Zimmermann, Y. C. Liang, "Physical-layer Network Coding for Uni-cast Applications," *IEEE VTC Spring 2008*, pp. 2291–2295, May 2008.
[105] H. Su, X. Zhang, "Modeling Throughput Gain of Network Coding in Multi-channel Multi-radio Wireless Ad hoc Networks," *IEEE J. Select. Areas Commun. Commun.*, vol. 27, no. 5, pp. 593–605, June 2009.
[106] http://en.wikipedia.org/wiki/Passive_optical_network.